\documentclass[reqno,12pt]{fcaa-var}  

\pdfoutput=1
 \usepackage{amsfonts,amssymb,amsmath,amsthm}
\usepackage{graphicx}
\usepackage{epsfig}

\usepackage{amsthm}
\usepackage{amsmath}
\usepackage{latexsym}
\usepackage{amsfonts}
\usepackage{amssymb}

 \usepackage[a4paper, top=2.5cm,bottom=3.5cm,right=2.5cm,left=2.5cm]{geometry}
\usepackage{tensor} 
\usepackage{color}
\theoremstyle{definition}
\newtheorem{definition}{Definition}[section]
\newtheorem{remark}{Remark}[section]

\newcommand{\Nset}{\mathbb{N}}

\newcommand{\Rset}{\mathbb{R}}
\newcommand{\Cset}{\mathbb{C}}

\def\texttiny#1{{\text{\tiny{#1}}}}
\def\DerCap#1#2#3{{{}_{#1}^{{\text{\tiny{C}}}}D^{#2}_{#3}}}
\def\eu{\ensuremath{\mathrm{e}}}
\def\iu{\ensuremath{\mathrm{i}}}
\def\du{\ensuremath{\mathrm{d}}}

 \def\theequation{\arabic{section}.\arabic{equation}}

 \def\themycopyrightfootnote{\vspace*{3pt}
{\footnotesize This paper is now published in \emph{Fractional Calculus and Applied Analysis}, Vol. 19, No 5 (2016), pp. 1105-1160,
DOI: \href{http://dx.doi.org/10.1515/fca-2016-0060}{10.1515/fca-2016-0060}, and is available online at
\href{http://www.degruyter.com/view/j/fca}{http://www.degruyter.com/view/j/fca}}
   \hfill  \vspace*{-36pt}
  }

\title[Models of dielectric relaxation based on \dots]
      {Models of dielectric relaxation based on \\ [3pt] completely monotone functions}  
 \author[\normalsize R. Garrappa, F. Mainardi, G. Maione]
        {\normalsize Roberto Garrappa $^1$, Francesco Mainardi $^2$, Guido Maione $^3$}

\begin{document}

 \vbox to 2.5cm { \vfill }

 \bigskip \medskip

 \begin{abstract}

The relaxation properties of dielectric materials are described, in the frequency domain, according to one of the several models proposed over the years: Kohlrausch-Williams-Watts, Cole-Cole, Cole-Davidson, Havriliak-Negami (with its modified version)  and Excess wing model are among the most famous. Their description in the time domain involves some mathematical functions whose knowledge is of fundamental importance for a full understanding of the models. In this work, we survey the main dielectric models and we illustrate the corresponding time-domain functions. In particular, we stress the attention on the completely monotone character of the relaxation and response functions. We also provide a characterization of the models in terms of differential operators of fractional order.  
 \medskip


 \smallskip


 \end{abstract}

 \maketitle

 \vspace*{-22pt}


\section{Introduction}\label{S:Introduction}

Dielectric materials (dielectrics for brevity) play a fundamental role in the accumulation and dissipation of electric and magnetic energy. Their molecular structure and properties have been widely analyzed since the early Faraday's works, especially for insulation in electrical and electronic circuits. Polarization by electric fields is typically used to store a large amount of energy and, to this aim, the polar dielectrics are preferred in many applications. However, modelling the behaviour and response of dielectric materials is important for various reasons. Namely, prediction of standard and anomalous phenomena is possible only by model analysis and model-based simulation. Moreover, in many applications, the models allow to evaluate performance indexes and to adjust the structure or parameters of dielectrics so that responses obey design specifications, reference behaviours, etc.

The main dielectric parameter affecting the models is the permittivity $\varepsilon(\omega)$ or the associated susceptibility $\chi(\omega)$, which is in fact depending on the frequency of the applied electric field. Also the time-domain descriptions expressed by the relaxation and response functions are relevant especially to perform simulations.

The standard and simplest model in the physics of dielectrics was provided by Debye in 1912 \cite{Debye1912} based on a relaxation function decaying exponentially in time with a characteristic relaxation time.

However, simple exponential models are often not satisfactory, while advanced non-exponential models (usually referred as ``anomalous relaxation'') are commonly required to better explain experimental observations of complex systems. Namely, the relaxation response of many dielectric materials cannot be explained by the standard Debye process and different models have been successively introduced.

Anomalous relaxation and diffusion processes are nowadays recognized in many complex or disordered systems that possess variable structures and parameters and show a time evolution different from the standard exponential pattern \cite{Cametti2009,FeldmanPuzenkoRyabov2005,Hilfer2002_JNCS,KhamzinNigmatullinPopov2014,MetzlerKlafter2000,UchaikinSibatov2013}. Biological tissues are an interesting example of complex systems with anomalous relaxation and diffusion processes \cite{CoffeyKalmykovYuTitov2002,MaginAbdullahBaleanuZhou2008} and they can be considered as dielectrics with losses.

Since the pioneering work of Kohlrausch in 1854 \cite{Kohlrausch_1854}, introducing a stretched exponential relaxation successively rediscovered by Williams and Watts \cite{WilliamsWatt1970}, important models where introduced by Cole and Cole \cite{Cole-Cole1941,Cole-Cole1942}, Davidson and Cole \cite{DavidsonCole1950_JCP,DavidsonCole1951_JCP}, Havriliak and Negami \cite{HavriliakNegami1967} and others.

The challenges are measuring or extrapolating the dielectric properties at high frequencies, fitting the experimental data from various tissues and from different samples of the same tissue, and representing the complex, nonlinear frequency-dependence of the permittivity \cite{FosterSchwan1989}. Cole-Cole relaxation models, for instance, are frequently used to model propagation in dispersive biological tissues (Cole-Cole media) because they represent the frequency-dependent relative permittivity better than classical Debye models and over a wide frequency range \cite{MaundyElwakil2012,Lin2010,SaidVaradan2009}. More generally, the universal relaxation response specified by a fractional power-law is used for electromagnetic fields propagation \cite{Tarasov2008_JPCM,Tarasov2009_TMP}.

Nowadays the aforementioned models, named after their proposers, are considered as the ``classical'' models for dielectrics but some other interesting models have been more recently introduced by Jurlewicz, Weron and Stanislavsky \cite{JurlewiczTrzmielWeron_ActaPhysPolB_2010,StanislavskyWeronTrzmiel2010} and Hilfer \cite{Hilfer2002_CP,Hilfer2002_JPCM} to better fit the experimental data in complex systems.

In the present paper, we try to survey the most common models existing in the literature to our best knowledge and describe their main properties under a mathematical point of view.

All the models have the common feature that for large times the relaxation and response functions generally decay with a power law that indeed is found in most experiments. This power law behaviour allows to derive the equations governing the evolution of the dielectric processes  by using non-local operators provided by the so-called fractional calculus, that are pseudo-differential operators interpretable as suitable integrals and derivative of non-integer order \cite{CaputoFabrizio2014}. Then certain high transcendental functions related to fractional calculus arise naturally in the description of the relaxation and response functions, mostly of the Mittag-Leffler type.

There is however a further feature which ties all these models and which we would like to highlight in this survey: relaxation and response are completely monotonic functions of time, which means that they are expressed as a continuous distributions of simple exponential functions with a non-negative spectrum of relaxation times. To our knowledge, the property of complete monotonicity has not been sufficiently outlined in the existing literature at variance with the power law asymptotic behaviour.

This paper is organized in the following way. In Section {\ref{S:Preliminary}} we provide a preliminary introduction to physical and mathematical aspects of dielectric relaxation by illustrating the main functions we will use to describe each model. In Section {\ref{S:MainModels}}, which constitutes the main part of this paper, we proceed to describe the principal dielectric models and, for each model, we present the characteristic functions (with their graphical representations) and study the associated evolution equations. Three appendixes complete the paper: the first one presents some results on Mittag-Leffler and related functions used to describe the characteristic functions of each model; the second appendix summarizes the main integral and derivative operators used in the analysis of the evolution equations; finally, few biographical notes are dedicated to the main authors who contributed to the introduction of the models considered in the paper.


\section{Preliminary physical and mathematical  introduction to dielectric relaxation}\label{S:Preliminary}

\setcounter{equation}{0}
\setcounter{theorem}{0}

Under the influence of the electric field, an electric polarization occurs to the matter. The electric displacement effect on free and bound charges is described by the displacement field $\mathbf{D}$ which is related to the electric field $\mathbf{E}$ and to the polarization $\mathbf{P}$ by
\begin{equation}
\mathbf{D} = \varepsilon_0 \mathbf{E} + \mathbf{P}\,,
\end{equation}
where $\varepsilon_0$ is the permittivity of the free space. For a perfect isotropic dielectric and for harmonic fields of frequency $\omega$,  the interdependence between $\mathbf{E}$ and $\mathbf{P}$ is described by a constitutive law
\begin{equation}\label{eq:ConstitutiveLaw}
\mathbf{P} = \varepsilon_0 [(\varepsilon_s - \varepsilon_\infty)(\hat{\varepsilon}(\iu \omega) - 1)] \mathbf{E} = \varepsilon_0 [(\varepsilon_s - \varepsilon_\infty)\hat{\chi}(\iu \omega)] \mathbf{E} \,,
\end{equation}
where $\varepsilon_s$ and $\varepsilon_\infty$ are the static and infinite dielectric constants. The normalized complex {\it permittivity} $\hat{\varepsilon}(\iu\omega)$ and the normalized complex {\it susceptibility} $\hat{\chi}(\iu\omega)$ are specific characteristics of the polarized medium and are usually determined by matching experimental data with an appropriate theoretical model.


From the physical point of view, the description of dielectrics, considered as passive and causal linear systems, is carried out also  in time by considering two causal functions of time (i.e., vanishing for $t<0$):
\begin{itemize}
	\item the {\it relaxation function} $\Psi(t)$,
	\item the {\it response function} $\phi(t)$.
\end{itemize}

The relaxation function describes the decay of polarization whereas the response function its decay rate  (the depolarization current).

\begin{remark}
Note that our notation $\{\Psi(t),\phi(t)\}$ for the relaxation and response functions is in conflict with a notation frequently used in the literature, where the relaxation function is denoted by $\phi(t)$ and the response function by $-\du\phi(t) /\du t $.
\end{remark}

 As a matter of fact, the relationship between response and relaxation functions can be better clarified by their probabilistic interpretation investigated in several papers by Karina Weron and her team (e.g., see \cite{Weron1991,WeronKlauzer2000,WeronKotulski1996}):  interpreting the {\it relaxation function} as a {\it survival probability} $\Psi(t)$, the {\it response function} turns out to be  the {\it probability density function} corresponding
to the {\it cumulative probability function} $\Phi(t)= 1 - \Psi(t)$.
Thus the two functions are interrelated as follows
\begin{equation}\label{eq:phi_der_Psi}
\phi(t) = - \frac{\du}{\du t}\Psi(t) =  \frac{\du}{\du t}\Phi(t)\,, \quad t\ge 0\,,
\end{equation}
 and
 \begin{equation}\label{eq:Psi_int_phi}
\Psi(t) = 1 - \int_{0}^{t} \phi(u) \, \du u\,, \quad t\ge 0\,.
\end{equation}

In view of their probabilistic meaning,  $\phi(t)$ and $\Psi(t)$ are both non-negative and non-increasing functions. In particular, we get the limit $\Psi(0^+)=1$ whereas $\phi(0^+)$ may be finite or infinite.

The response function $\phi(t)$ is obtained as the inverse Laplace transform of the normalized complex susceptibility by setting the Laplace parameter  $s = \iu \omega$, that is
\begin{equation}\label{eq:Response}
	\phi(t) = {\mathcal L}^{-1} \left( \widetilde{\chi}(s)  ; t \right)\,,
\end{equation}
where we have used the superscript $\, \widetilde {\null}\,$ to denote a Laplace transform, i.e. $\widetilde{\chi}(s)=\hat{\chi}(\iu \omega)$; then, for  the relaxation function $\Psi(t)$ we have
\begin{equation}\label{eq:Relaxation}
	\Psi(t) = 1 - 	{\mathcal L}^{-1} \left( \frac{1}{s} \widetilde{\chi}(s) ; t \right) = {\mathcal L}^{-1} \left( \frac{1}{s} - \frac{1}{s} \widetilde{\chi}(s) ; t \right)\,.
\end{equation}

We can thus outline  the basic Laplace transforms pairs as follows
\begin{equation}
	\phi(t) \div \widetilde\phi(s) = \widetilde \chi(s) \, ,
	\quad \Psi(t) \div \widetilde \Psi(s) = \frac{1-\widetilde\phi(s)}{s}\, ,
\end{equation}
where we have adopted  the notation $\div$ to denote the juxtaposition of a function of time $f(t)$ with
its Laplace transform
$\widetilde f(s) =  \int_0^\infty \eu^{-st} f(t) \, \du t$.

The standard model in the physics of dielectrics was  provided by Debye \cite{Debye1912} according to which the normalized complex susceptibility, depending on the frequency of the external field, is provided, unless a proper multiplicative constant,
as
\begin{equation} \label{Debye}
	\hat{\chi}(\iu\omega) =  \frac{1}{1 + \iu\omega\tau_{\texttiny{D}}} \,
	,
\end{equation}
where $\tau_\texttiny{D}$ is the only expected relaxation time. In this case, both relaxation and response functions turn out to be purely exponential. In fact, recalling standard  results in Laplace transforms, we get
\begin{equation} \label{DebyeRelResp}
\Psi_\texttiny{D}(t) = \eu^{-t/\tau_\texttiny{D}} \,, \quad
\phi_\texttiny{D}(t) = \frac{1}{\tau_\texttiny{D}} \, \eu^{-t/\tau_\texttiny{D}}\,.
\end{equation}

Even though the Debye relaxation model was the first derived on the basis of statistical mechanics, it  finds a little application in complex systems where it is more reasonable to have a  discrete  or a continuous  distribution of  Debye models with different relaxation times,
so that the complex susceptibility reads
\begin{equation}
\hat{\chi}(\iu\omega) = \frac{\rho_1}{1 + \iu\omega\tau_1} + \frac{\rho_2}{1 + \iu\omega\tau_2} + \dots
\end{equation}
with ${\rho_1}, {\rho_2}, \dots$ non-negative constants
or,  more generally,
\begin{equation}\label{rho-tau}
\hat{\chi}(\iu\omega) = \int_0^{\infty} \frac{\rho(\tau) }{1 + \iu\omega\tau}\, \du \tau \,,
\end{equation}
with $\rho(\tau) \ge 0$. In mathematical language the above  properties are achieved requiring relaxation and response to be locally integrable and completely monotone (LICM) functions \cite{HanygaSeredynska2008}. The local integrability is requested to be Laplace transformable in the classical sense. The complete monotonicity means that the functions are non-negative with infinitely many derivatives for $t>0$ alternating in sign; we provide here its formal definition.

\begin{definition}
A function $f:(0,+\infty) \to \Rset$ is said completely monotonic (CM) on $(0,+\infty)$ if $f$ has derivatives of all orders and $(-1)^{k}f^{(k)}(t) \ge 0$ for any $k\in\Nset$ and $t > 0$.
\end{definition}

The above definition can be extended to $[0,\infty)$ when the limits of $f^{(k)}(t)$ as $t \to 0$ are finite.

As discussed by Hanyga \cite{Hanyga2005b}, CM is essential to ensure the monotone decay of the energy in isolated systems (as it appears reasonable from physical considerations); thus, restricting to CM functions is essential for the physical acceptability and realizability of the dielectric models (see \cite{AnhMcVinish2003}).

For the basic Bernstein theorem for LICM functions \cite{Widder1941}, $\Psi(t)$ and $ \phi(t) $ are represented as real Laplace transforms of non-negative spectral functions (of frequency)
\begin{equation}\label{eq:K_Psi_K_phi}
\Psi(t)= \int_0^\infty \!\! \eu^{-rt} K^\Psi(r) \, \du r\,,
\quad
\phi(t) =  \int_0^\infty \!\!\eu^{-rt} K^\phi(r) \, \du r\,.
\end{equation}

Due to  the interrelation between $\Psi(t)$ and $\phi(t)$,  the corresponding spectral functions are obviously related and indeed
\begin{equation}
 K^\Psi(r) = K^\phi(r)/r\,, \quad K^\phi(r)= r\, K^\Psi(r)\,.
 \end{equation}

As a matter of fact, the Laplace transform $\widetilde \Psi(s)$ of $\Psi(t)$ and  $\widetilde \phi(s)$ of $\phi(t)$ turn to be iterated Laplace transforms (that are, Stieltjes transforms)  of the corresponding frequency spectral  functions $K^\Psi(r)$, $K^\phi(r)$. In fact, by exchanging order in the Laplace integrals, we get
 \begin{equation}\label{K-Psi-phi}
 \widetilde\Psi(s) =  \int_0^\infty
\frac{K^\Psi(r)}{s+r} \, \du r\,, \quad
\widetilde\phi(s) =  \int_0^\infty
\frac{K^\phi(r)}{s+r} \, \du r\,.
 \end{equation}

As a consequence, the frequency spectral functions can be derived from $\widetilde\Psi(s)$ and $\widetilde\phi(s)$ as their inverse Stieltjes transforms thanks to  the Titchmarsh inversion formula \cite{Titchmarsh1937},
\begin{equation}\label{eq:InversionSpectral}
	K^{\Psi}(r) = \mp \frac{1}{\pi}
		{\rm Im} \left[ \widetilde{\Psi}(s) \bigl|_{s=re^{\pm \iu \pi}} \right]\, , \;
	K^{\phi}(r) = \mp \frac{1}{\pi}
		{\rm Im} \left[ \widetilde{\phi}(s) \bigl|_{s=re^{\pm \iu \pi}} \right] \,.
\end{equation}

For a physical viewpoint, it may be more interesting to deal with spectral functions expressed in terms of relaxation times $\tau=1/r$ rather than frequencies $r$. Then we write
\begin{equation}\label{eq:H_Psi_phi}
	\Psi(t) =\int_0^\infty \eu^{-t/\tau} \, H^\Psi(\tau)\, \du \tau \, ,\quad
	\phi(t) =\int_0^\infty \eu^{-t/\tau} \, H^\phi(\tau)\, \du \tau \, ,
\end{equation}
 so that the time spectral functions are obtained from the corresponding frequency spectral ones by the variable change
\begin{equation}\label{eq:H_K}
  H^{\Psi,\phi}(\tau) = \frac{K^{\Psi,\phi}(1/\tau)}{\tau^2}\,
\end{equation}
(we also refer to \cite{NigmatullinKhamzinBaleanu2016} for a method based on the generalized multiplication Efros theorem to derive spectral functions).
	
We have to keep in mind that, even though we have distributions of multiple Debye relaxation processes characterized by time exponentials of negative argument, the characteristic functions can decay for small and large times as power laws with negative exponents due to certain asymptotic behaviours of the corresponding frequency or time spectral functions. In these cases the dielectric relaxation is told to be {\it anomalous}. This power-law behaviour is usually found, for instance, in complex bodies so that suitable models of anomalous relaxation have been introduced in the literature to properly account for experimental data.

We close this section  with  a few further  considerations on spectral functions used to characterize the processes of anomalous dielectric relaxation.

At first, we note that in Eq. (\ref{rho-tau}) we have
\begin{equation} \label{H-rho-tau}
\rho(\tau) =  \tau H^\phi(\tau) \, ;
\end{equation}
indeed, from (\ref{eq:Response}) and (\ref{eq:H_Psi_phi}) we get
\begin{equation*}
	\begin{aligned}
	\widetilde \chi(s) := \widetilde \phi(s)
	&= \int_0^\infty \! \eu^{-st}\, \phi(t)\, \du t =
		\int_0^\infty \!  \eu^{-st} \int_0^\infty\! \!\eu^{-t/\tau} H^\phi(\tau)\, \du \tau \, \du t  \\
	&= \int_0^\infty \!\frac{H^\phi(\tau)}{s+1/\tau}\, \du \tau =
			\int_0^\infty \! \frac{\tau H^\phi(\tau)}{1+s\tau}\, \du \tau =
			\int_0^\infty \!\frac{\rho(\tau)}{1 +s\tau}\, \du \tau\, . \
	\end{aligned}		
\end{equation*}
and (\ref{H-rho-tau}) follows after setting  $s= \iu\omega$ in Eq. (\ref{rho-tau}).

Finally, we consider the logarithmic spectral function of relaxation times based on the scaling $u = \log(\tau)= -\log(r)$ running from $-\infty$ to $+\infty$ for both the functions $\Psi(t)$ and $\phi(t)$. Setting in Eq. (\ref{K-Psi-phi}) $r = \exp (-u)$ we get
\[
	\widetilde \chi(s) := \widetilde \phi(s) = \int_{0}^{+\infty} \frac{K^{\phi}(r)}{s+r} \, \du r
	= \int_{-\infty}^{+\infty} \frac{\eu^{-u} K^{\phi}(\eu^{-u})}{s+\eu^{-u}} \, \du u
\]
and similarly for $\widetilde{\Psi}(s)$. Thus we may introduce the required logarithmic spectral functions, that we denote by $L^{\Psi,\phi}(u)$, which are related to frequency spectral functions $K^{\Psi,\phi}(r)$ by
\begin{equation}\label{eq:L_K}
	L^{\Psi,\phi}(u) = e^{-u} K^{\Psi,\phi}(e^{-u})
\end{equation}
and, consequently, the time relaxation spectral functions $H^{\Psi,\phi}(\tau)$ by
\begin{equation}\label{eq:L_H}
	L^{\Psi,\phi}(u) = e^{u} H^{\Psi,\phi}(e^{u}) \, .
\end{equation}

These spectral functions  are mostly used in experiments because they cover several decades of relaxation times accounting in an equal measure for the smaller and the larger times.


\section{Main  models for anomalous dielectric relaxation}\label{S:MainModels}
\setcounter{equation}{0}
\setcounter{theorem}{0}

The Debye model \cite{Debye1912} is one of the first models introduced to describe physical properties of dielectrics and, as discussed in Section {\ref{S:Preliminary}}, involves relaxation and response functions of exponential type; we refer to this purpose to equations (\ref{Debye}) and (\ref{DebyeRelResp}).

As revealed by a number of experiments, a broad variety of dielectric materials exhibits relaxation behaviours which strongly deviate from the exponential Debye law. The observation of ``anomalous'' phenomena such as broadness, asymmetry and excess in the dielectric dispersion has motivated the proposition of new empirical laws in order to modify the Debye relaxation and match experimental data in a more accurate way.

In 1970s, after analyzing the dielectric properties of several materials, Jonscher and his co-workers propounded the existence of a universal relaxation law (URL)  following a fractional power law dependence and capable of fitting most of the experimental data \cite{Jonscher1977,NgaiJonscherWhite1979}. The URL proposed by Jonscher is an empirical law in which the ratio of the imaginary to the real part of the susceptibility is assumed to be constant and characterized by two power law exponents, say $m$ (for low frequencies) and $n$ (for high frequencies), with $0<m,n<1$ \cite{Jonscher1983,Jonscher1996}.

In particular, once we distinguish the real and imaginary part of the complex susceptibility, namely $\hat{\chi}(\iu\omega) = \chi^{\prime}(\omega) - \iu \chi^{\prime\prime}(\omega)$, it is assumed that
\[
	\chi'(0) - \chi'(\omega) \sim \omega^{m} , \quad \chi''(\omega) \sim \omega^{m} , \quad \omega \tau_{\star} \ll 1
\]
and
\[
	\chi'(\omega) \sim \omega^{n-1} , \quad \chi''(\omega) \sim \omega^{n-1} , \quad \omega \tau_{\star} \gg 1 ,
\]
where $\tau_{\star}$ is the characteristic relaxation time.

It is nowadays well established that the relaxation properties of a large variety of materials obey to this law and some models proposed in literature (before and after the work of Jonscher) fit in an excellent way the Jonscher's URL; this is the case, for instance, of the
\begin{itemize}
	\item Cole-Cole (CC) model,
	\item Havriliak-Negami (HN) model,
	\item modified HN or Jurlewicz-Weron-Stanislavsky (JWS) model.
\end{itemize}

There also exist materials whose relaxation data can be interpreted in an accurate way only by means of different types of experimental fitting functions which do not completely fit the Jonscher's URL, as for the
\begin{itemize}
	\item Kohlrausch-Williams-Watts (KWW) model,
	\item Davidson-Cole (DC) model,
	\item excess wing (EW) model.
\end{itemize}

As it will be better illustrated further on, the DC, HN, JWS and EW models are closely related to the CC model since they are obtained from similar complex susceptibilities, mainly based on the insertion of real powers in the susceptibility of the Debye model. Conversely, the derivation of the KWW model is made in a completely different way and a closed-form representation of its complex susceptibility is not available (a comparison of DC and KWW models is discussed in \cite{LindseyPatterson1980}).

To some extent, the CC model (together with the DC, HN, JWS and EW models) and the KWW model represent two large and distinct families of relaxation models for dielectrics \cite{Fu2014}. Nevertheless, an attempt of creating a mathematical bridge between these two distinct families has been proposed by Capelas, Mainardi and Vax in \cite{CapelasMainardiVaz2014}, where a more general model combining CC and KWW models was introduced; we will outline also this model which is indicated as the
\begin{itemize}
  \item Capelas-Mainardi-Vaz (CMV) model.
\end{itemize}

The above list of dielectric models is obviously not exhaustive. Other models have been discussed in literature, but most of them can be obtained from one of the above models. This is the case, for instance, of the model proposed in 1999 by Raicu \cite{Raicu1999}, which is not treated in this survey since it can be reverted in a special case of the HN model, or the model based on fractional order operators of hyper-Bessel type (for details one can see in \cite{Kiryakova2014}), and still having CM relaxation functions, investigated in \cite{GarraGiustiMainardiPagnini2014}.

In the remainder of this section, we illustrate each model and we discuss their main features under a mathematical perspective.

\subsection{The Cole-Cole model}

The Cole-Cole model, named after the brothers K.S. Cole and R.H. Cole, was introduced in 1941 \cite{Cole-Cole1941} (see also \cite{Cole-Cole1942}). As described in \cite{BotcherBordewijk1978}, it finds applications in ``systems with rather small deviations from a single relaxation time, e.g. many compounds with rigid molecules in the pure liquid state and in solution in non-polar, non-viscous solvents''.  Nowadays this model is still used to represent impedance of biological tissues, to describe relaxation in polymers, to represent anomalous diffusion in disordered systems and so on \cite{KalmykovCoffeyCrothersTitov2004,Lin2010,MaundyElwakil2012}.

\begin{figure}[ht]
\centering
\includegraphics[width=.60\textwidth]{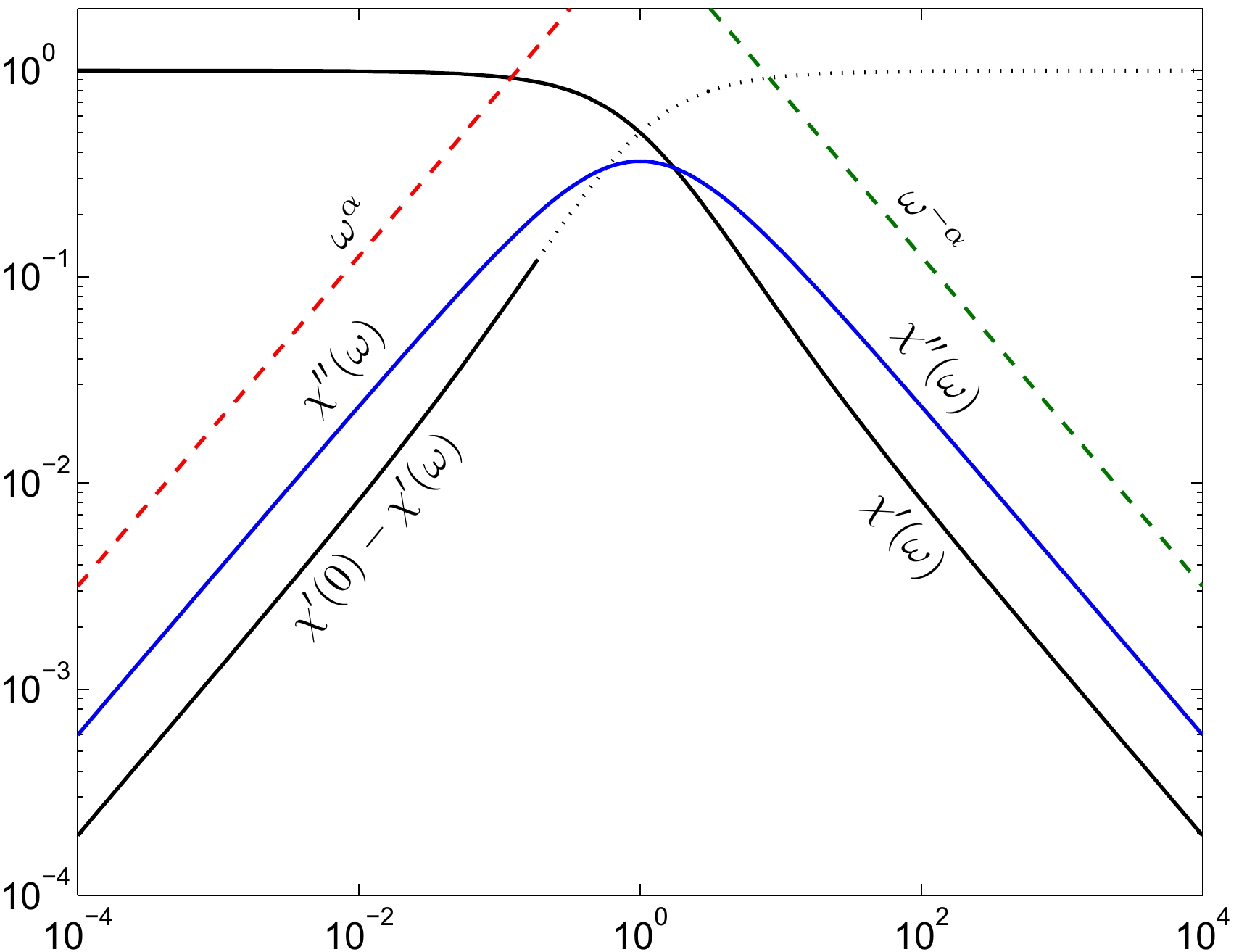}
\caption{CC susceptibility for $\alpha=0.8$.}
\label{fig:Chi_URL_CC}
\end{figure}

The complex susceptibility of the Cole-Cole model is derived by inserting a real power in the original Debye model, thus to fit data presenting a broader loss peak, and it is given by
\begin{equation}\label{eq:Chi_CC}
	\hat{\chi}_{\texttiny{CC}}(\iu\omega) = \frac{1}{1 + \bigl(\iu \omega \tau_{\star}\bigr)^{\alpha}}\,, \quad  0<\alpha \le 1\,,
\end{equation}
where $\tau_{\star}$ denotes a reference relaxation time. It is an elementary task to verify that $\hat{\chi}_{\texttiny{CC}}(\iu\omega)$ fits the Jonscher's URL with $m=\alpha$ and $n=1-\alpha$, as enhanced in Figure \ref{fig:Chi_URL_CC}; unless otherwise specified, in all the subsequent plots a normalized relaxation time $\tau_{\star}=1$ is assumed.

By operating the Laplace inversion of (\ref{eq:Chi_CC}) we get the corresponding response and relaxation functions respectively as
\begin{equation}
	\phi_{\texttiny{CC}}(t) =
	{\mathcal L}^{-1} \left( \frac{1}{1+ (s\tau_{\star})^\alpha} \right)
	=
	\frac{1}{\tau_{\star}} \bigl( t/\tau_{\star} \bigr)^{\alpha-1} E_{\alpha,\alpha}\left(- \bigl( t/\tau_{\star} \bigr)^{\alpha} \right)\,,
\end{equation}
and
\begin{equation}\label{eq:CC_RelaxationFunction}
	\begin{aligned}
			\Psi_{\texttiny{CC}}(t)
			&= {\mathcal L}^{-1} \left( \frac{1}{s} - \frac{1}{s \bigl(1 + \bigl(s \tau_{\star}\bigr)^{\alpha}\bigr)} \right) \\
		  &= 1 - \left( {t}/{\tau_{\star}}\right)^{\alpha} E_{\alpha,\alpha+1} \left(- \left({t}/{\tau_{\star}}\right)^{\alpha}\right)
			= E_{\alpha,1}\left(- \bigl( t/\tau_{\star} \bigr)^{\alpha} \right) .
  \end{aligned}
\end{equation}

\begin{figure}[ht]
\centering
\includegraphics[width=0.46\textwidth]{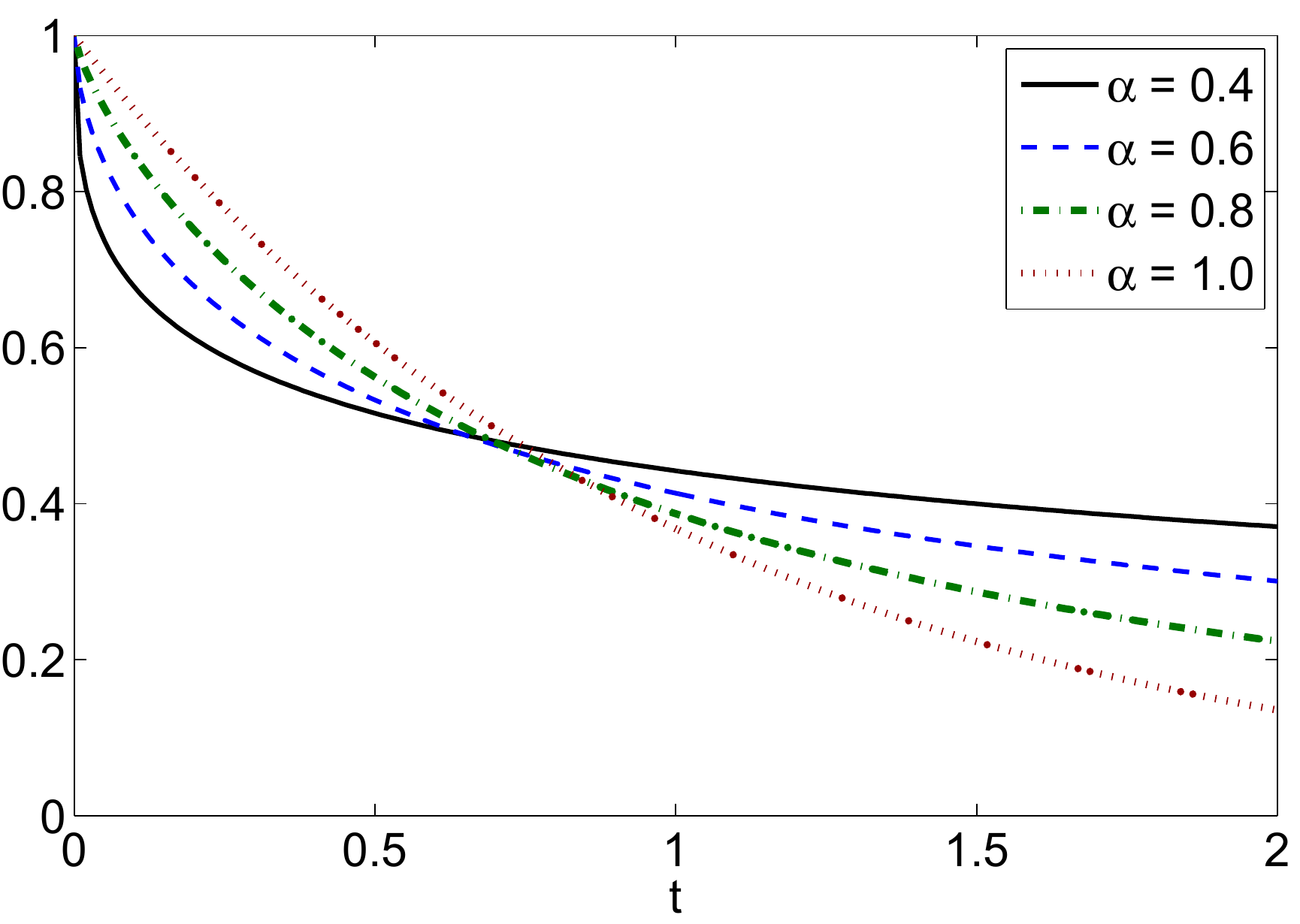}
\includegraphics[width=0.46\textwidth]{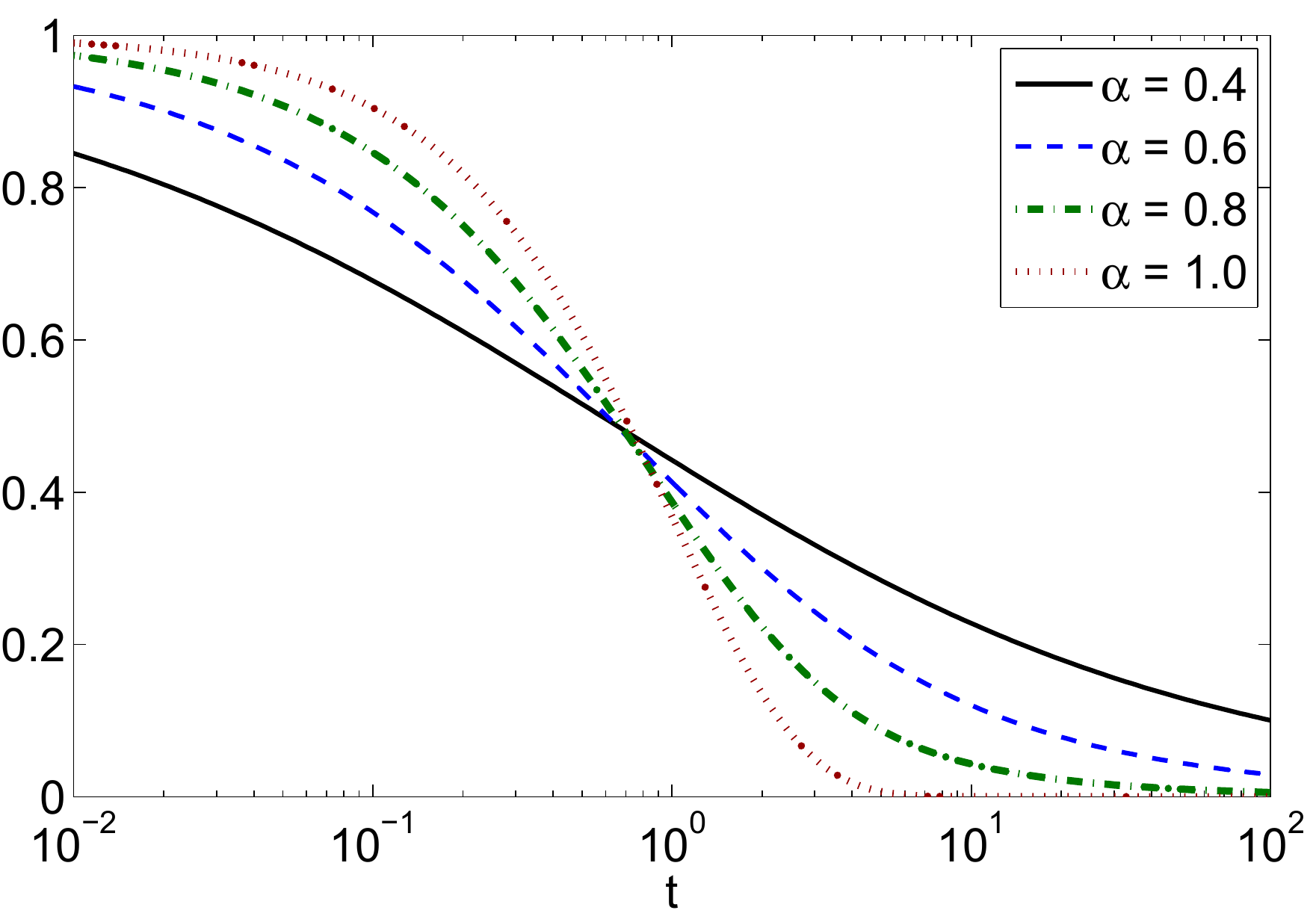}
\caption{Relaxation function $\Psi_{\texttiny{CC}}(t)$ on varying $\alpha$.}
\label{fig:Rel_CC}
\end{figure}

We report in Figure \ref{fig:Rel_CC} the plots of the relaxation function $\Psi_{\texttiny{CC}}(t)$ using linear (left plot) and logarithmic (right plot) scales.

The plots of the relaxation and response functions are also found in  Mainardi's  book \cite{Mainardi2010} along with their asymptotic representations. Here we just recall that by using standard results on the asymptotic behaviour of the Mittag-Leffler function it is possible to verify that
\begin{equation}
	\phi_{\texttiny{CC}}(t) \sim \left\{ \begin{array}{ll}
		\displaystyle\frac{1}{\tau_{\star} \Gamma(\alpha)}\bigl( t/\tau_{\star} \bigr)^{\alpha-1} \, , \, & \text{for } t \ll \tau_{\star}  , \\
		-\displaystyle\frac{1}{\tau_{\star} \Gamma(-\alpha)} \bigl( t/\tau_{\star} \bigr)^{-\alpha-1} \, , \, & \text{for } t \gg \tau_{\star} ,\
	\end{array} \right.
\end{equation}
and
\begin{equation}
	\Psi_{\texttiny{CC}}(t) \sim \left\{ \begin{array}{ll}
		1 - \displaystyle\frac{1}{\Gamma(\alpha+1 )}\bigl( t/\tau_{\star} \bigr)^{\alpha} \, , \, & \text{for } t \ll \tau_{\star} , \\
		\displaystyle\frac{1}{\Gamma(1-\alpha)} \bigl( t/\tau_{\star} \bigr)^{-\alpha} \, , \, & \text{for } t \gg \tau_{\star}. \\
	\end{array} \right.
\end{equation}

It is interesting to note that the  Cole brothers were not initially interested to express relaxation and response in terms of Mittag-Leffler functions in \cite{Cole-Cole1941}, but, soon one year later \cite{Cole-Cole1942}, they made reference for these functions to the treatise by Davis of 1936 \cite{Davis1936}.  Indirect references to Mittag-Leffler functions in anomalous dielectric relaxation can be found in the  works by Gross of 1937 \cite{Gross1937}, of 1938 \cite{Gross1938}, of 1941  \cite{Gross1941}, but  more explicitly in the 1939 papers by his student F.M de Oliveira Castro \cite{Oliveira-Castro1939b,Oliveira-Castro1939}.

 Later, in 1947  Gross \cite{Gross1947} has explicitly proposed the Mittag-Leffler function in mechanical relaxation in the framework of linear viscoelasticity. This argument was revisited in 1971 by Caputo and Mainardi \cite{Caputo-Mainardi1971b,Caputo-Mainardi1971a} in order to propose the so-called fractional Zener model making use of the time fractional derivative in the Caputo sense. A function strictly related to the Mittag-Leffler function  was introduced by Rabotnov in 1948 \cite{Rabotnov1948} and soon later numerical tables of the Rabotnov function appeared by his collaborators. However, the first plots of the Mittag-Lefller function appeared only in the 1971 papers by Caputo and Mainardi \cite{Caputo-Mainardi1971b,Caputo-Mainardi1971a}. Nowadays, because of the relevance of this function in Fractional Calculus as solution of differential equations of fractional order, a number of computing routines are available, by  Gorenflo et al. \cite{GorenfloLoutchkoLuchko2002}, by Seybold and Hilfer \cite{HilferSeybold2008}, by Podlubny et al. \cite{PodlubnyKacenak2012} and, more recently, by Garrappa et al. \cite{GarrappaPopolizio2013,Garrappa2015}.

Consequently, the Cole brothers, even though they have not explicitly used  fractional derivatives or integrals, can be considered as ``indirect'' pioneers of this mathematical branch (for an historical perspective  see \cite{Valerio2014}).

By the way,
the Cole brothers are famous because  of their idea to plot the locus of the (normalized) complex permittivity
 $\hat{\varepsilon}(\iu\omega)
= \varepsilon^\prime(\omega) + \iu \varepsilon^{\prime\prime}(\omega)$
in the complex plane, henceforth named after them Cole-Cole plots.

\begin{figure}[ht]
\centering
\includegraphics[width=0.58\textwidth]{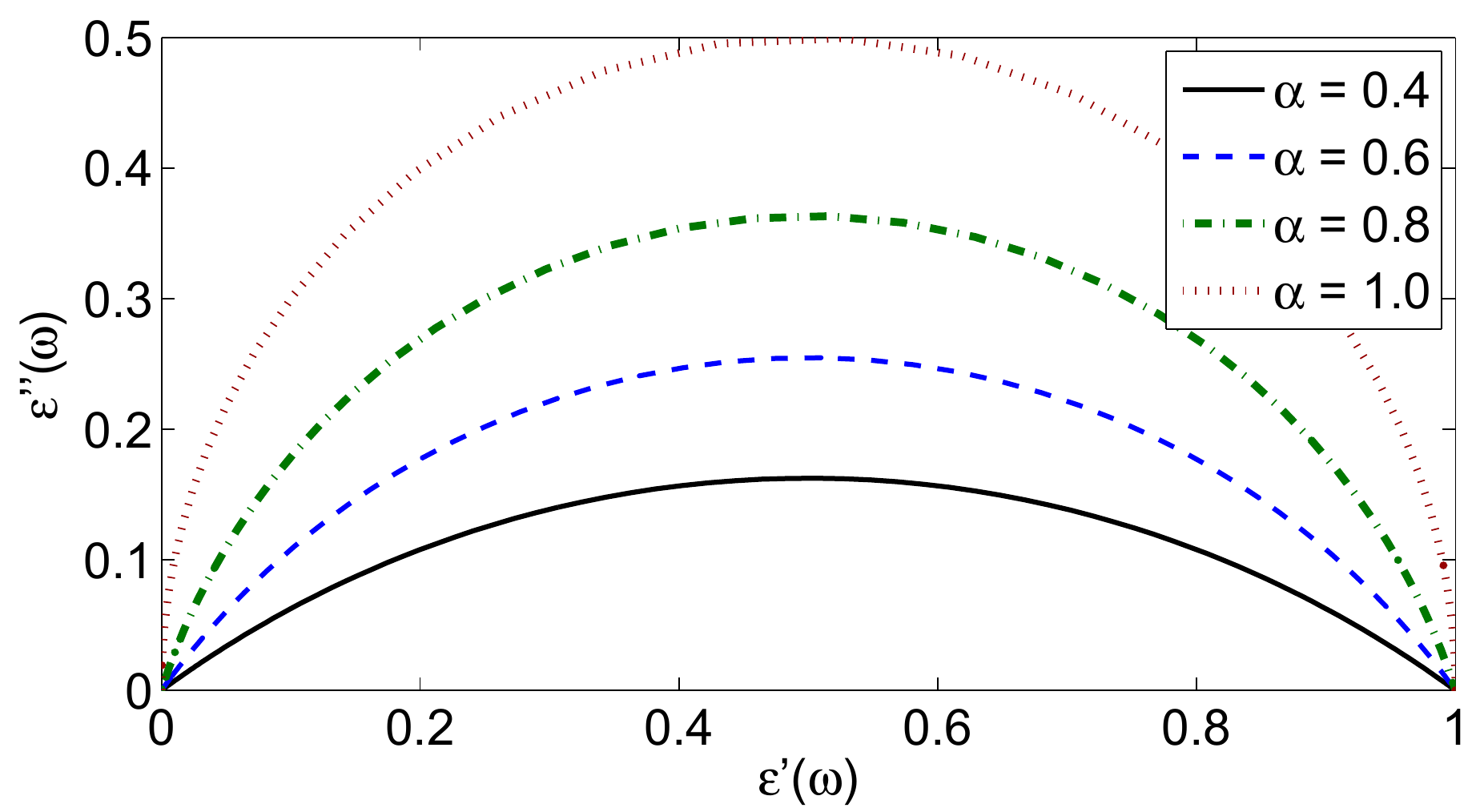}
\caption{Cole-Cole plots for the CC model.}
\label{fig:ColeCole_CC}
\end{figure}

In Figure \ref{fig:ColeCole_CC} we exhibit the Cole-Cole plots for the CC model with varying $\alpha$,  showing a semicircle with its center on the real axis in the Debye case ($\alpha=1$) and an arc of a circle with its center below the real axis, namely at $C=(\frac{1}{2},-\frac{\cos(\alpha\pi/2)}{2\sin(\alpha\pi/2)})$, and radius $\rho = \frac{1}{2\sin(\alpha\pi/2)}$, in the anomalous case ($0<\alpha<1$).

Let us now consider the spectral functions related to the Cole-Cole model restricting  our attention to the relaxation function
$\Psi_{\texttiny{CC}}(t)$. From (\ref{eq:InversionSpectral}) and (\ref{eq:CC_RelaxationFunction}) the frequency spectral function for $\Psi_{\texttiny{CC}} (t)$ turns out to be
\begin{equation}
	  K^\Psi_{\texttiny{CC}}(r)
		= \frac{\tau_{\star}}{\pi }\, \frac{ (r \tau_{\star})^{\alpha -1}\, \sin \,(\alpha \pi)}
    {(r \tau_{\star})^{2\alpha} + 2\, (r \tau_{\star})^{\alpha} \, \cos \, (\alpha \pi) +1} \ge 0\,.
\end{equation}

With the change of variable $\tau=1/r$ we get the corresponding spectral representation $H^\Psi_{\texttiny{CC}}(\tau) = \tau^{-2}  K^\Psi_{\texttiny{CC}}(1/\tau)$ in relaxation times (as already outlined in Section {\ref{S:Preliminary}}) from which it is immediate to evaluate
\begin{equation}\label{eq:H-Psi-tau}
 H^\Psi_{\texttiny{CC}}(\tau)
= \frac{1}{\pi \tau_{\star}}\,
   \frac{ (\tau/\tau_{\star})^{\alpha -1} \sin (\alpha \pi)}
    {(\tau/\tau_{\star})^{2\alpha} + 2 (\tau/\tau_{\star})^{\alpha}  \cos (\alpha \pi) +1}
\end{equation}
and thus one easily recognizes the identity $K^\Psi_{\texttiny{CC}}(r) = H^\Psi_{\texttiny{CC}}(\tau)$ between the two spectral functions when the relaxation time is normalized to $\tau_{\star}=1$.

The coincidence between the two spectral functions is a  surprising fact pointed out for the Mittag-Leffler function $E_{\alpha,1}(-t^\alpha)$ with $0<\alpha<1$ by Mainardi  in his 2010 book \cite{Mainardi2010} and in his recent paper \cite{Mainardi2014}. This kind of universal/scaling property seems therefore peculiar for the Cole-Cole relaxation function $\Psi_{\texttiny{CC}}(t)$.

\begin{figure}[ht]
\centering
\includegraphics[width=0.46\textwidth]{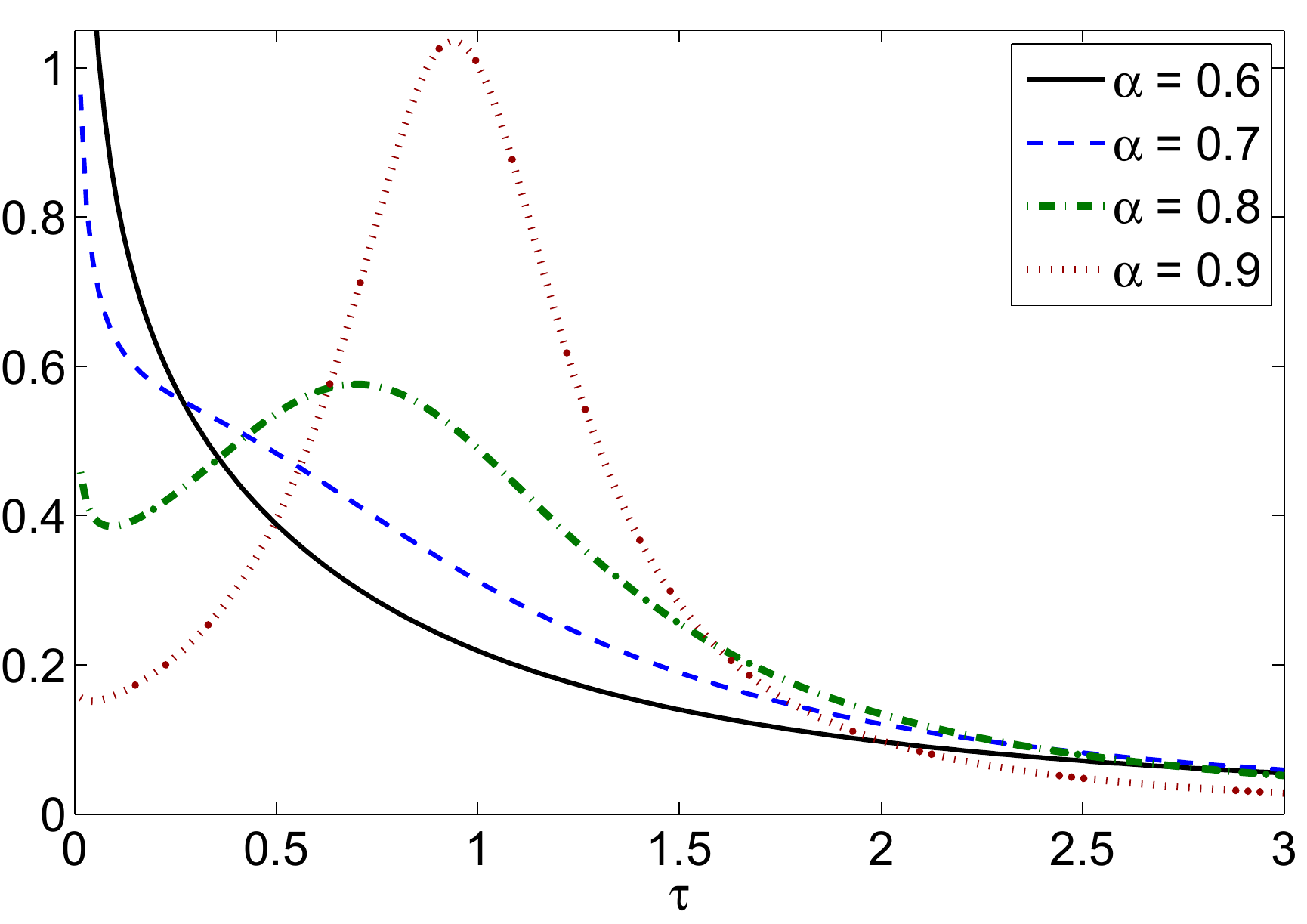}
\includegraphics[width=0.46\textwidth]{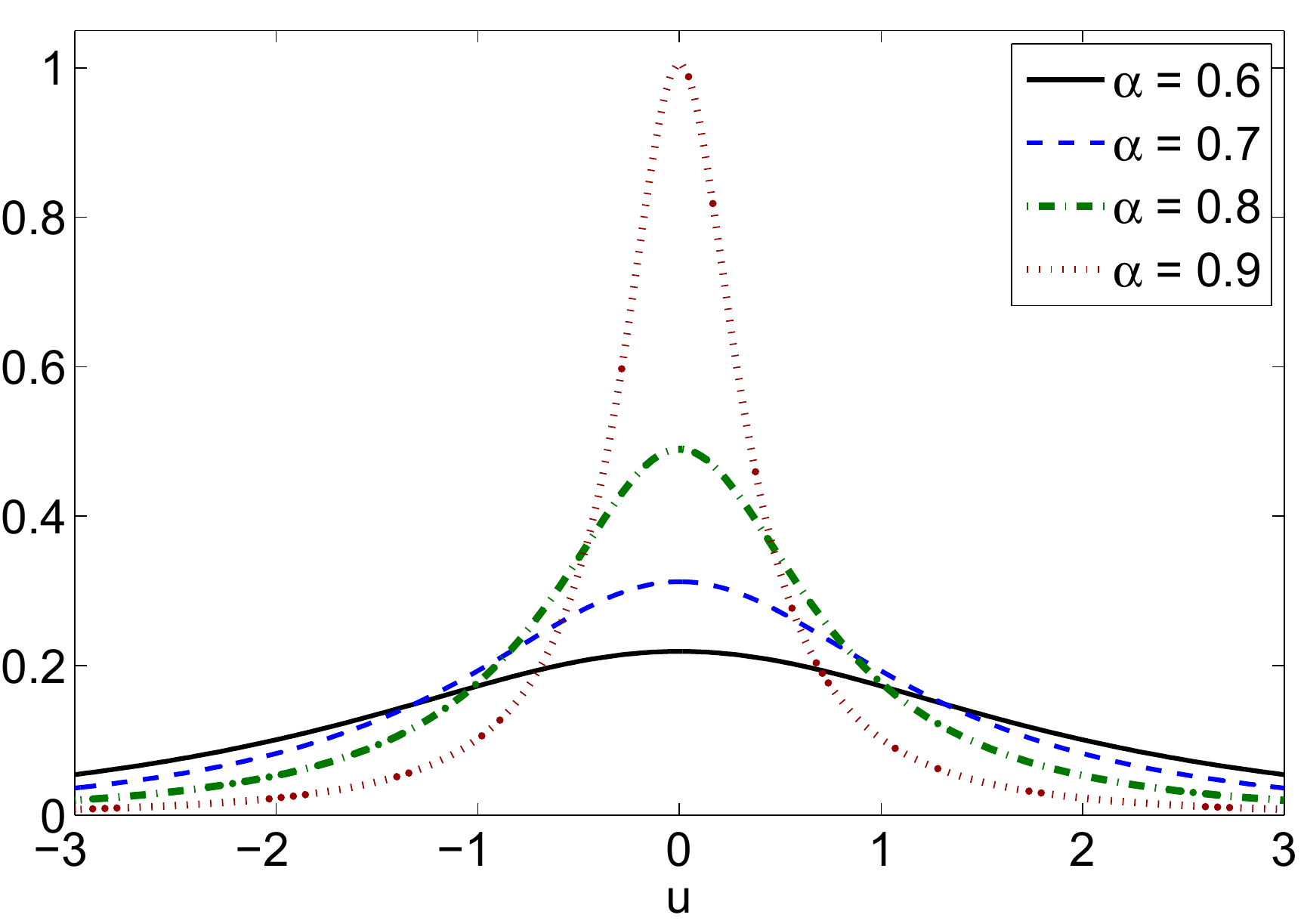}
\caption{Spectral distributions $H^{\Psi}_{\texttiny{CC}}(\tau)$ (left) and $L^{\Psi}_{\texttiny{CC}}(u)$ (right).}
\label{fig:Spectral_CC}
\end{figure}

For some values of the parameter $\alpha$ and with respect to the  relaxation function $\Psi_{\texttiny{CC}}(t)$ of the CC model, we show in Figure {\ref{fig:Spectral_CC}} the time spectral distribution $H^\Psi_{\texttiny{CC}}(\tau)$ given by (\ref{eq:H-Psi-tau}) and its logarithmic representation $L^\Psi_{\texttiny{CC}}(u) = e^{u} H^\Psi_{\texttiny{CC}}(e^{u})$, i.e.
\begin{equation}
	L^\Psi_{\texttiny{CC}} (u)
	= \frac{1}{2\pi} \,
		\frac{\sin(\alpha\pi)}{\cosh[\alpha (u - \log \tau_{\star})] + \cos(\alpha\pi)} \,, \quad u = \log(\tau) \,.
\end{equation}

Of course, for $\alpha=1$ the Mittag-Leffler function in (\ref{eq:CC_RelaxationFunction}) reduces to the exponential function $\exp(-t/\tau_{\star})$ and the corresponding spectral distributions are both equal to the Dirac delta generalized function centred, respectively, at $\tau=\tau_{\star}$ and $u=\log(\tau_{\star})$.

Note that both spectral functions were formerly outlined in 1947 by Gross \cite{Gross1947} and later revisited, in 1971, by Caputo and Mainardi
\cite{Caputo-Mainardi1971b,Caputo-Mainardi1971a}.

The response and relaxation functions of the CC model satisfy some evolution equations expressed by means of fractional differential operators. In particular, for the response $\phi_{\texttiny{CC}}(t)$ after applying some basic properties of the Riemann-Liouville derivative ${}_{0} D^{\alpha}_t$ (see the Appendix \ref{S:FractionalIntegralsDerivatives} at the end of this paper and \cite[Section 2.2]{Diethelm2010}) it is straightforward to derive
\begin{equation}\label{eq:CC_EvolEq_phi}
	{}_{0} D^{\alpha}_t  \phi_{\texttiny{CC}}(t) = -\frac{1}{\tau_{\star}^{\alpha}}  \phi_{\texttiny{CC}}(t)
	, \quad
	\lim_{t\to 0^{+}} {}_{0}J_{t}^{1-\alpha} \phi_{\texttiny{CC}}(t) = \frac{1}{\tau_{\star}^{\alpha}}
	\, ,
\end{equation}
and, correspondingly, the application of the Caputo fractional derivative  $\DerCap{0}{\alpha}{t}$ leads to
\begin{equation}\label{eq:CC_EvolEq_Psi}
	\DerCap{0}{\alpha}{t}  \Psi_{\texttiny{CC}}(t) = -\frac{1}{\tau_{\star}^{\alpha}}  \Psi_{\texttiny{CC}}(t)
	, \quad
	\Psi_{\texttiny{CC}}(0) = 1 \, .
\end{equation}

\subsection{The Davidson-Cole model} 

After a decade from the introduction of the CC model, another  dielectric model, still depending on one real parameter, was  proposed to generalize the standard Debye model. The introduction in 1950-1951 of the new model by D.W. Davidson and R. H. Cole \cite{DavidsonCole1950_JCP,DavidsonCole1951_JCP} was motivated by the need of fitting the broader range of dispersion observed at high frequencies in some organic compounds such as glycerine, glycerol, propylene glycol, and $n$-propanol.

\begin{figure}[ht]
\centering
\includegraphics[width=0.58\textwidth]{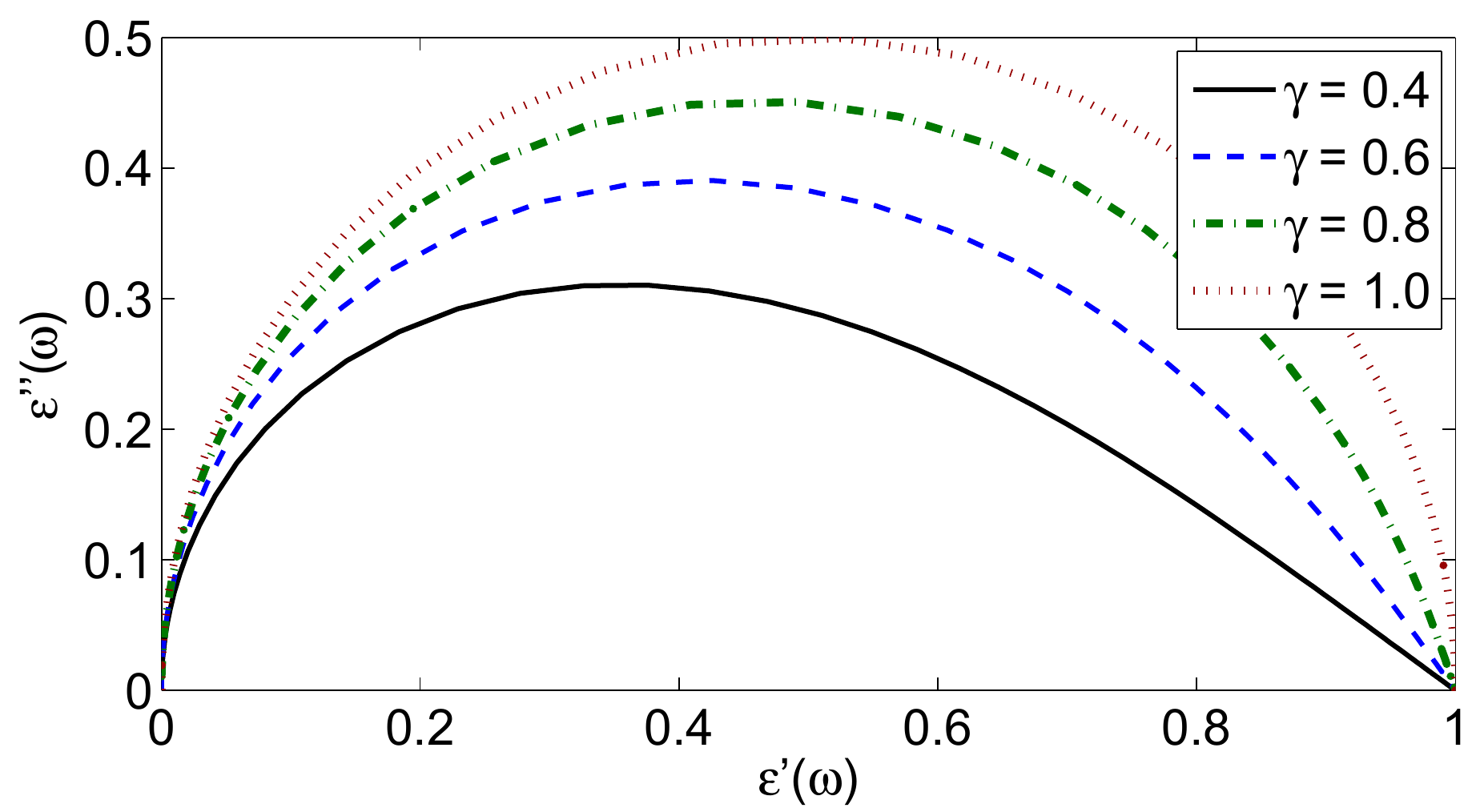}
\caption{Cole-Cole plots for the DC model.}
\label{fig:ColeCole_DC}
\end{figure}

This asymmetry is obtained in the Davidson-Cole (DC) model by considering the following complex susceptibility
\begin{equation}\label{eq:DC}
	\hat{\chi}_{\texttiny{DC}}(\iu \omega) = \frac{1}
	{(1 + \iu \omega \tau_{\star})^\gamma}\,, \quad  0<\gamma \le 1\,,
\end{equation}
and it is clearly reflected in the corresponding Cole-Cole plots presented in Figure \ref{fig:ColeCole_DC}.

\begin{figure}[ht]
\centering
\includegraphics[width=.60\textwidth]{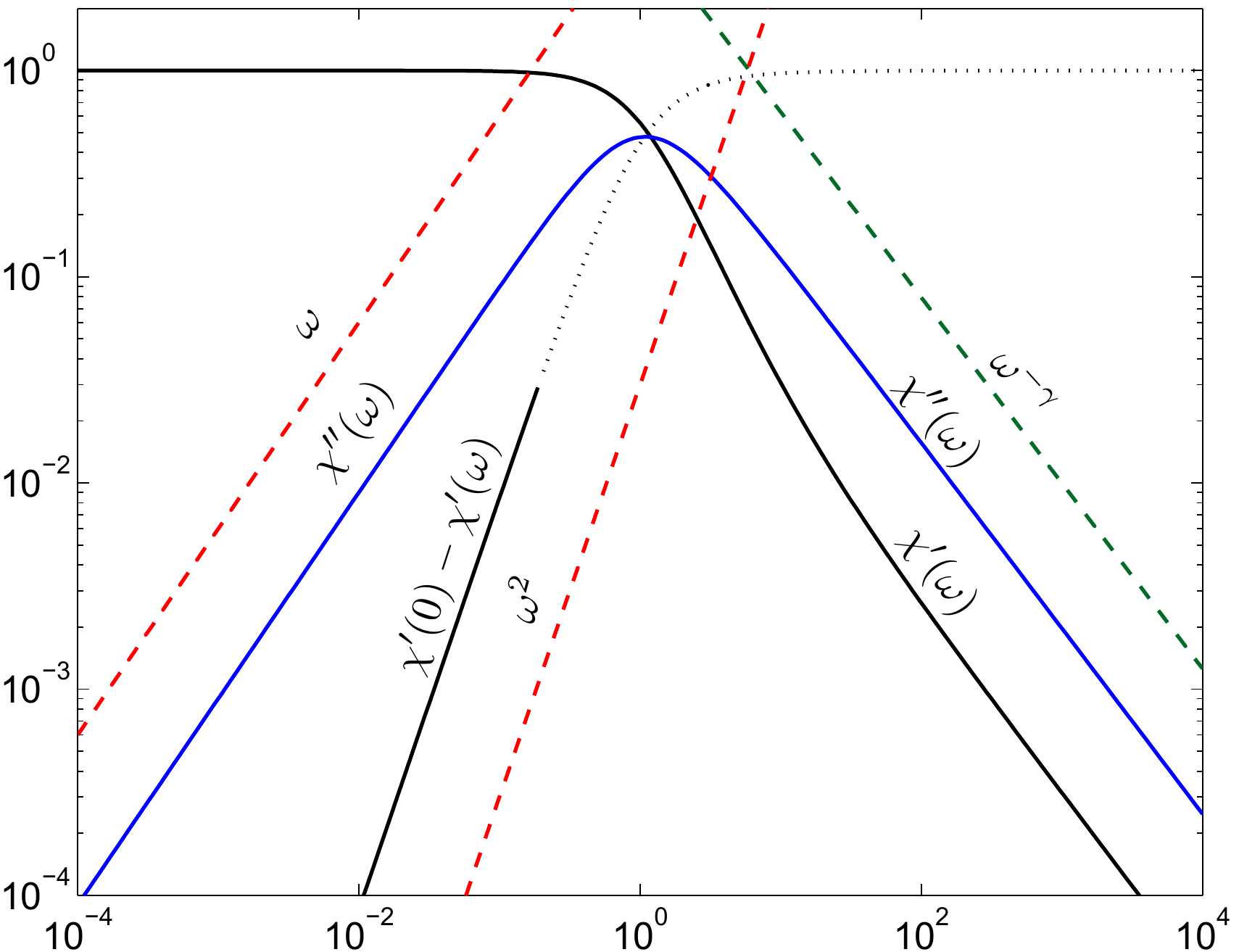}
\caption{CD susceptibility for $\gamma=0.9$.}
\label{fig:Chi_URL_DC}
\end{figure}

This model does not completely fit the Jonscher's URL. Indeed, after writing the complex susceptibility in the equivalent formulation
\begin{equation}
	\hat{\chi}_{\texttiny{DC}}(\iu \omega) = \frac{\cos\bigl[\gamma \theta(\omega)\bigr] - \iu \sin\bigl[\gamma\theta(\omega)\bigr]}{(1+\omega^2\tau_{\star}^2)^{\gamma/2}} \,
	, \quad
	\theta(\omega) = \frac{\pi}{2} - \arctan \Bigl[\frac{1}{\omega \tau_{\star}}\Bigr] \, ,
\end{equation}
and evaluating the asymptotic expansions
\begin{equation*}
	\cos\bigl[\gamma \theta(\omega)\bigr] \sim \left\{ \begin{array}{l}
		1 - \frac{1}{2}\gamma^2 \tau_{\star}^2 \omega^2  \\
		C_{\gamma} + \frac{\gamma}{\tau_{\star}\omega} S_{\gamma}  \
	\end{array} \right. \quad
	\sin\bigl[\gamma \theta(\omega)\bigr] \sim \left\{ \begin{array}{ll}
		\gamma \tau_{\star} \omega \, , \, & \omega \tau_{\star} \ll 1   \\
		S_{\gamma} - \frac{\gamma}{\tau_{\star}\omega} C_{\gamma}  \quad \, & \omega \tau_{\star} \gg 1 \
	\end{array} \right.	
\end{equation*}
with $C_{\gamma}=\cos\bigl[\gamma\pi/2\bigr]$ and $S_{\gamma}=\sin\bigl[\gamma\pi/2\bigr]$, we are able to infer the following asymptotic behaviour (graphically illustrated in Figure \ref{fig:Chi_URL_DC}) of $\hat{\chi}_{\texttiny{DC}}(\iu \omega)=\chi'_{\texttiny{DC}}(\omega)-\iu \chi''_{\texttiny{DC}}(\omega)$ for low and high frequencies
\[
	\begin{array}{lll}
		\chi'_{\texttiny{DC}}(0) - \chi'_{\texttiny{DC}}(\omega) \sim \frac{\gamma^2\tau_{\star}^2}{2}\omega^{2} \, , \quad &\chi_{\texttiny{DC}}''(\omega) \sim \gamma \tau_{\star} \omega \, , \quad & \omega \tau_{\star} \ll 1\, ,  \\
		\chi'_{\texttiny{DC}}(\omega) \sim C_{\gamma} \tau_{\star}^{-\gamma} \omega^{-\gamma} \, , \quad &\chi''_{\texttiny{DC}}(\omega) \sim S_{\gamma} \tau_{\star}^{-\gamma} \omega^{-\gamma} \, , \quad & \omega \tau_{\star} \gg 1 \, ,
	\end{array}
\]

\begin{figure}[ht]
\centering
\includegraphics[width=0.46\textwidth]{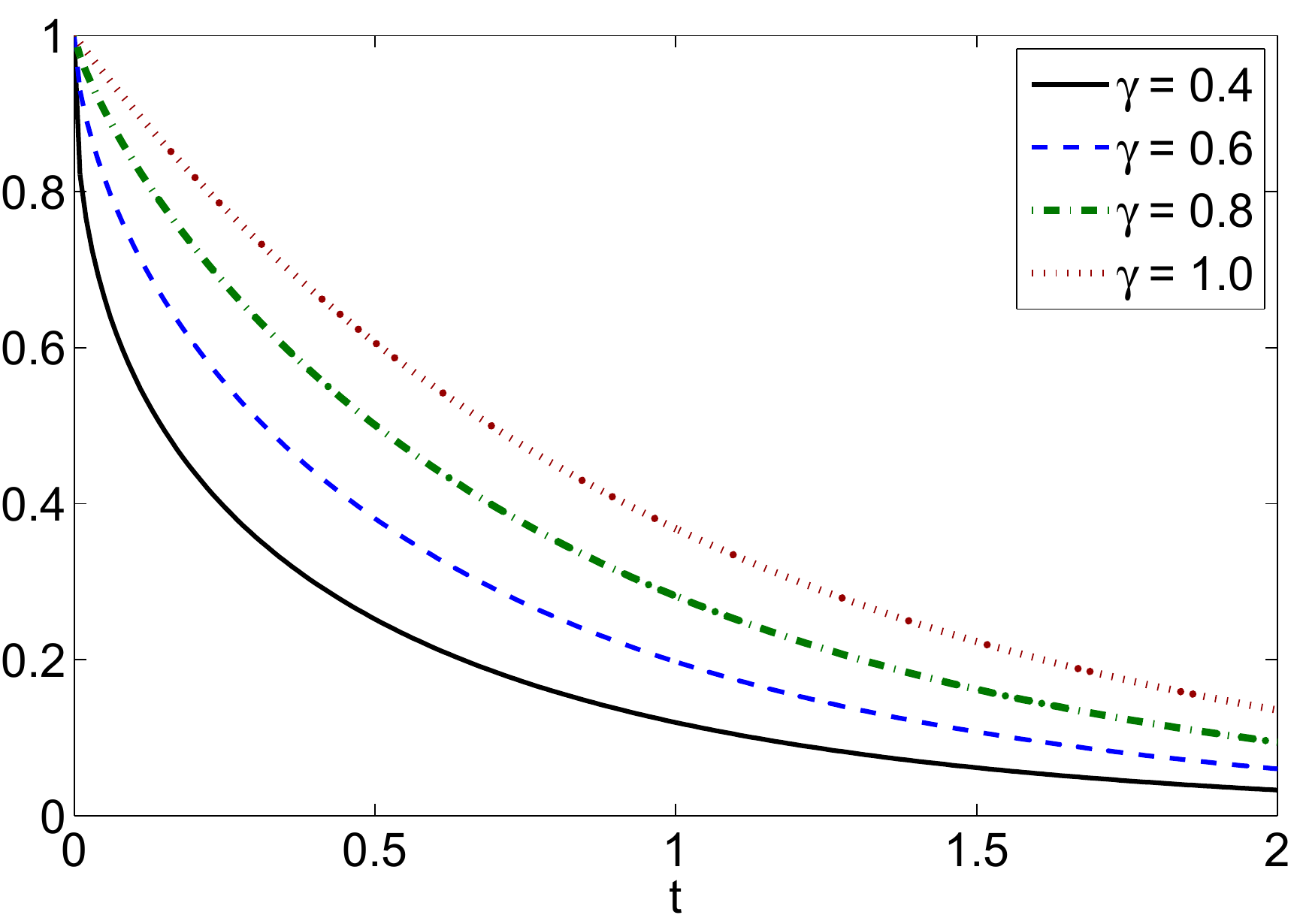}
\includegraphics[width=0.46\textwidth]{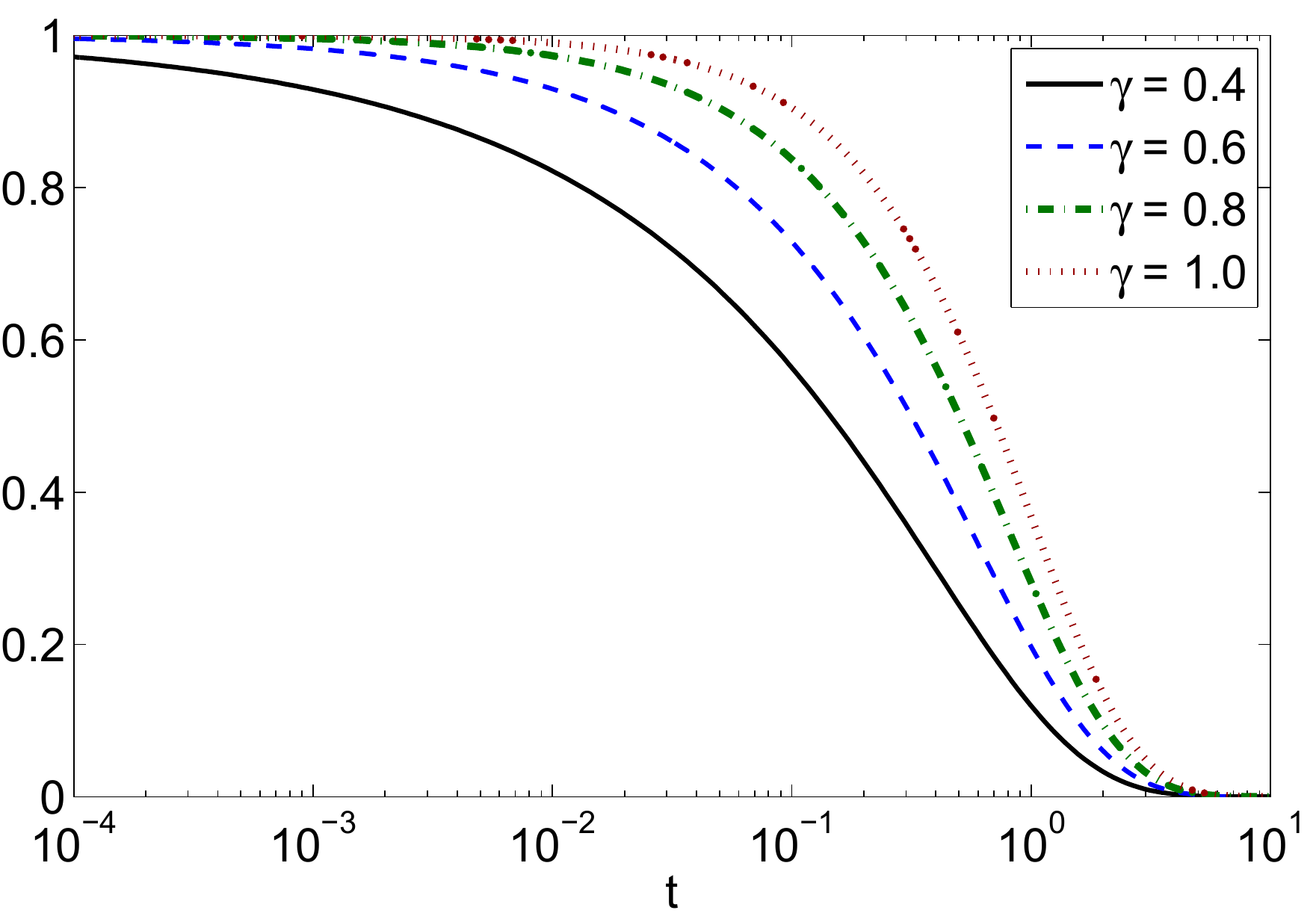}
\caption{Relaxation function $\Psi_{\texttiny{DC}}(t)$ on varying $\gamma$.}
\label{fig:Rel_DC}
\end{figure}

By operating the Laplace transform inversion, we get the corresponding response and relaxation functions
\begin{equation}
	\begin{aligned}
			\phi_{\texttiny{DC}}(t)
			= {\mathcal L}^{-1} \left( \frac{1}{(1+ s\tau_{\star})^\gamma} \right)
		  &= \frac{1}{\tau_{\star}} \, \left( {t}/{\tau_{\star}}\right)^{\gamma-1} E_{1,\gamma}^\gamma \left(- {t}/{\tau_{\star}}\right) \\
			&= \frac{1}{\tau_{\star}} \, \frac{\left({t}/{\tau_{\star}}\right)^{\gamma-1}} 	{\Gamma(\gamma)} \, \exp(-t/\tau_{\star})
  \end{aligned}
\end{equation}
and
\begin{equation}
	\begin{aligned}
		\Psi_{\texttiny{DC}}(t)
		 = {\mathcal L}^{-1} \left( \frac{1}{s} - \frac{1}{s \left(1 + s\tau_{\star} \right)^\gamma }\right)
		&= 1 - \left( {t}/{\tau_{\star}}\right)^{\gamma} E_{1,{\gamma+1}}^\gamma \left(- {t}/{\tau_{\star}}\right) \\
		&= \frac{1}{\Gamma(\gamma)} \Gamma(\gamma,t/\tau_{\star})	
 	\end{aligned}
\end{equation}
where, for $\Re(\gamma)>0$, $\Gamma(a,z)=\int_{z}^{\infty}t^{a-1}\eu^{-t} \, \du t$ is the incomplete gamma function and the last equality for $\Psi_{\texttiny{DC}}(t)$ is obtained by integration of the response function $\phi_{\texttiny{DC}}(t)$, namely after applying (\ref{eq:Psi_int_phi}). The plots of the relaxation function $\Psi_{\texttiny{DC}}(t)$ using linear and logarithmic scales in the normalized time $\tau_{\star}=1$ are reported in Figure \ref{fig:Rel_DC}.

The well-know asymptotic expansion for the incomplete gamma function (e.g., see \cite[Eq. 6.5.32]{AbramowitzStegun1964}) with real and positive argument $z$
\begin{equation}
	\Gamma(a,z) \sim z^{a-1} \eu^{-z} 
	, \quad z \to \infty \, ,
\end{equation}
allows to see that, at variance with the CC model, the characteristic functions $\Psi_{DC}(t)$ and $\phi_{DC}(t)$
decay both exponentially for large times, so more rapidly of any power law, namely
 \begin{equation}
	\phi_{\texttiny{DC}}(t) \sim \left\{ \begin{array}{ll}
		\displaystyle\frac{1}{\tau_{\star} \Gamma(\gamma)}\bigl( t/\tau_{\star} \bigr)^{\gamma-1} \, , \, & \text{for } t \ll \tau_{\star}  , \\
		\displaystyle\frac{1}{\tau_{\star} \Gamma(\gamma)} \bigl( t/\tau_{\star} \bigr)^{\gamma-1} \exp\bigl( - t/\tau_{\star} \bigr)  \, , \, & \text{for } t \gg \tau_{\star} ,\
	\end{array} \right.
\end{equation}
and
\begin{equation}
	\Psi_{\texttiny{DC}}(t) \sim \left\{ \begin{array}{ll}
		1 - \displaystyle\frac{1}{\Gamma(\gamma+1 )}\bigl( t/\tau_{\star} \bigr)^{\gamma} \, , \, & \text{for } t \ll \tau_{\star} , \\
		\displaystyle\frac{1}{\Gamma(\gamma)} \bigl( t/\tau_{\star} \bigr)^{\gamma-1} \exp\bigl( - t/\tau_{\star} \bigr) \, , \, & \text{for } t \gg \tau_{\star}. \\
	\end{array} \right.
\end{equation}

Furthermore the spectral distribution functions exhibit a cut off, so they vanish at low frequencies $r < 1/\tau_{\star}$ and indeed we get the following expressions
\begin{equation}
	K^{\phi}_{\texttiny{DC}}(r)
	= \left\{ \begin{array}{ll}
			0 \, , \quad & r < 1/\tau_{\star} \, ,\\
			\displaystyle\frac{1}{\pi} \, \frac{\sin (\gamma\pi)}{(r\tau_{\star}-1)^{\gamma}} \, , \quad & r>1/\tau_{\star} \, , \
	  \end{array}	\right.
\end{equation}
and
\begin{equation}
	K^{\Psi}_{\texttiny{DC}}(r)
	= \left\{ \begin{array}{ll}
			0 \, , \quad & r < 1/\tau_{\star} \, , \\
			\displaystyle\frac{1}{\pi} \, \frac{\sin (\gamma\pi)}{r(r\tau_{\star}-1)^{\gamma}} \, , \quad & r>1/\tau_{\star} \, . \
	  \end{array}	\right.
\end{equation}

\begin{figure}[ht]
\centering
\includegraphics[width=0.46\textwidth]{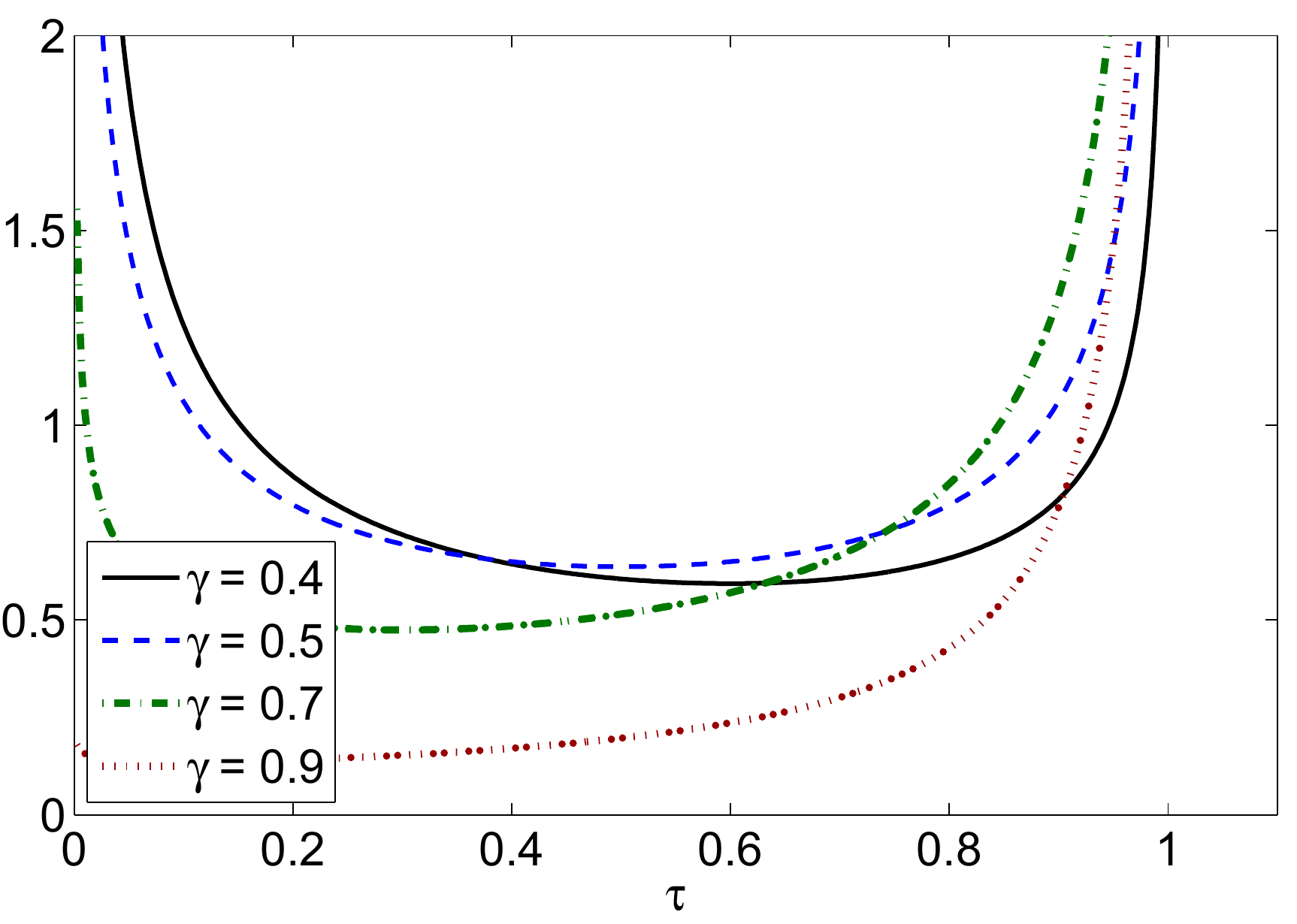}
\includegraphics[width=0.46\textwidth]{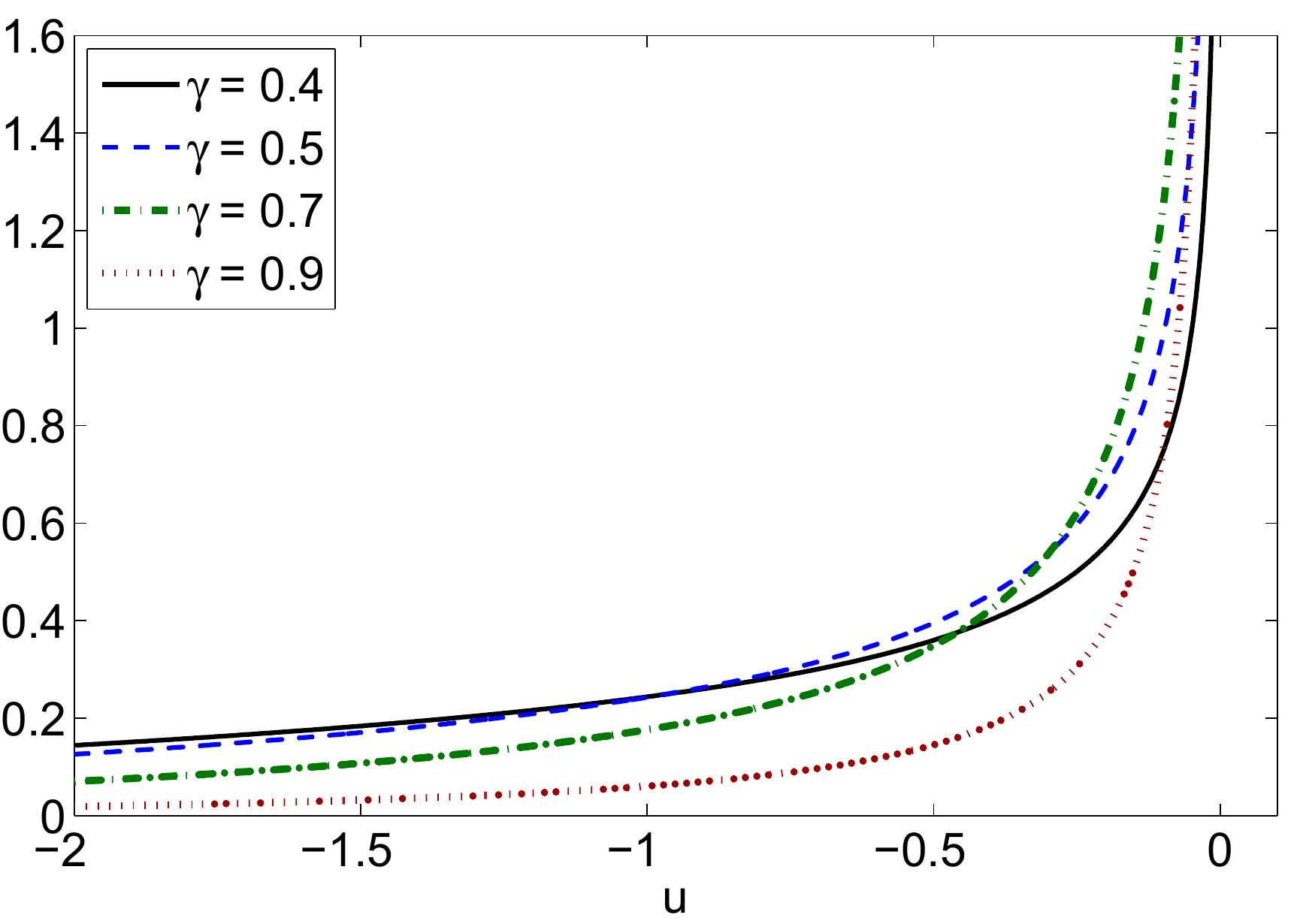}
\caption{Spectral distributions $H^{\Psi}_{\texttiny{DC}}(\tau)$ (left) and $L^{\Psi}_{\texttiny{DC}}(u)$ (right).}
\label{fig:Spectral_DC}
\end{figure}

For the plots of spectral distributions, in Figure \ref{fig:Spectral_DC} we limit to exhibit (as for CC model) those corresponding to the relaxation function $\Psi_{\texttiny{DC}}(t)$, that is
\begin{equation}
	H^{\Psi}_{\texttiny{DC}}(\tau)
	= \left\{ \begin{array}{ll}
		0 \, , \quad & \tau>\tau_{\star} \, , \\
		\displaystyle \frac{1}{\pi \tau} \, \frac {\sin (\gamma\pi)}{(\tau_{\star}/\tau - 1)^{\gamma}} \, , \quad & \tau<\tau_{\star} \, ,
	\end{array}	\right.
\end{equation}
and $L^\Psi_{\texttiny{DC}}(u)=\eu^{-u}K^\Psi_{\texttiny{DC}}(\eu^{-u})$, where $u =\log (\tau)$.

By standard derivation it is elementary to see that the response $\phi_{\texttiny{DC}}(t)$ satisfies the equation
\begin{equation}
	D_{t} \, \phi_{\texttiny{DC}}(t) = - \frac{1}{\tau_{\star}} \left[ 1 - (\gamma-1)\frac{\tau_{\star}}{t} \right] \phi_{\texttiny{DC}}(t)
\end{equation}
but by taking into account the composite operator (\ref{eq:DC_Operator}) it is also possible to obtain
\begin{equation*}
	\left(D_t + \tau_{\star}^{-1} \right)^{\gamma} \phi_{\texttiny{DC}}(t) = \eu^{-t/\tau_{\star}} {}_{0}D^{\gamma}_t \eu^{t/\tau_{\star}} \phi_{\texttiny{DC}}(t) = \frac{1}{\tau_{\star}} \eu^{-t/\tau_{\star}} {}_{0} D^{\gamma}_t \frac{\left({t}/{\tau_{\star}}\right)^{\gamma-1}}{\Gamma(\gamma)} = 0 \, ,
\end{equation*}
where for the last equality we refer to \cite[Example 2.4]{Diethelm2010}; hence $\phi_{\texttiny{DC}}(t)$ satisfies the following equation \begin{equation}
	\left(D_t + \tau_{\star}^{-1} \right)^{\gamma} \phi_{\texttiny{DC}}(t) = 0
	, \quad
	\lim_{t\to 0^{+}}  {}_{0}J_t^{1-\gamma} \bigl[ \eu^{t/\tau_{\star}} \phi_{\texttiny{DC}}(t) \bigr] = \frac{1}{\tau_{\star}^{\gamma}}
	\, ,
\end{equation}
where the initial condition is obtained from (\ref{eq:DC_Operator_LT}). Observe that, as $t\to0^{+}$, the contribution of the exponential $\eu^{t/\tau_{\star}}$ in the fractional integral ${}_{0}J_t^{1-\gamma}$ is always equal 1, i.e.
\[
	\lim_{t\to 0^{+}}  {}_{0}J_t^{1-\gamma} \bigl[ \eu^{t/\tau_{\star}} \phi_{\texttiny{DC}}(t) \bigr]
	= \lim_{t\to 0^{+}}  {}_{0}J_t^{1-\gamma} \phi_{\texttiny{DC}}(t) \, ,
\]
thus providing the same initial condition associated to the equation (\ref{eq:CC_EvolEq_phi}) for the CC model.

For the relaxation function $\Psi_{\texttiny{DC}}(t)$ we use the operator
${}^{{\mathcal C}}{}{\bigl(D_t + \lambda \bigr)^{\gamma}}$
defined in (\ref{eq:DC_Operator_Caputo})
and standard derivations allow us to compute
\begin{equation}\label{eq:DC_EvolEq_Psi}
	{}^{{\mathcal C}}{}{\bigl(D_t + \tau_{\star}^{-1} \bigr)^{\gamma}} \Psi_{\texttiny{DC}}(t)
	= - \frac{1}{\tau_{\star}^{\gamma}} \left[ 1 - \left( {t}/{\tau_{\star}}\right)^{1-\gamma} E_{1,{2-\gamma}}^{1-\gamma} \left(- {t}/{\tau_{\star}}\right)  \right]
	\, ,
\end{equation}
with the usual initial condition $\Psi_{\texttiny{DC}}(0) = 1$. It is worthwhile to note that when $\gamma=1$ it is $E_{1,{2-\gamma}}^{1-\gamma} \left(- {t}/{\tau_{\star}}\right)  \equiv 1$ and, as expected, (\ref{eq:DC_EvolEq_Psi}) returns the standard evolution equation for the relaxation function of the Debye model.

Alternatively, we can consider the particular case, for $\alpha=1$, of the operator ${}^{\text{\tiny{C}}}{}{\bigl({}_{0}D^{\alpha}_{t} + \tau_{\star}^{-1} \bigr)^{\gamma}}$ introduced in (\ref{eq:HN_Caputo}) and the corresponding equation would read as
\begin{equation}\label{eq:DC_EvolEq_Psi2}
	{}^{\text{\tiny{C}}}{}{\bigl(D_{t} + \tau_{\star}^{-1} \bigr)^{\gamma}} \Psi_{\texttiny{DC}}(t) = - \frac{1}{\tau_{\star}^{\gamma}}
	\, , \quad
	\Psi_{\texttiny{DC}}(0) = 1 \, .
\end{equation}

The difference between (\ref{eq:DC_EvolEq_Psi}) and (\ref{eq:DC_EvolEq_Psi2}) is clearly related to the use of different operators ${}^{{\mathcal C}}{}{\bigl(D_t + \tau_{\star}^{-1} \bigr)^{\gamma}}$ and ${}^{\text{\tiny{C}}}{}{\bigl(D_{t} + \tau_{\star}^{-1} \bigr)^{\gamma}}$; for a full understanding of the difference between operators marked by the symbols ``$\mathcal C$'' and ``$\text{C}$'' we refer to Appendix {\ref{S:FractionalIntegralsDerivatives}} and in particular to Remark {\ref{rem:HN_Caputo2}}.

\subsection{The Havriliak-Negami model}\label{SS:HavriliakNegami} 

In 1967 the American S.J. Havriliak and the Japanese-born S. Negami proposed a new model \cite{HavriliakNegami1967} with two real powers to take into account, at the same time, both the asymmetry and the broadness observed in the shape of the permittivity spectrum of some polymers.

\begin{figure}[ht]
\centering
\includegraphics[width=.60\textwidth]{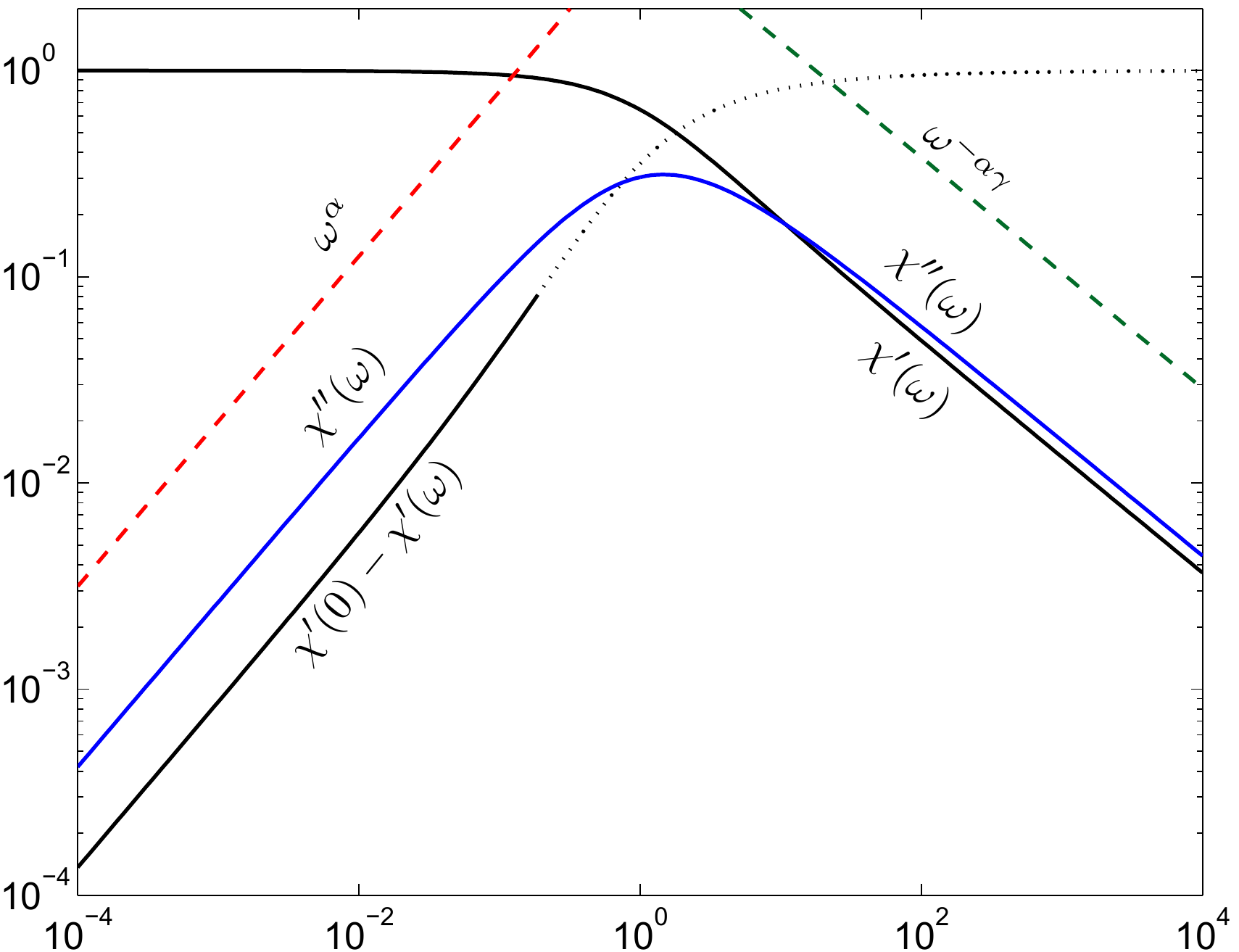}
\caption{HN susceptibility for $\alpha=0.8$ and $\gamma=0.7$.}
\label{fig:Chi_URL_HN}
\end{figure}

The normalized complex susceptibility proposed with the Havriliak-Negami (HN) model is given by
\begin{equation}\label{eq:HN}
	\hat{\chi}_{\texttiny{HN}}(\iu \omega) = \frac{1}{\left( 1 + \bigl( \iu \tau_{\star} \omega \bigr)^{\alpha} \right)^{\gamma}}
\end{equation}
and it is immediate to verify that, since
\begin{equation}
	\begin{array}{r@{\hspace{2pt}}c@{\hspace{2pt}}lr}
		\hat{\chi}_{\texttiny{HN}}(\iu\omega) &\sim&  \bigl( \iu \tau_{\star} \omega)^{-\alpha\gamma} , \, \, &
		 \tau_{\star} \omega \gg 1 , \\
		\Delta \hat{\chi}_{\texttiny{HN}}(\iu\omega) = \chi_{\texttiny{HN}}(0) - \hat{\chi}_{\texttiny{HN}}(\iu\omega) &\sim& \gamma \bigl( \iu \tau_{\star} \omega)^{\alpha} , \, \, &
		\tau_{\star} \omega \ll 1 ,\\
	\end{array}
\end{equation}
the model fits in the Jonscher's universal law with $m=\alpha$ and $n=1-\alpha\gamma$. The representation of the HN susceptibility, together with its asymptotic behaviour at low and high frequencies, is illustrated in Figure \ref{fig:Chi_URL_HN}

\begin{figure}[ht]
\centering
\includegraphics[width=0.48\textwidth]{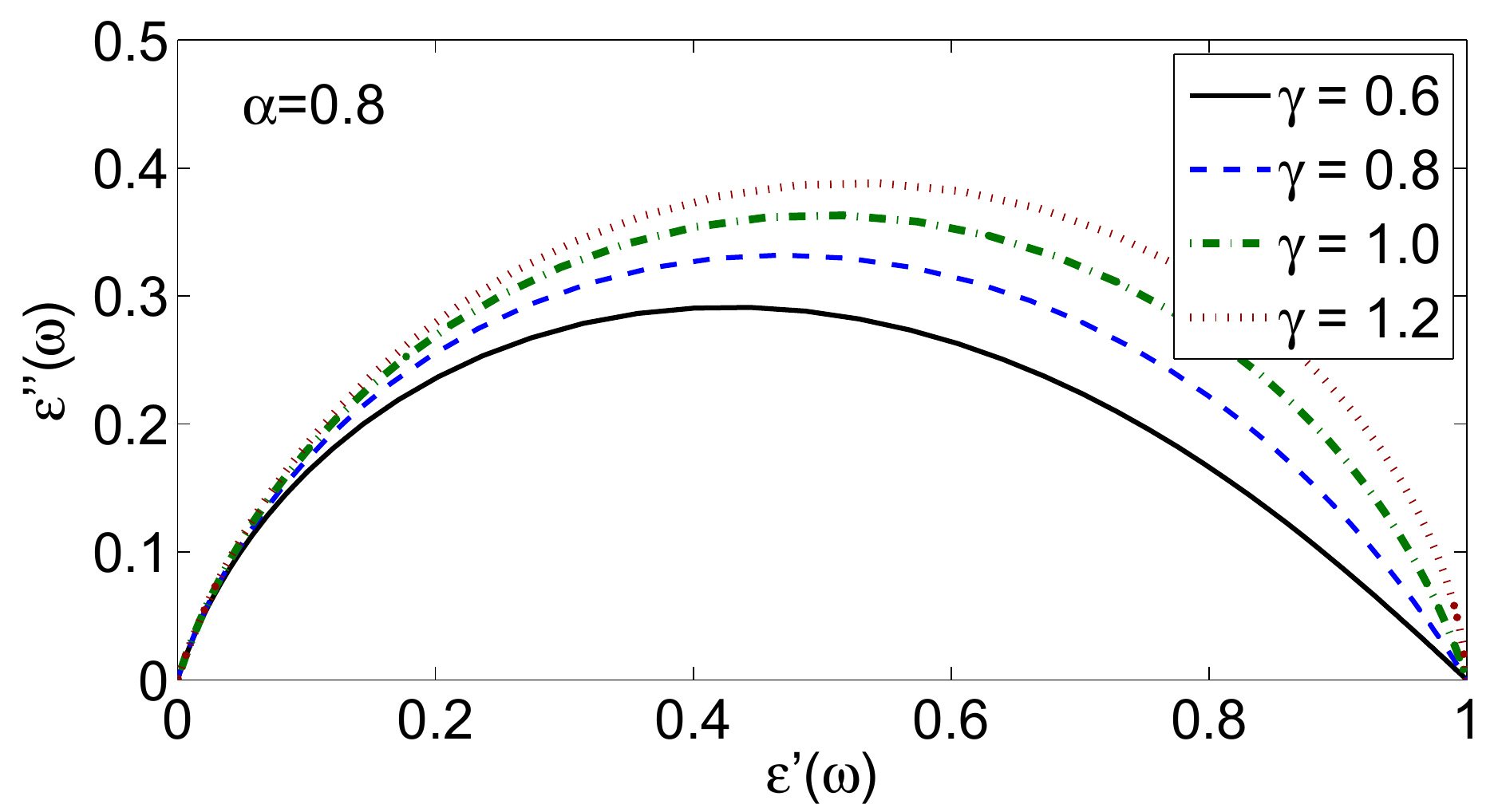}
\includegraphics[width=0.48\textwidth]{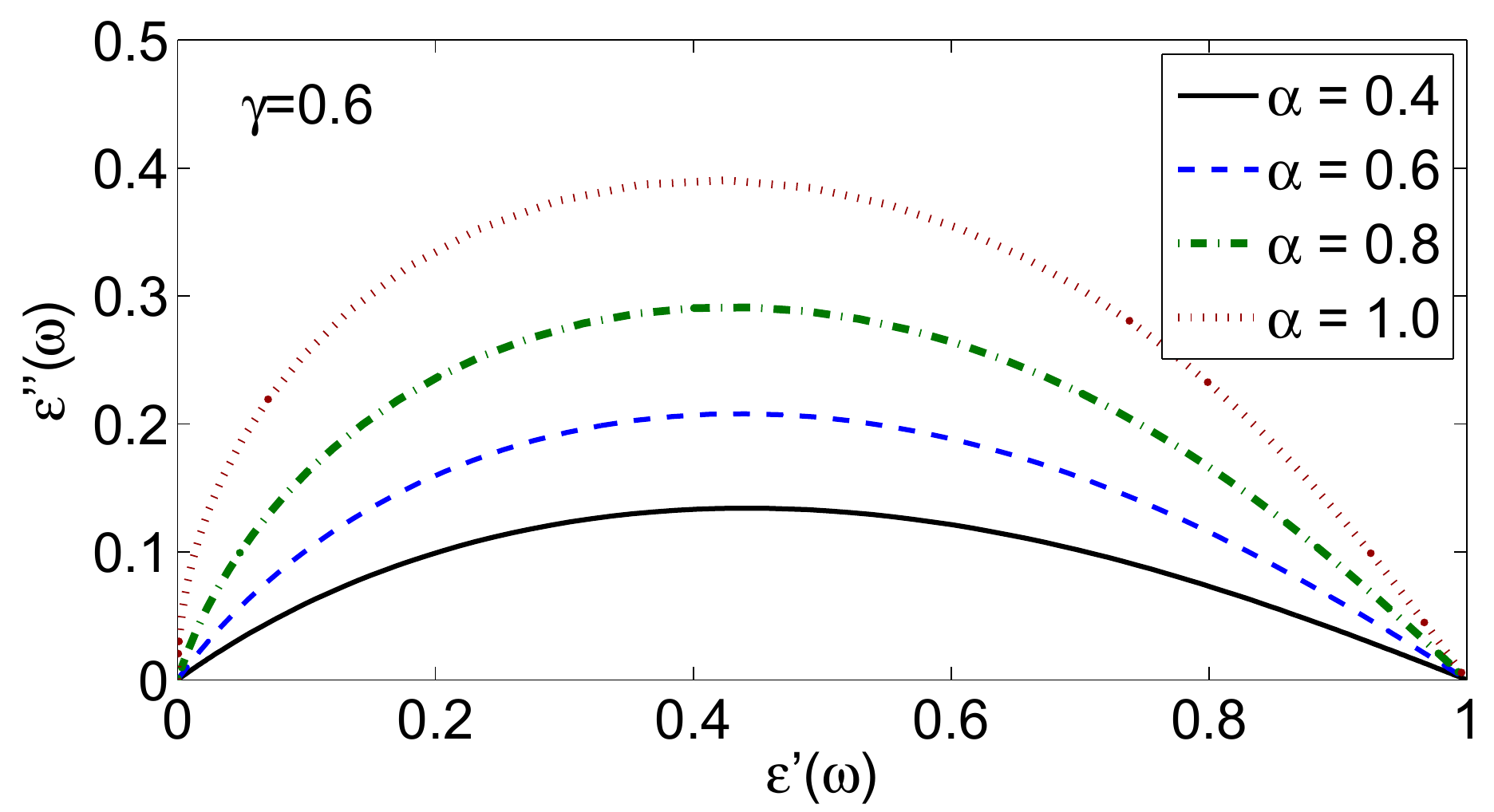}
\caption{Cole-Cole plots for the HN model.}
\label{fig:ColeCole_HN}
\end{figure}

In Figures \ref{fig:ColeCole_HN} we show as the Cole-Cole plots of the HN model changes with respect to changes in $\gamma$ (left plots) and to changes of $\alpha$ (right plots).

The time-domain response and the time-domain relaxation of the HN model are respectively
\begin{equation}
	\phi_{\texttiny{HN}}(t) = \frac{1}{\tau_{\star}} \bigl( t/\tau_{\star} \bigr)^{\alpha\gamma-1} E_{\alpha,\alpha\gamma}^{\gamma} \left(- \bigl( t/\tau_{\star} \bigr)^{\alpha} \right)
\end{equation}
and
\begin{equation}
	\Psi_{\texttiny{HN}}(t) = 1 - \bigl( t/\tau_{\star} \bigr)^{\alpha\gamma} E_{\alpha,\alpha\gamma+1}^{\gamma} \left(- \bigl( t/\tau_{\star} \bigr)^{\alpha} \right) ,
\end{equation}
where $E_{\alpha,\beta}^{\gamma}(z)$ is the three-parameter Mittag-Leffler (ML) function, also known as the Prabhakar function, described in Appendix \ref{S:ML}.

\begin{figure}[ht]
\centering
\includegraphics[width=0.46\textwidth]{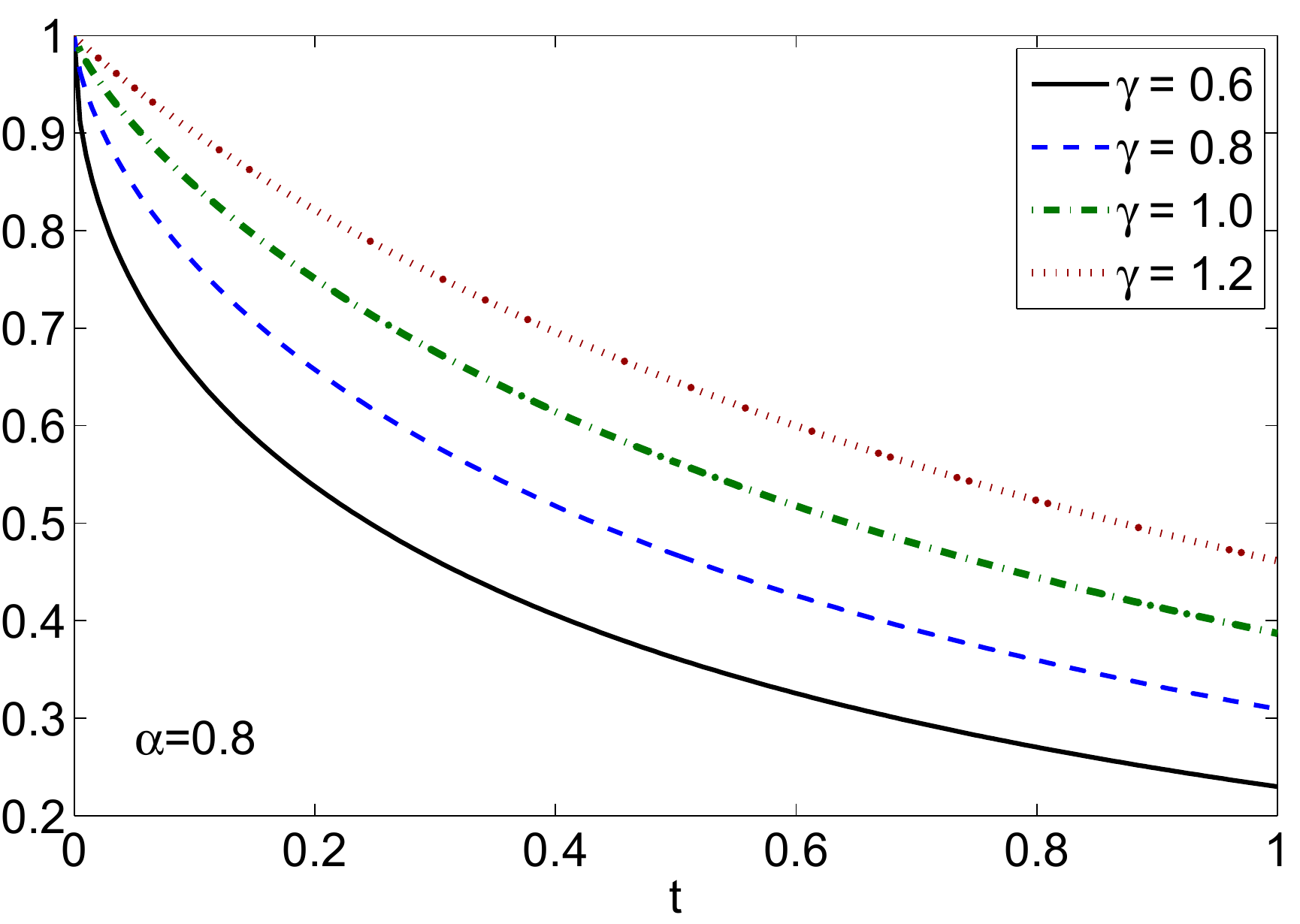}
\includegraphics[width=0.46\textwidth]{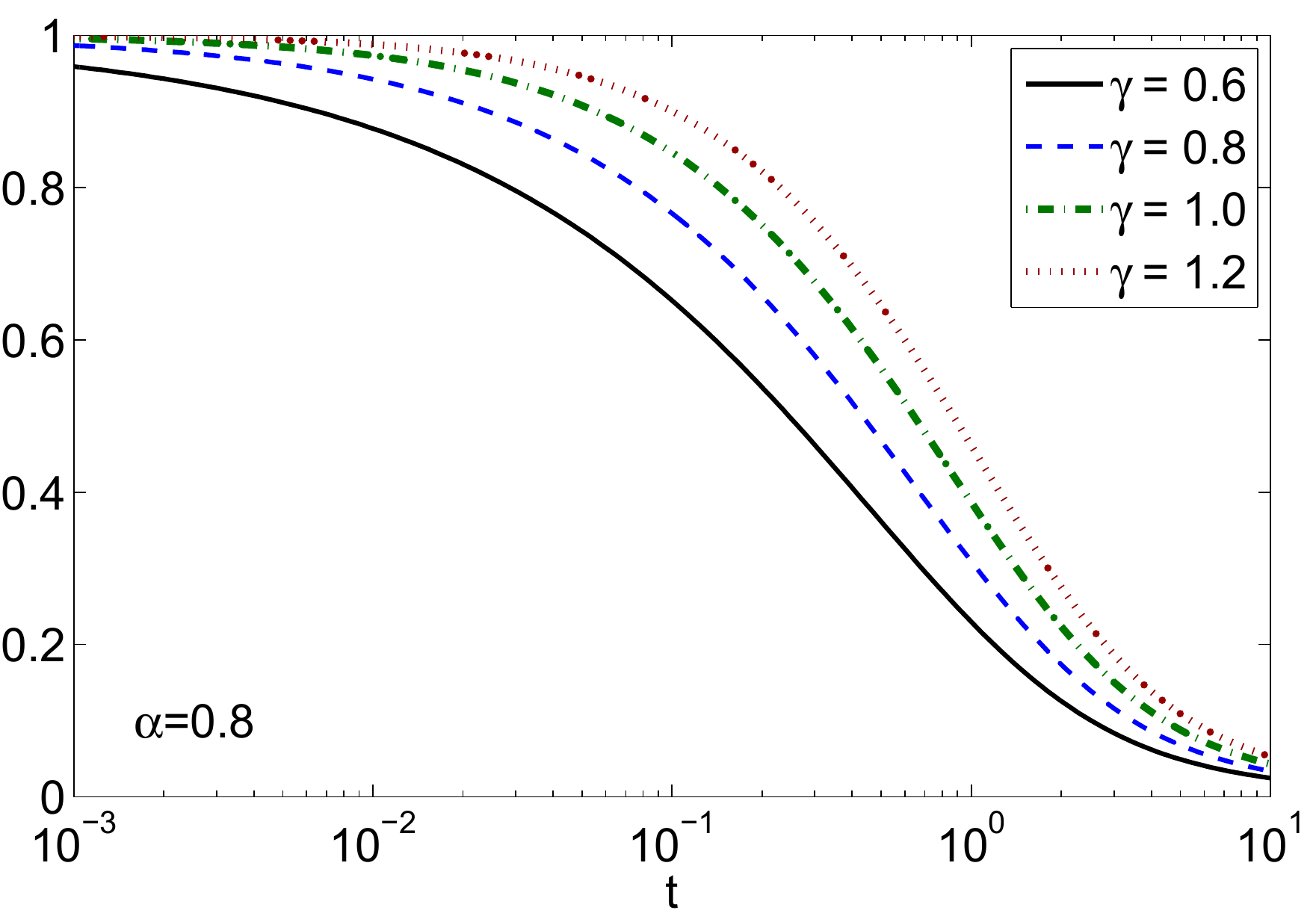}
\caption{Relaxation functions $\Psi_{\texttiny{HN}}(t)$ for $\alpha=0.8$.}
\label{fig:Rel_Hng}
\end{figure}

\begin{figure}[ht]
\centering
\includegraphics[width=0.46\textwidth]{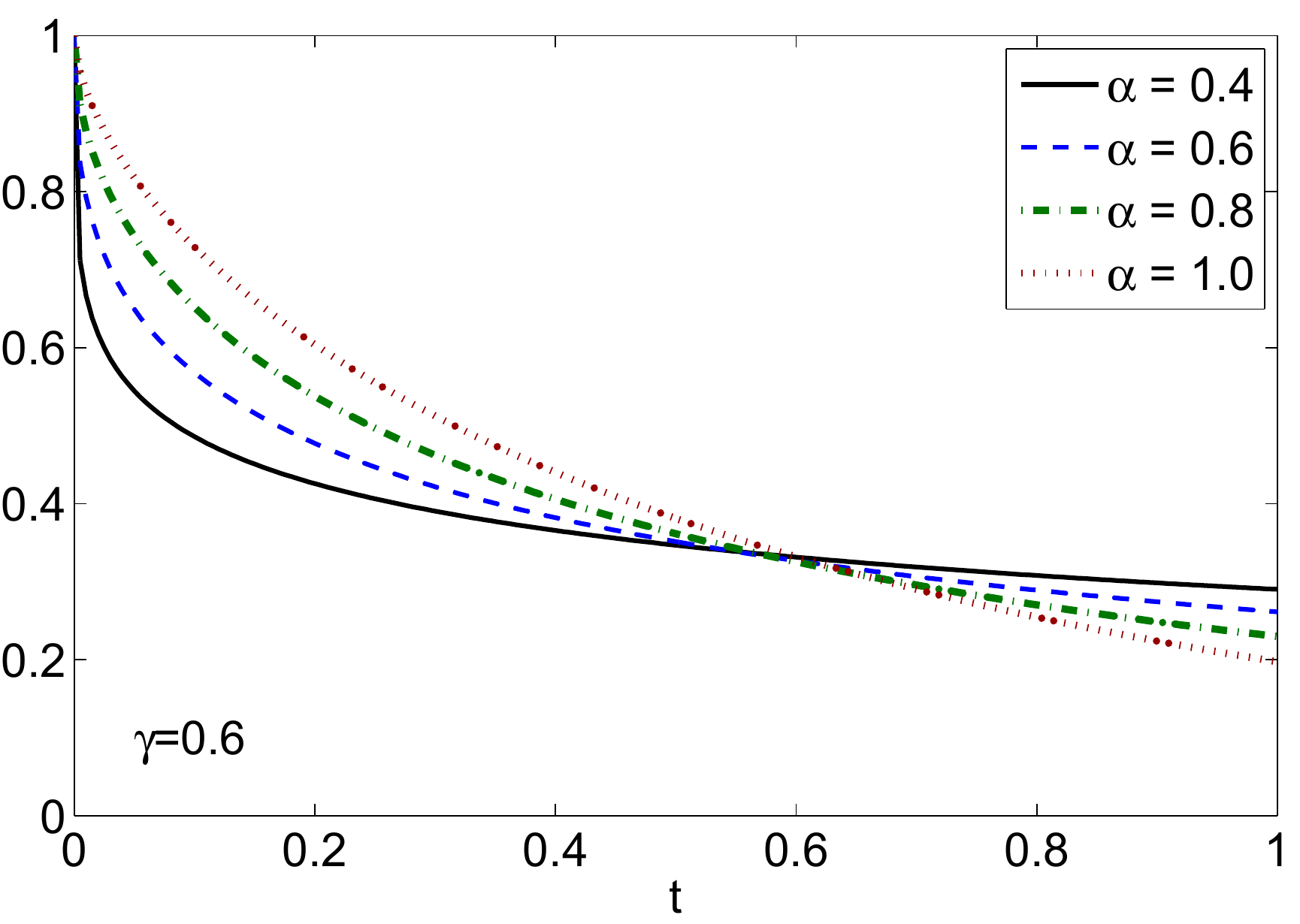}
\includegraphics[width=0.46\textwidth]{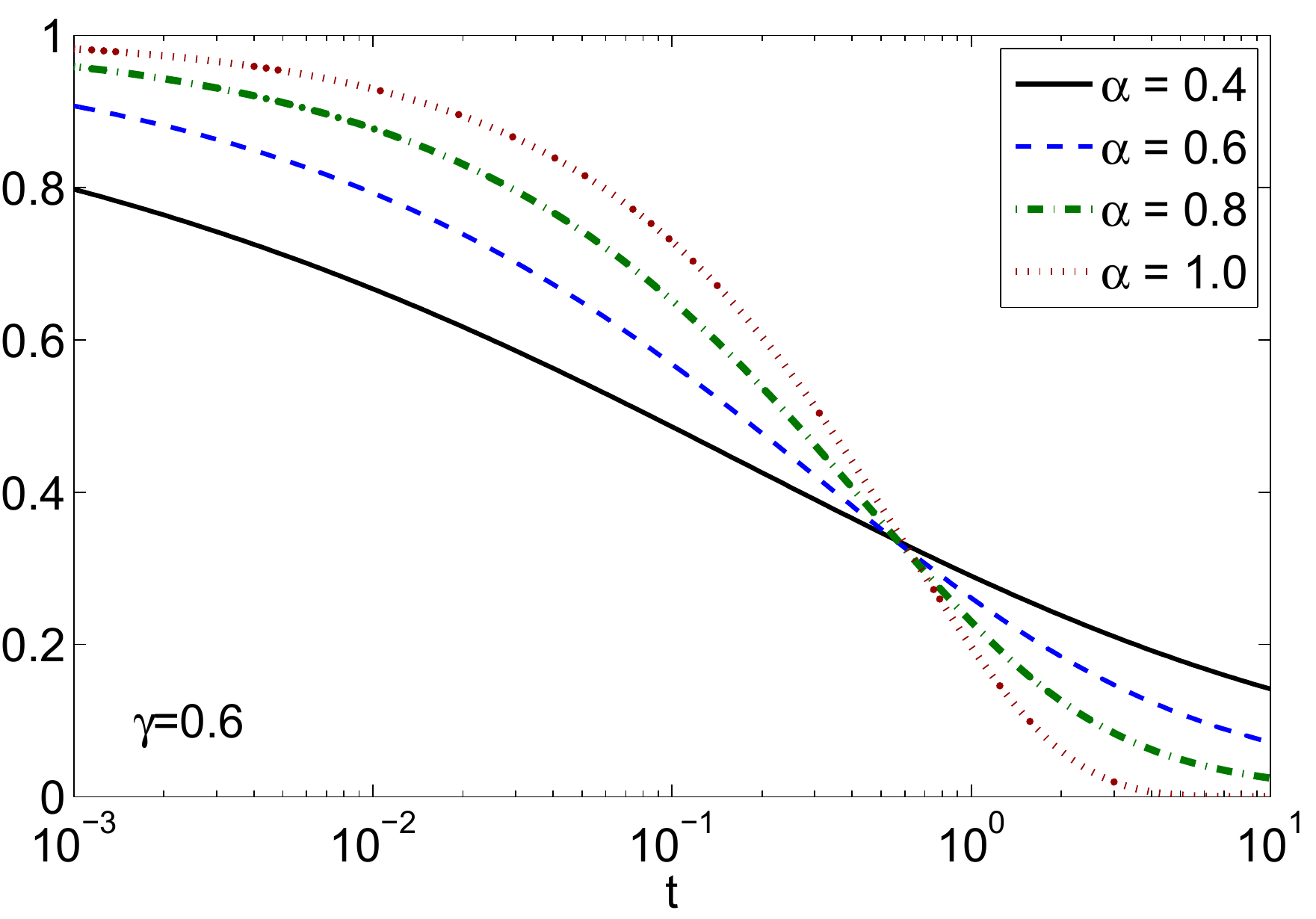}
\caption{Relaxation functions $\Psi_{\texttiny{HN}}(t)$ for $\gamma=0.6$.}
\label{fig:Rel_Hna}
\end{figure}

Some plots of the relaxation function $\Psi_{HN}(t)$ on varying $\gamma$ are reported in Figure \ref{fig:Rel_Hng} and on varying $\alpha$ in Figure \ref{fig:Rel_Hna}; as usual, the left plots are in normal scale, the right plots in logarithmic scale and $\tau_{\star}=1$.


By considering the series definition (\ref{eq:ML3}) for small $t$ and the asymptotic expansion (\ref{eq:ML3_exp_large}) for $t\to \infty$, it is possible to verify that the HN response has the following short- and long-time power-law dependencies
\begin{equation}
	\phi_{\texttiny{HN}}(t) \sim \left\{ \begin{array}{ll}
		\displaystyle\frac{1}{\tau_{\star} \Gamma(\alpha\gamma)}\bigl( t/\tau_{\star} \bigr)^{\alpha\gamma-1} , \, & \text{for } t \ll \tau_{\star}\, , \\
		-\displaystyle\frac{\gamma}{\tau_{\star} \Gamma(-\alpha)} \bigl( t/\tau_{\star} \bigr)^{-\alpha-1} , \, & \text{for } t \gg \tau_{\star} \, ,\
	\end{array} \right.
\end{equation}
while, thanks to (\ref{eq:ML3}) and (\ref{eq:ML3_exp_large_compl}) respectively, the short- and long-time power law dependencies of the HN relaxation for $0<\alpha<1$ are
\begin{equation}
	\Psi_{\texttiny{HN}}(t) \sim \left\{ \begin{array}{ll}
		1- \displaystyle\frac{1}{\Gamma(\alpha\gamma+1 )}\bigl( t/\tau_{\star} \bigr)^{\alpha\gamma} , \, & \text{for } t \ll \tau_{\star} \, ,\\
		\displaystyle\frac{\gamma}{\Gamma(1-\alpha)} \bigl( t/\tau_{\star} \bigr)^{-\alpha} , \, & \text{for } t \gg \tau_{\star} \, .\
	\end{array} \right.
\end{equation}

There is a lively debate in the literature about the range of admissibility of the parameters $\alpha$ and $\gamma$. Usually it is assumed $0<\alpha,\gamma \le 1$ but in \cite{HavriliakHavriliak1994}, on the basis of the observation of a large amount of experimental data, it was proposed an extension to $0<\alpha,\alpha\gamma \le 1$. The completely monotonicity of the relaxation and response functions (which is considered an essential feature for the admissibility of the model \cite{Hanyga2005b}) has been recently proved in \cite{CapelasMainardiVaz2011,MainardiGarrappa2015} also for this extended range of parameters.

To this purpose we observe that the inversion formulas (\ref{eq:InversionSpectral}) leads  to
\begin{equation}
	K^{\Psi}_{{\texttiny{HN}}}(r) = \frac{\tau_{\star}}{\pi}
	\frac{ (\tau_{\star} r)^{-1}  \sin\left[ \gamma \, \theta_\alpha(r) \right]}{\left( (\tau_{\star} r)^{2\alpha} + 2 (\tau_{\star} r)^{\alpha}\,\cos(\alpha\pi)+1 \right)^{\gamma/2}}
\end{equation}
and
\begin{equation}
	K^{\phi}_{{\texttiny{HN}}}(r) = \frac{1}{\pi}
	\frac{\sin\left[ \gamma \, \theta_\alpha(r) \right]}{\left( (\tau_{\star} r)^{2\alpha} + 2 (\tau_{\star} r)^{\alpha}\,\cos(\alpha\pi)+1 \right)^{\gamma/2}} \, ,
\end{equation}
where
\begin{equation}\label{eq:Theta_HN}
	\theta_\alpha(r) =
	\frac{\pi}{2} - \arctan\left[ \frac{ \cos(\pi \alpha) + (\tau_{\star} r)^{-\alpha} }{ \sin (\pi \alpha)} \right]  \in [0,\pi] ,
\end{equation}
and since $(\tau_{\star} r)^{-\alpha} \ge 0$ the argument of the arctangent function is clearly $\ge 1/ \tan \pi \alpha$ and hence $\theta_\alpha(r) \le \alpha \pi$ from which it follows that $K^{\phi}_{{\texttiny{HN}}}(r) \ge 0$ for any $r \ge 0$ and for $0<\alpha,\alpha\gamma \le 1$.

We can therefore, by means of (\ref{eq:H_K}), consider for the relaxation function $\Psi_{\texttiny{HN}}(t)$ the time spectral distribution
\begin{equation}
	H^{\Psi}_{\texttiny{HN}}(\tau) =
 \frac{1}{\pi \tau} \frac{\sin\left[ \gamma \, \theta_\alpha(1/\tau) \right]}{\left( (\tau/\tau_{\star})^{-2\alpha} + 2 (\tau/\tau_{\star})^{-\alpha}\,\cos(\alpha\pi)+1 \right)^{\gamma/2}} \, ,
\end{equation}
and, thanks to (\ref{eq:L_K})-(\ref{eq:L_H}), its representation on logarithmic scale $u= \log(\tau)$
\begin{equation}
	L^{\Psi}_{\texttiny{HN}}(u) =
	\frac{1}{\pi} \frac{\sin\left[ \gamma \, \theta_\alpha(\eu^{-u}) \right]}{\left( \tau_{\star}^{2\alpha} \eu^{-2 \alpha u}  + 2  \tau_{\star}^{\alpha} \eu^{-\alpha u} \,\cos(\alpha\pi) + 1 \right)^{\gamma/2}} \, .
\end{equation}

\begin{figure}[ht]
\centering
\includegraphics[width=0.46\textwidth]{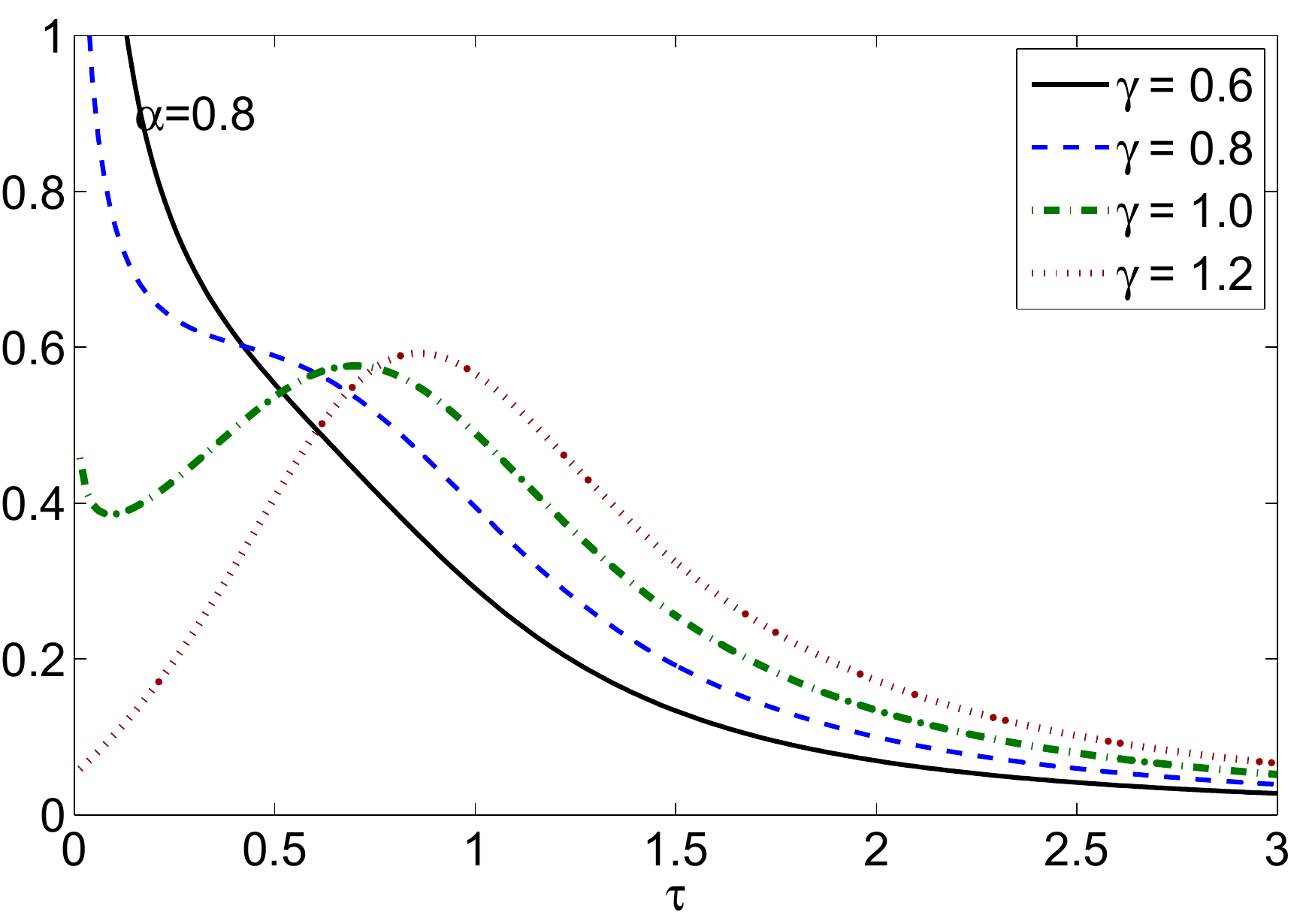}
\includegraphics[width=0.46\textwidth]{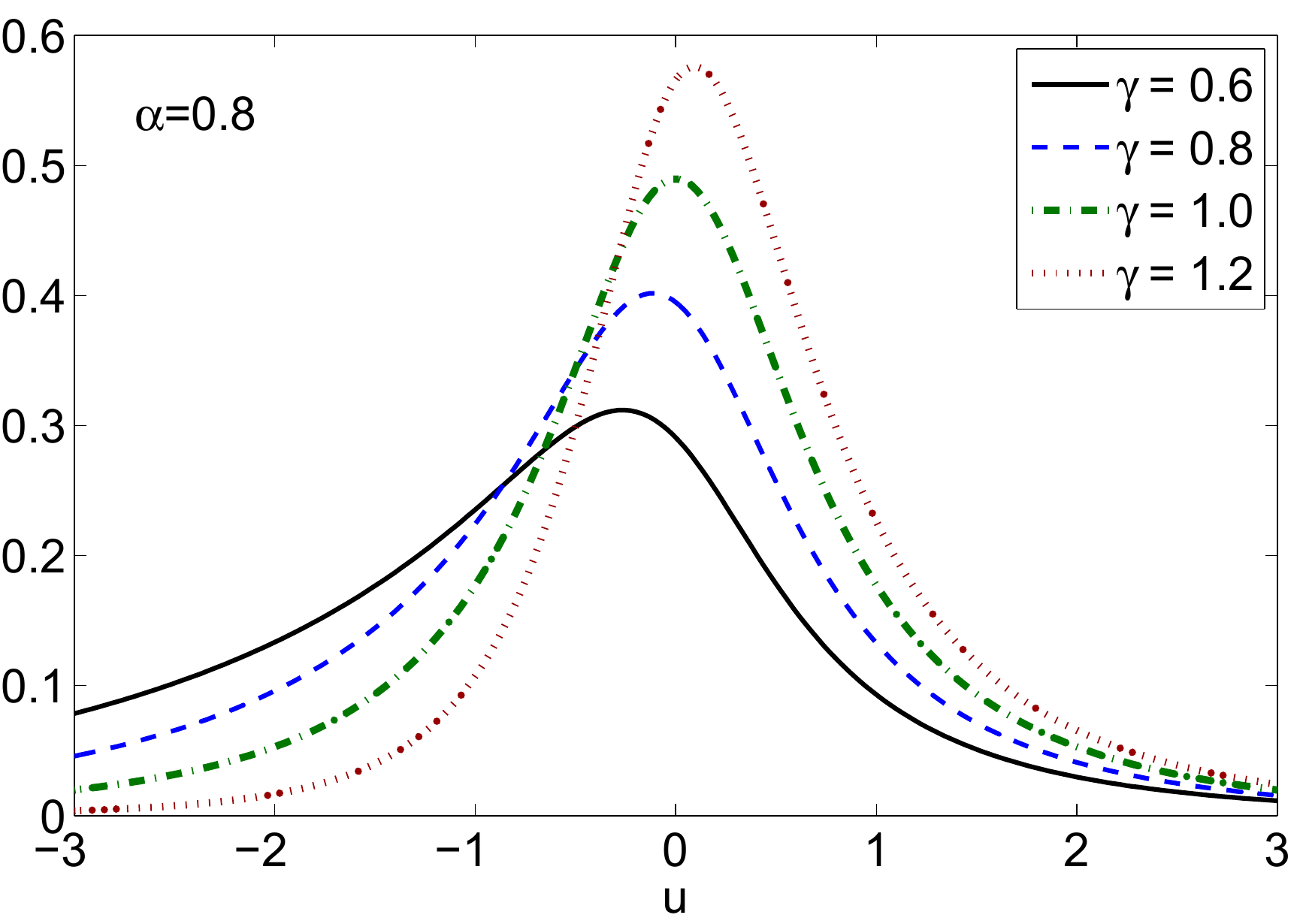}
\caption{Spectral distributions $H^{\Psi}_{\texttiny{HN}}(\tau)$ (left) and $L^{\Psi}_{\texttiny{HN}}(u)$ (right) for $\alpha=0.8$.}
\label{fig:Spectral_Hng}
\end{figure}

\begin{figure}[ht]
\centering
\includegraphics[width=0.46\textwidth]{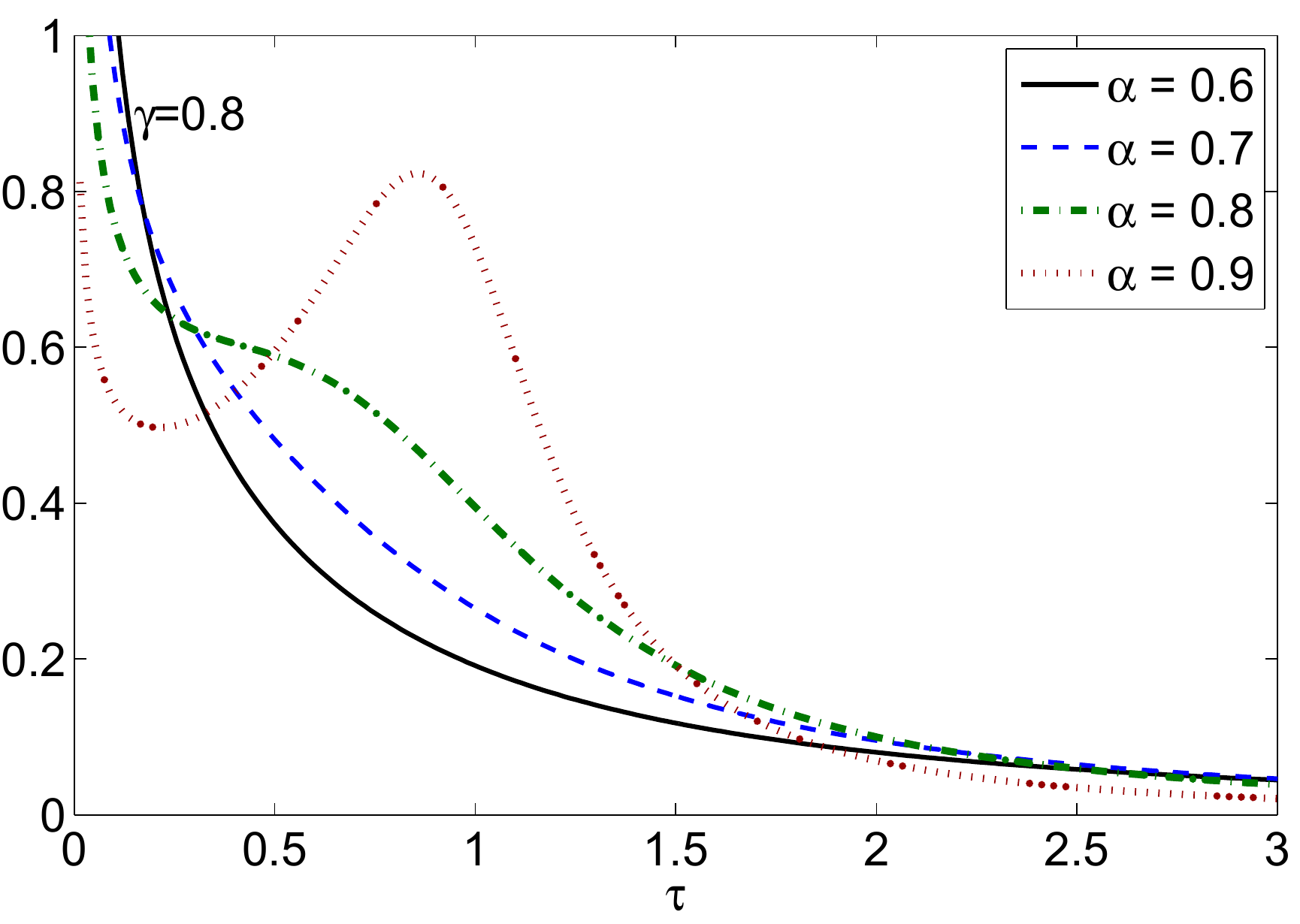}
\includegraphics[width=0.46\textwidth]{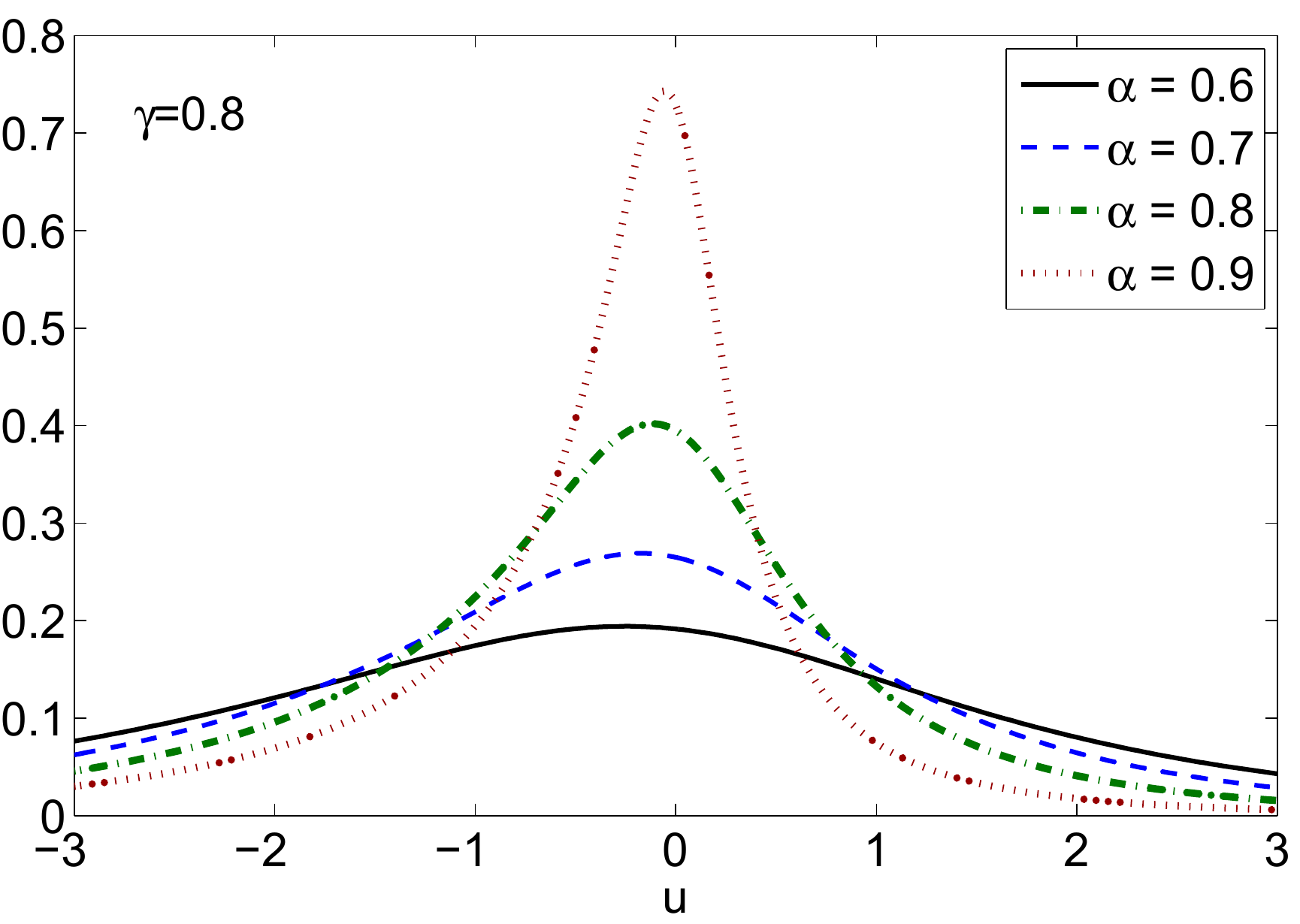}
\caption{Spectral distributions $H^{\Psi}_{\texttiny{HN}}(\tau)$ (left) and $L^{\Psi}_{\texttiny{HN}}(u)$ (right) for $\gamma=0.8$.}
\label{fig:Spectral_Hna}
\end{figure}

Few instances of the time spectral distributions $H^{\Psi}_{\texttiny{HN}}(\tau)$ and $L^{\Psi}_{\texttiny{HN}}(u)$, on varying $\gamma$ and $\alpha$ respectively, are presented in Figures \ref{fig:Spectral_Hng} and \ref{fig:Spectral_Hna}.


To derive the evolution equations for the HN characteristic functions, we start by recalling that since the Laplace transform of the response $\phi_{\texttiny{HN}}(t)$ is
\[
	\widetilde{\phi}_{\texttiny{HN}}(s) = \frac{1}{\tau_{\star}^{\alpha\gamma}(s^{\alpha} + \tau_{\star}^{-\alpha})^{\gamma}} ,
\]
after using (\ref{eq:HN_der_LT}) it is straightforward to obtain
\begin{equation}
	{\mathcal L} \left( \bigl({}_{0}D^{\alpha}_{t} +\tau_{\star}^{-\alpha} \bigr)^{\gamma}  \phi_{\texttiny{HN}}(t) \, ; \, s \right)
	= \frac{1}{\tau_{\star}^{\alpha\gamma}}
	- \lim_{t\to0^+} {\mathbf E}_{\alpha,1-\alpha\gamma,-\tau_{\star}^{-\alpha},0^{+}}^{-\gamma} \phi_{\texttiny{HN}}(t) \, .
\end{equation}

Moreover, after evaluating the limit and transforming back to the time domain, one easily obtains the equation
\begin{equation}\label{eq:HN_Evolution_RL}
	\bigl({}_{0}D^{\alpha}_{t} +\tau_{\star}^{-\alpha} \bigr)^{\gamma} \phi_{\texttiny{HN}}(t) = 0 , \quad
		\lim_{t\to0^+} {\mathbf E}_{\alpha,1-\alpha\gamma,-\tau_{\star}^{-\alpha},0^{+}}^{-\gamma} \phi_{\texttiny{HN}}(t) = \frac{1}{\tau_{\star}^{\alpha \gamma}} \, ,
\end{equation}
(we refer to Appendix {\ref{SS:OperatorPrabhakar}} for a description of the operators $\bigl({}_{0}D^{\alpha}_{t} +\tau_{\star}^{-\alpha} \bigr)^{\gamma}$ and ${\mathbf E}_{\alpha,1-\alpha\gamma,-\tau_{\star}^{-\alpha},0^{+}}^{-\gamma}$ involved in the above equation). We note that the slight difference with respect to \cite[Eq. (24)]{WeronJurlewiczMagdziarz_ActaPhysPolB_2005} is due to the different way in which the operator $\bigl({}_{0}D^{\alpha}_{t} +\tau_{\star}^{-\alpha} \bigr)^{\gamma}$ is introduced. Indeed, we use the approach proposed in \cite{GarraGorenfloPolitoTomovski2014}, which has been published successively with respect to \cite{WeronJurlewiczMagdziarz_ActaPhysPolB_2005}. We can easily verify that, if $\gamma=1$, then the equivalence ${\mathbf E}_{\alpha,1-\alpha\gamma,-\tau_{\star}^{-\alpha},0^{+}}^{-\gamma} \equiv {}_{0}J_{t}^{1-\alpha}$ holds. Hence, as expected, the evolution equation (\ref{eq:CC_EvolEq_phi}) for the response in the CC model is just the particular case, for $\gamma=1$, of (\ref{eq:HN_Evolution_RL}) in light of the initial condition in (\ref{eq:HN_Evolution_RL}).

In a similar way, the equation for the HN relaxation function $\Psi_{\texttiny{HN}}(t)$ is derived by first recalling its Laplace transform
\begin{equation}
	\widetilde{\Psi}_{\texttiny{HN}}(s)  = \frac{1}{s} - \frac{1}{\tau_{\star}^{\alpha\gamma} s \bigl( s^{\alpha} + \tau_{\star}^{-\alpha}\bigr)^{\gamma}}
\end{equation}
and by considering the operator ${}^{\text{\tiny{C}}}{}{\bigl({}_{0}D^{\alpha}_{t} + \tau_{\star}^{-\alpha} \bigr)^{\gamma}}$ obtained after a regularization (in the Caputo sense) of $\bigl({}_{0}D^{\alpha}_{t} +\tau_{\star}^{-\alpha} \bigr)^{\gamma}$ (we refer again to Appendix {\ref{SS:OperatorPrabhakar}}). Thanks to (\ref{eq:HN_der_Cap_LT}), and since $\Psi_{\texttiny{HN}}(0)=1$, we have in this case
\begin{equation}
	{\mathcal L} \left( {}^{\text{\tiny{C}}}{}{\bigl({}_{0}D^{\alpha}_{t} + \tau_{\star}^{-\alpha} \bigr)^{\gamma}} \Psi_{\texttiny{HN}}(t) \, ; \, s \right)
	= -\frac{1}{\tau_{\star}^{\alpha\gamma}s} ,
\end{equation}
from which it is an immediate task to obtain
\begin{equation}\label{eq:HN_Evolution_Psi}
	{}^{\text{\tiny{C}}}{}{\bigl({}_{0}D^{\alpha}_{t} + \tau_{\star}^{-\alpha} \bigr)^{\gamma}} \Psi_{\texttiny{HN}}(t) = - \frac{1}{\tau_{\star}^{\alpha\gamma}} , \quad  \Psi_{\texttiny{HN}}(0)=1 \, .
\end{equation}

It can be a bit surprising that, with $\gamma=1$, the above equation slightly differs from the evolution equation (\ref{eq:CC_EvolEq_Psi}) of the CC model. This difference is due to the fact that the operator ${}^{\text{\tiny{C}}}{}{\bigl({}_{0}D^{\alpha}_{t} + \tau_{\star}^{-\alpha} \bigr)^{\gamma}}$ is not actually the same as $\bigl( \DerCap{0}{\alpha}{t} + \tau_{\star}^{-\alpha}\bigr)^{\gamma}$, as one could expect. For a clearer understanding of this issue the reader is referred to Remark \ref{rem:HN_Caputo} in Appendix {\ref{S:FractionalIntegralsDerivatives}}.

\subsection{The modified Havriliak-Negami or JWS model} 

A modified version of the HN model has bee recently derived, in the diffusion framework, by A. Jurlewicz, K. Weron and A. Stanislavsky \cite{JurlewiczTrzmielWeron_ActaPhysPolB_2010,StanislavskyWeronTrzmiel2010} with the aim of fitting with the Jonscher's URL some experimental data \cite{TrzmielJurlewiczWeron2010_JPCM,TrzmielMarciniszynKomar2011} exhibiting a less typical two-power-law relaxation pattern with frequency power law exponents $m$ and $n$ satisfying $m < 1 - n$.

We recall that, when fitted with the Jonscher's universal law, the HN model is characterized by exponents $m=\alpha$ and $1-n=\alpha \gamma$. Thus fitting data with $m<1-n$ requires $\gamma > 1$, a range which is not deemed admissible by some authors \cite{StanislavskyWeronTrzmiel2010,WeronJurlewiczMagdziarzWeronTrzmiel2010,StanislavskyWeronWeron2015} since the corresponding HN function can not be derived within theoretical approaches based on subdiffusion mechanisms (such as the fractional Fokker-Planck equation) or the continuous-time random walk, as for the CC relaxation process.

\begin{figure}[ht]
\centering
\includegraphics[width=.60\textwidth]{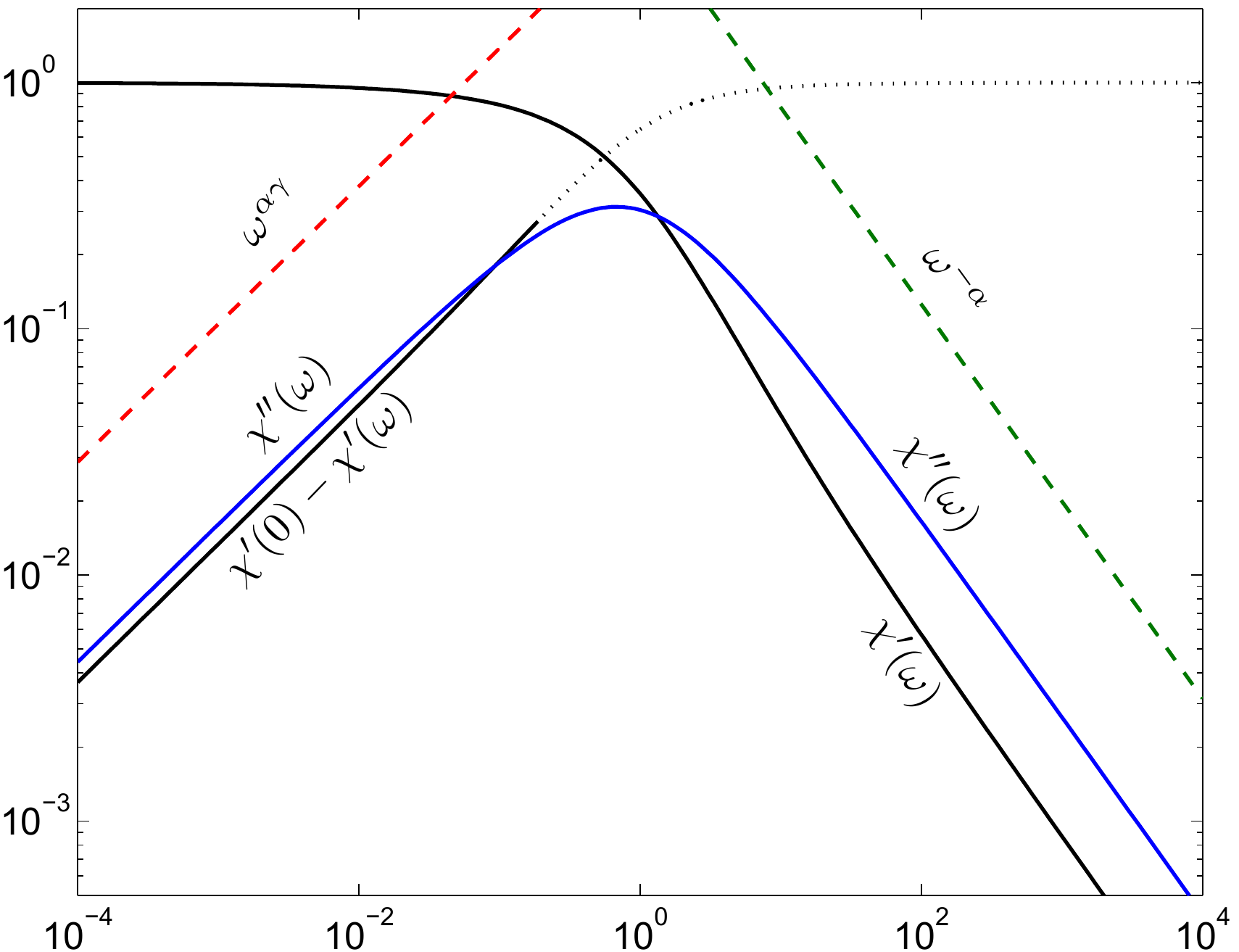}
\caption{JWS susceptibility for $\alpha=0.8$ and $\gamma=0.7$.}
\label{fig:Chi_URL_JWS}
\end{figure}

The modified HN model proposed in \cite{JurlewiczTrzmielWeron_ActaPhysPolB_2010,StanislavskyWeronTrzmiel2010}, and termed as JWS in \cite{TrzmielMarciniszynKomar2011} earlier and in \cite{StanislavskyWeron2016_FCAA} later, is formulated according to
\begin{equation}\label{eq:JWS_HN}
	\hat{\chi}_{\texttiny{JWS}}(\iu \omega)
	= 1- \frac{1}{\left( 1 + \bigl( \iu \tau_{\star} \omega \bigr)^{-\alpha} \right)^{\gamma}}
	= 1 - (\iu \tau_{\star} \omega)^{\alpha\gamma} \chi_{\texttiny{HN}}(\iu \omega)
\end{equation}
and it is immediate to verify that
  \begin{equation}
	\begin{aligned}
		\hat{\chi}_{\texttiny{JWS}}(\iu \omega) &\sim \gamma \bigl( \iu \tau_{\star} \omega)^{-\alpha} , \, &
		 \tau_{\star} \omega \gg 1 \\
		\Delta \hat{\chi}_{\texttiny{JWS}}(\iu \omega) = \hat{\chi}_{\texttiny{JWS}}(0) - \hat{\chi}_{\texttiny{JWS}}(\iu \omega) &\sim \bigl( \iu \tau_{\star} \omega)^{\alpha\gamma} , \,  &
		\tau_{\star} \omega \ll 1 \\
	\end{aligned}
\end{equation}
thus fitting the Jonscher's universal law with $m=\alpha\gamma$ and $1-n=\alpha$ and hence allowing $m<1-n$ under the restriction $\gamma < 1$. The plot of the JWS susceptibility is shown in Figure \ref{fig:Chi_URL_JWS}.

\begin{figure}[ht]
\centering
\includegraphics[width=0.48\textwidth]{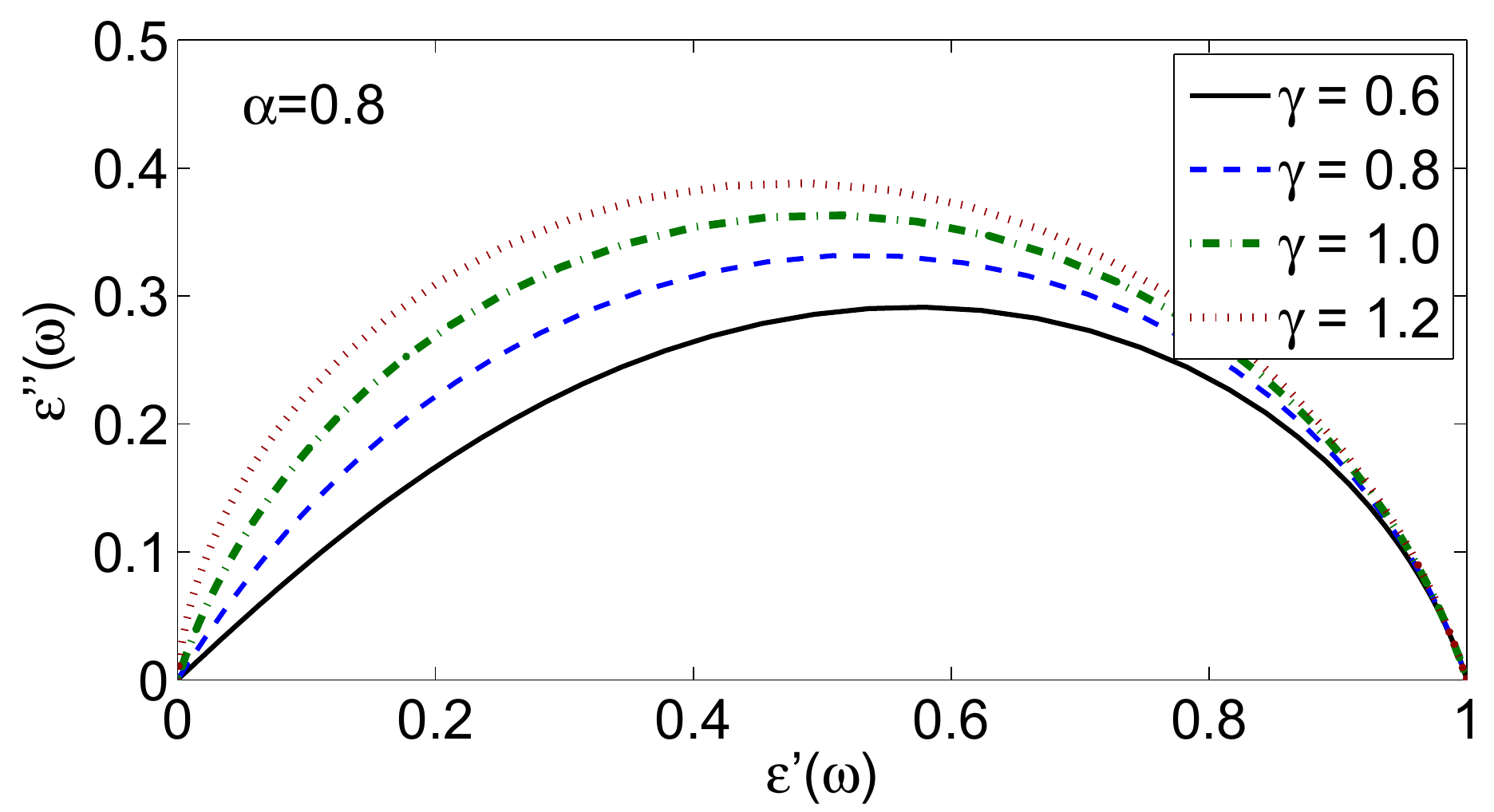}
\includegraphics[width=0.48\textwidth]{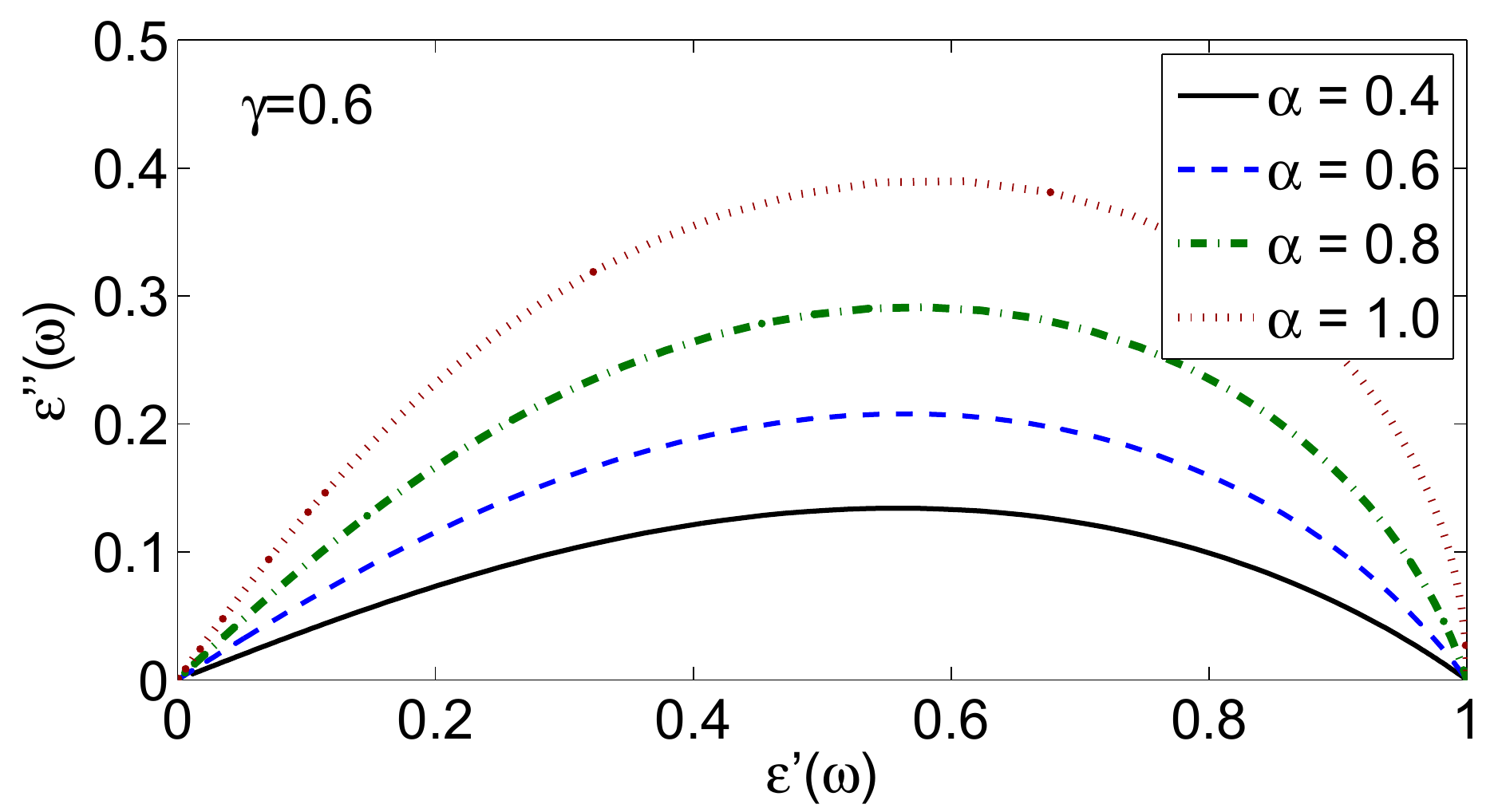}
\caption{Cole-Cole plots for the JWS model.}
\label{fig:ColeCole_JWS}
\end{figure}

From the above plot, and from the Cole-Cole plots of the susceptibility (\ref{eq:JWS_HN}) presented in Figure \ref{fig:ColeCole_JWS}, we observe that the JWS model appears as the specular reflection of the HN model (see Figures \ref{fig:Chi_URL_HN} and \ref{fig:ColeCole_HN}).

\begin{figure}[ht]
\centering
\includegraphics[width=0.46\textwidth]{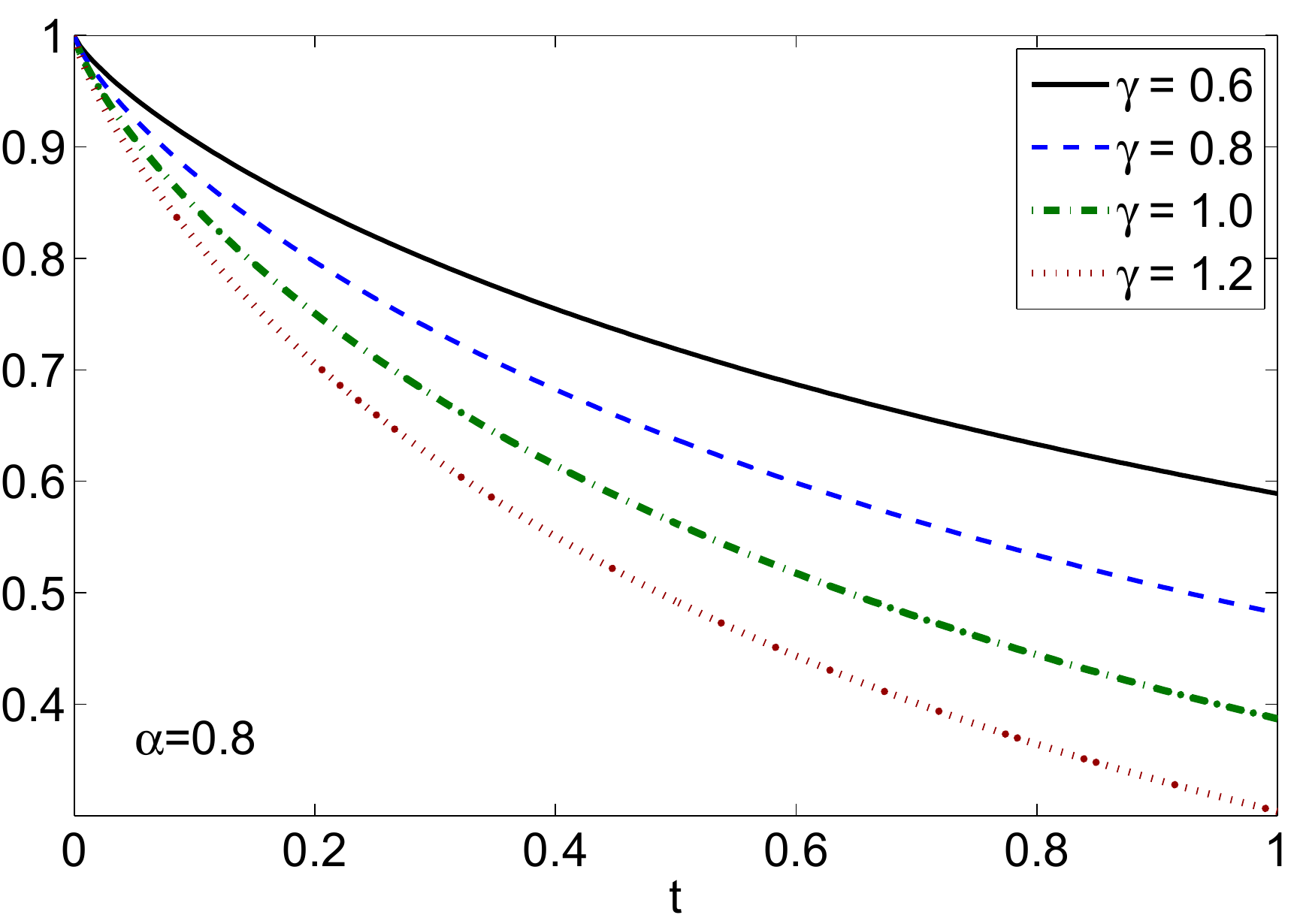}
\includegraphics[width=0.46\textwidth]{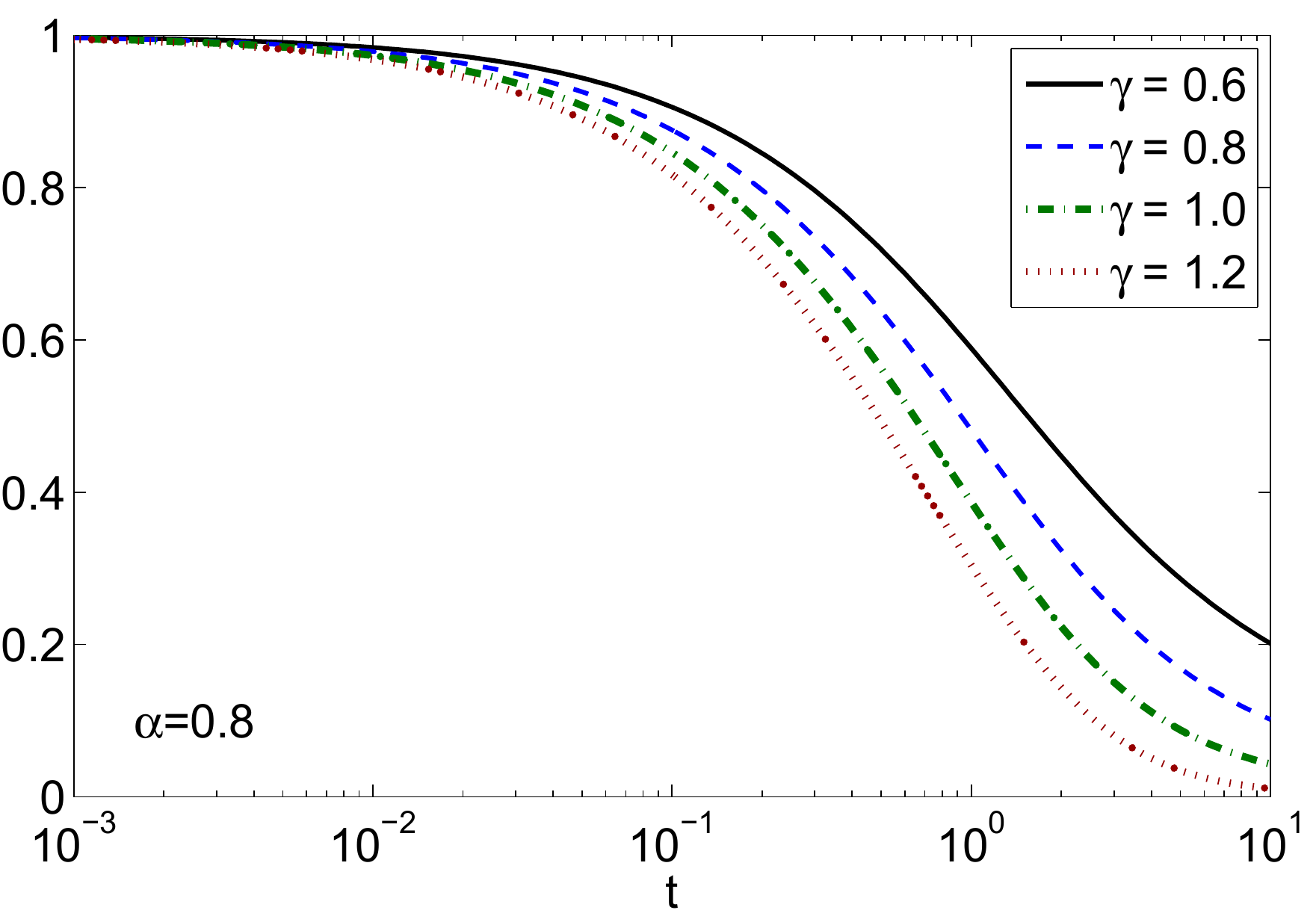}
\caption{Relaxation functions $\Psi_{\texttiny{JWS}}(t)$ on varying $\gamma$.}
\label{fig:Rel_JWSg}
\end{figure}

\begin{figure}[ht]
\centering
\includegraphics[width=0.46\textwidth]{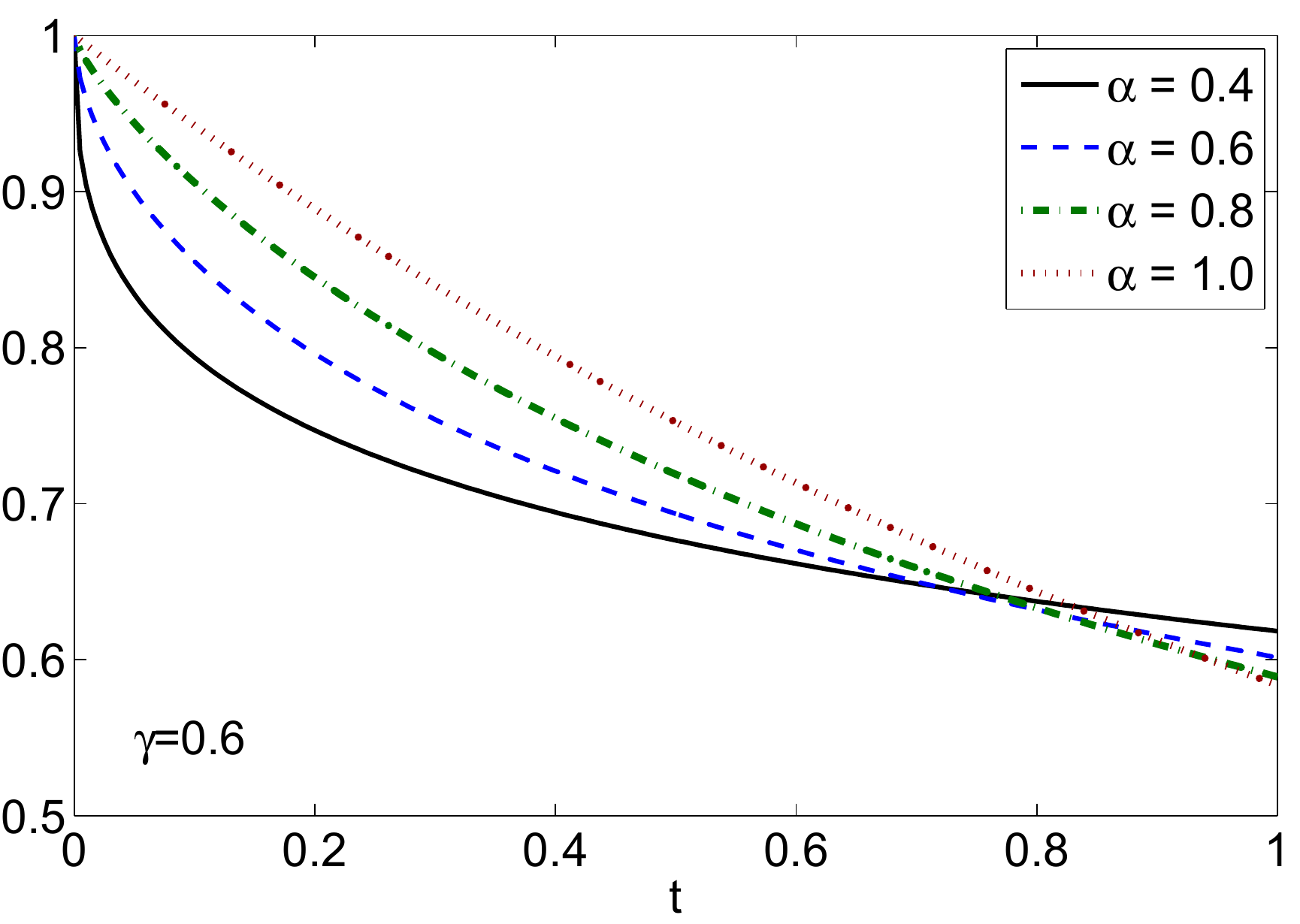}
\includegraphics[width=0.46\textwidth]{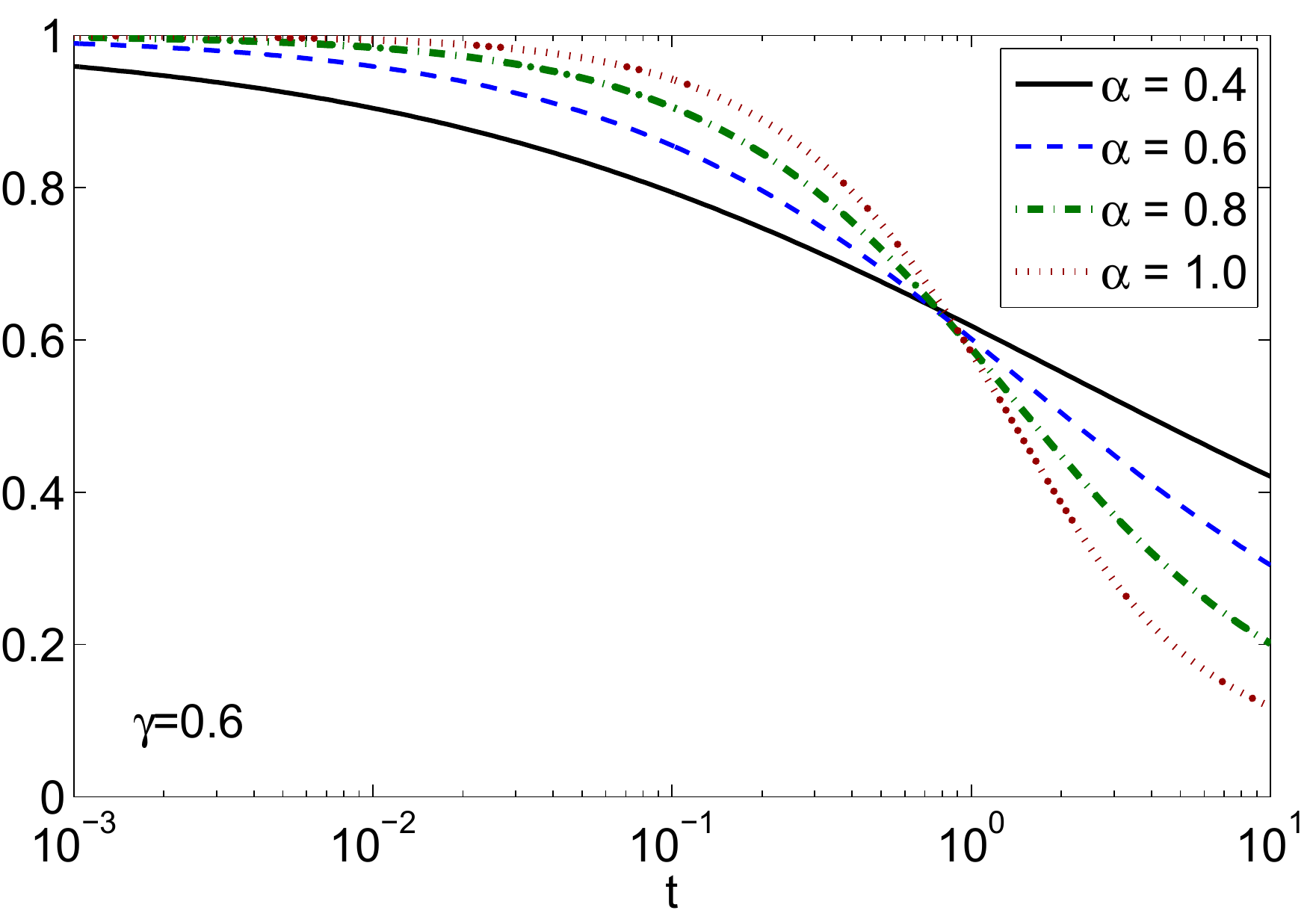}
\caption{Relaxation functions $\Psi_{\texttiny{JWS}}(t)$ on varying $\alpha$.}
\label{fig:Rel_JWSa}
\end{figure}

The corresponding time-domain response and relaxation are respectively given by
\begin{equation}
	\phi_{\texttiny{JWS}}(t) = \delta(t) - \frac{1}{\tau_{\star}} \bigl( t/\tau_{\star} \bigr)^{-1} E_{\alpha,0}^{\gamma}\left(- \bigl( t/\tau_{\star} \bigr)^{\alpha} \right)
\end{equation}
and
\begin{equation}
	\Psi_{\texttiny{JWS}}(t) = E_{\alpha,1}^{\gamma}\left(- \bigl( t/\tau_{\star} \bigr)^{\alpha} \right)
\end{equation}
and the plots of $\Psi_{\texttiny{JWS}}(t)$ are presented on varying $\gamma$ and $\alpha$ respectively in Figure \ref{fig:Rel_JWSg} and \ref{fig:Rel_JWSa}.


From the above representation one could be tempted to infer the presence of a non-integrable singularity in $\phi_{\texttiny{JWS}}(t)$. However, the analysis of the asymptotic behaviour of the Prabhakar function $E_{\alpha,\beta}^{\gamma}(z)$ with $\beta=0$ shows that the singularity is only apparent. Indeed, by exploiting the asymptotic expansions (\ref{eq:ML3_Series0}) and (\ref{eq:ML3_exp_large}) for small $t$ and  for $t\to +\infty$ respectively, we have the following short- and long-time power-law dependencies
\begin{equation}
	\phi_{\texttiny{JWS}}(t) \sim \left\{ \begin{array}{ll}
		\displaystyle\frac{\gamma}{\tau_{\star} \Gamma(\alpha)} \bigl( t/\tau_{\star} \bigr)^{\alpha-1} \, , \, & \text{for } t \ll \tau_{\star} \,  \\
		-\displaystyle\frac{1}{\tau_{\star} \Gamma(-\alpha\gamma)} \bigl( t/\tau_{\star} \bigr)^{-\alpha\gamma-1} \, , \, & \text{for } t \gg \tau_{\star} .\
	\end{array} \right.
\end{equation}

Similarly, the short- and long-time power-law dependencies for $\Psi_{\texttiny{JWS}}(t)$ can be explicitly given by
\begin{equation}
	\Psi_{\texttiny{JWS}}(t)  \sim \left\{ \begin{array}{ll}
		1- \displaystyle\frac{\gamma}{\Gamma(\alpha+1 )}\bigl( t/\tau_{\star} \bigr)^{\alpha} \, , \, & \text{for } t \ll \tau_{\star} \,  \\
		\displaystyle\frac{\gamma}{\Gamma(1-\alpha\gamma)} \bigl( t/\tau_{\star} \bigr)^{-\alpha\gamma} \, , \, & \text{for } t \gg \tau_{\star} .\
	\end{array} \right.
\end{equation}

\begin{figure}[ht]
\centering
\includegraphics[width=0.46\textwidth]{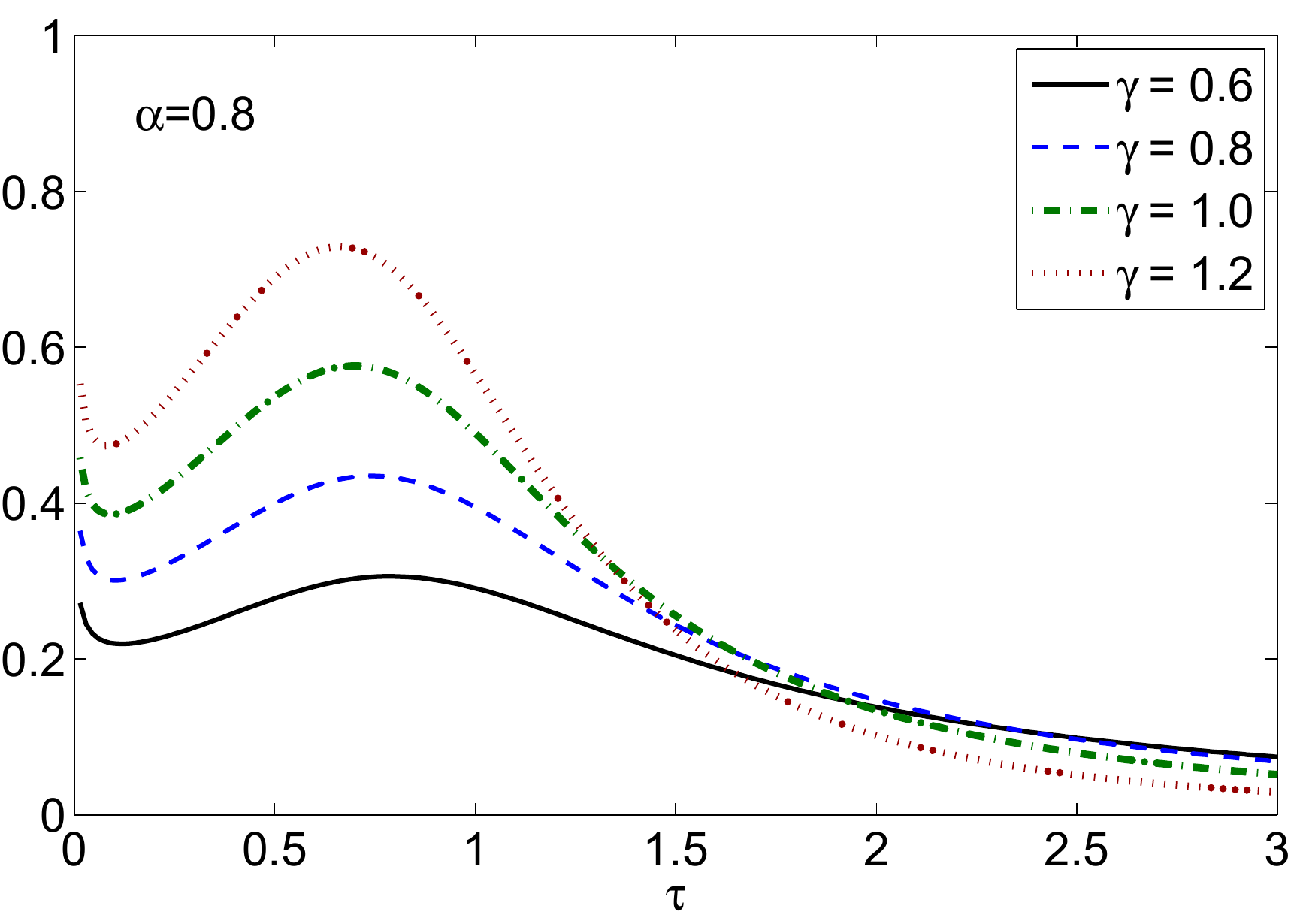}
\includegraphics[width=0.46\textwidth]{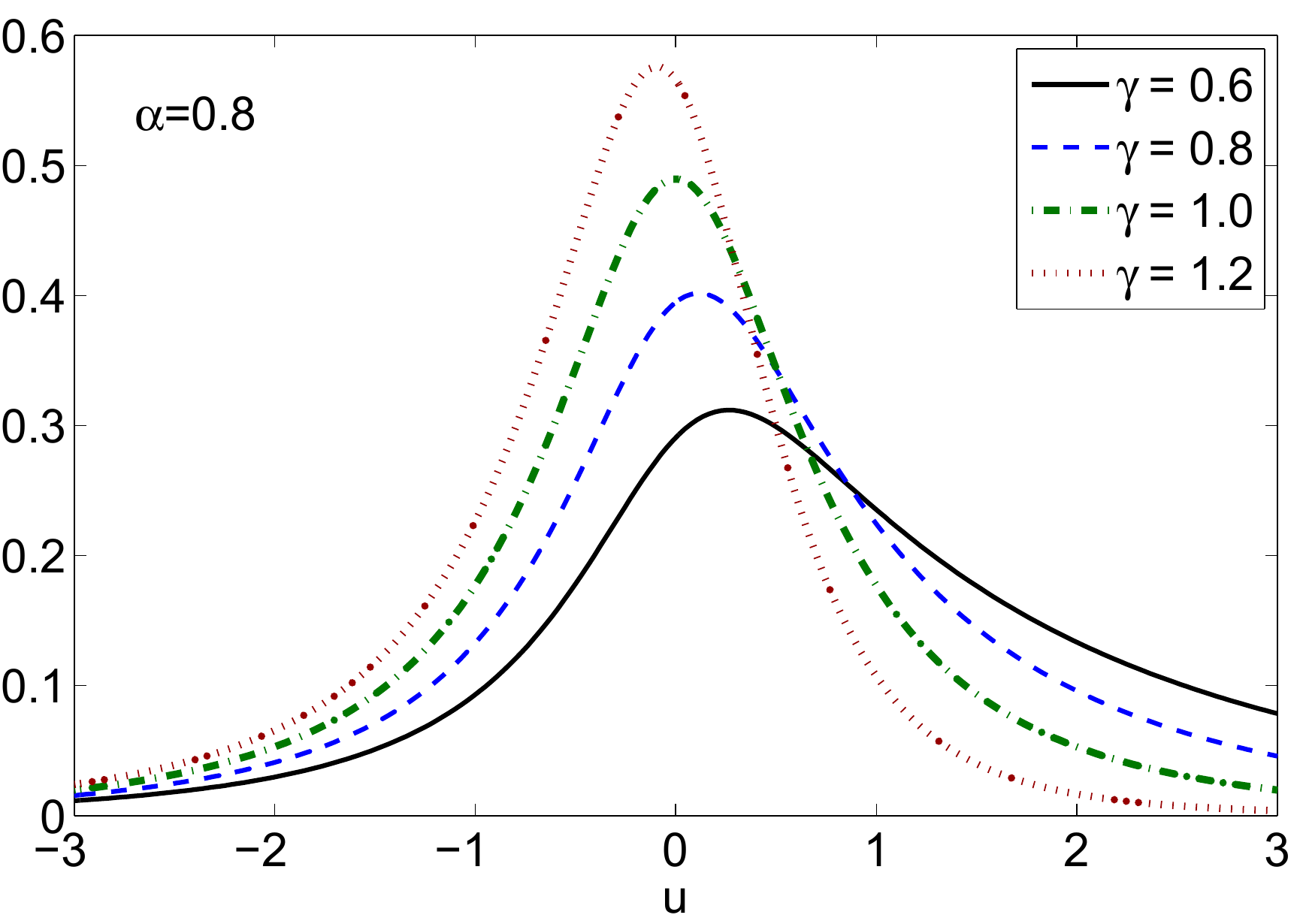}
\caption{Spectral distributions $H^{\Psi}_{\texttiny{JWS}}(\tau)$ (left) and $L^{\Psi}_{\texttiny{JWS}}(u)$ (right) on varying $\gamma$.}
\label{fig:Spectral_JWSg}
\end{figure}

\begin{figure}[ht]
\centering
\includegraphics[width=0.46\textwidth]{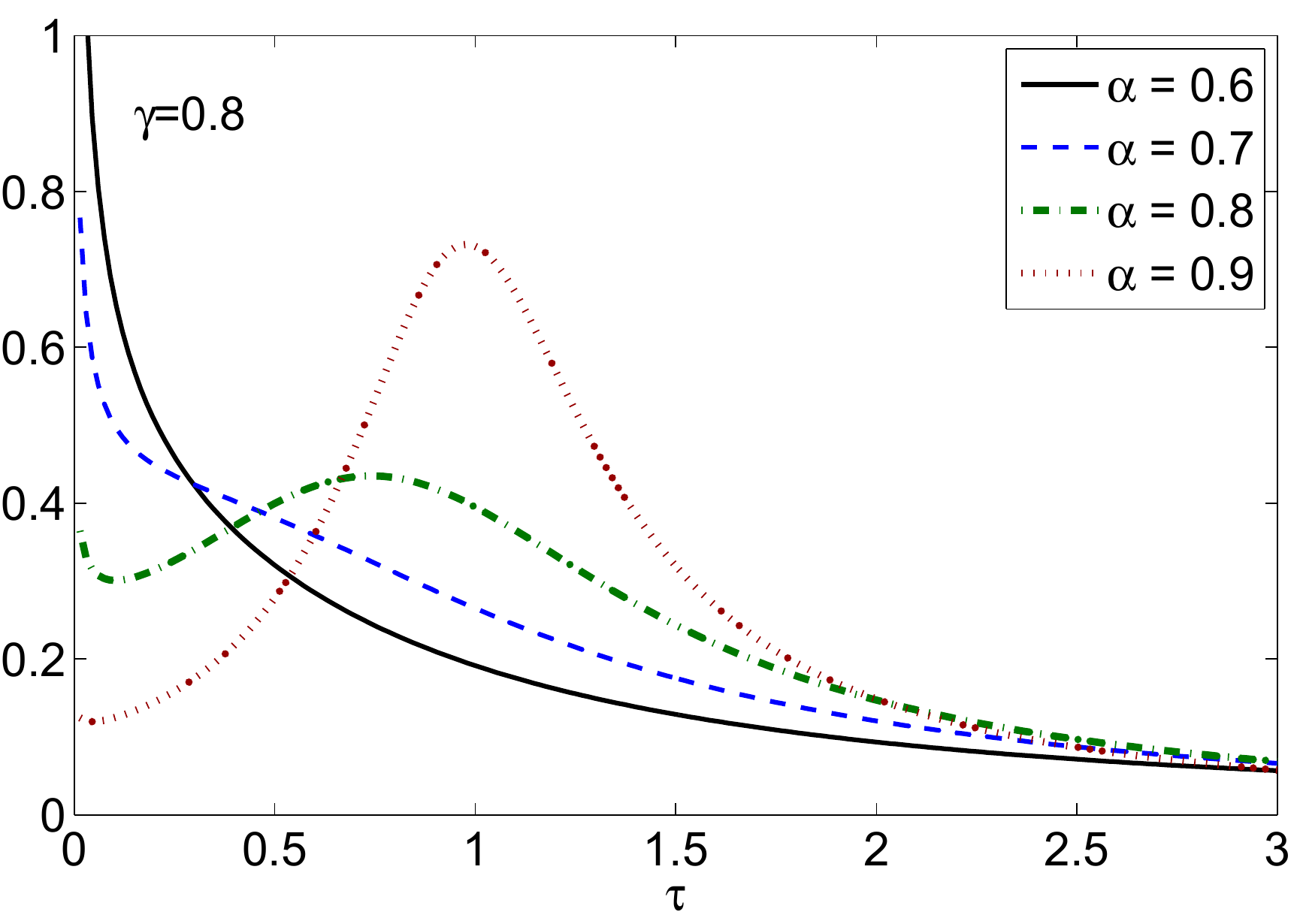}
\includegraphics[width=0.46\textwidth]{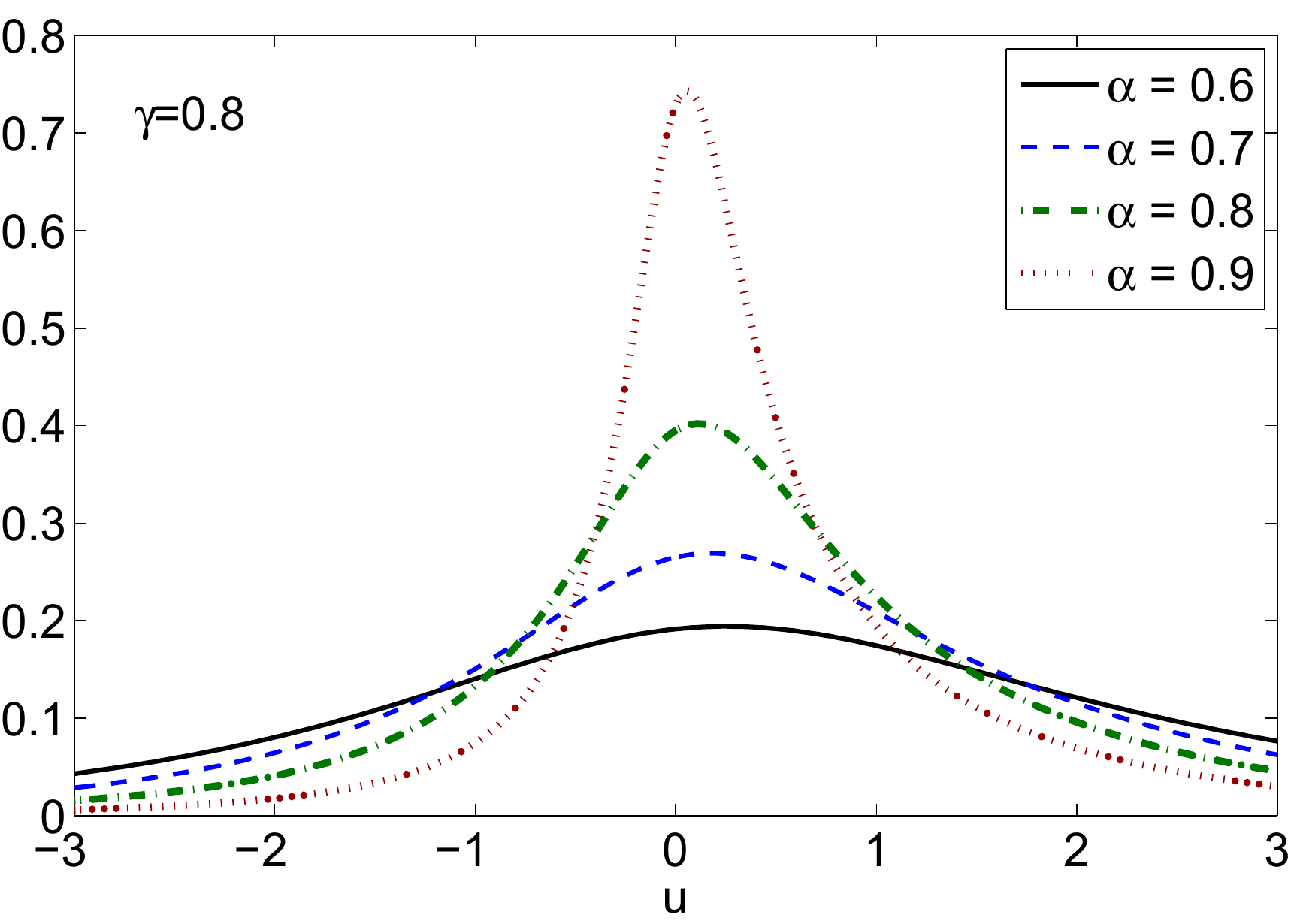}
\caption{Spectral distributions $H^{\Psi}_{\texttiny{JWS}}(\tau)$ (left) and $L^{\Psi}_{\texttiny{JWS}}(u)$ (right) on varying $\alpha$.}
\label{fig:Spectral_JWSa}
\end{figure}

The spectral functions are obtained in a similar way to those of the HN model and, indeed, it is sufficient to observe that
\begin{equation}
	K^{\Psi}_{\texttiny{JWS}}(r) = \frac{\tau_{\star}}{\pi}
	\frac{ (\tau_{\star} r)^{\alpha\gamma-1}  \sin\left[ \gamma \, \bigl(\alpha \pi - \theta_\alpha(r)\bigr) \right]}{\left( (\tau_{\star} r)^{2\alpha} + 2 (\tau_{\star} r)^{\alpha}\,\cos(\alpha\pi)+1 \right)^{\gamma/2}} ,
\end{equation}
where $\theta_\alpha(r)$ is the same function given in (\ref{eq:Theta_HN}). Since $0\le\theta_\alpha(r)<\alpha\pi$, it is $0<\alpha \pi - \theta_\alpha(r)\le \alpha \pi$ and hence $K^{\Psi}_{{\texttiny{JWS}}}(r) \ge 0$, for any $r \ge 0$ whenever $0<\alpha,\alpha\gamma \le 1$, thus assuring the CM property of $\Psi_{\texttiny{JWS}}(t)$.  The plots of the corresponding time spectral distributions $H^{\Psi}_{\texttiny{JWS}}(\tau)$ and $L^{\Psi}_{\texttiny{JWS}}(u)$, on varying $\gamma$ and $\alpha$ respectively, are presented in Figures \ref{fig:Spectral_JWSg} and \ref{fig:Spectral_JWSa}.

An evolution equation expressed in terms of the same operators introduced for the HN model has been recently presented in \cite{StanislavskyWeron2016_FCAA}. It can be obtained by the Laplace transform of the relaxation function
\begin{equation}
	\widetilde{\Psi}_{\texttiny{JWS}}(s)  = \frac{s^{\alpha\gamma-1}}{\bigl( s^{\alpha} + \tau_{\star}^{-\alpha}\bigr)^{\gamma}}
\end{equation}
after applying (\ref{eq:HN_der_LT}) in order to obtain
\begin{equation}\label{eq:LT_Evol_Eq_RL_JWS}
	{\mathcal L} \left( \bigl({}_{0}D^{\alpha}_{t} +\tau_{\star}^{-\alpha} \bigr)^{\gamma} \Psi_{\texttiny{JWS}}(t) \, ; \, s \right)
	= s^{\alpha\gamma-1} ,
\end{equation}
and hence by transforming back (\ref{eq:LT_Evol_Eq_RL_JWS}) to the time domain once the appropriate initial condition is used
\begin{equation}
		\bigl({}_{0}D^{\alpha}_{t} +\tau_{\star}^{-\alpha} \bigr)^{\gamma} \Psi_{\texttiny{JWS}}(t)
	= \frac{t^{-\alpha\gamma}}{\Gamma(1-\alpha\gamma)}
	, \quad
	\lim_{t\to0^+} {\mathbf E}_{\alpha,1-\alpha\gamma,-\tau_{\star}^{-\alpha},0^{+}}^{-\gamma} \Psi_{\texttiny{JWS}}(t) = 0
\end{equation}

With respect to the HN derivative regularized in the Caputo sense, thanks to (\ref{eq:HN_der_Cap_LT}) and since $\Psi_{\texttiny{JWS}}(0)=1$, we instead have
\begin{equation}
	{\mathcal L} \left( {}^{\text{\tiny{C}}}{}{\bigl({}_{0}D^{\alpha}_{t} + \lambda \bigr)^{\gamma}}  \Psi_{\texttiny{JWS}}(t) \, ; \, s \right)
	= s^{\alpha\gamma-1} - \frac{\bigl(s^{\alpha} + \tau_{\star}^{-\alpha}\bigr)^{\gamma}}{s},
\end{equation}
and hence we easily obtain
\begin{equation}
	{}^{\text{\tiny{C}}}{}{\bigl({}_{0}D^{\alpha}_{t} + \lambda \bigr)^{\gamma}}  \Psi_{\texttiny{JWS}}(t)
	= \left[ \frac{ t^{-\alpha\gamma}}{\Gamma(1-\alpha\gamma)} - t^{-\alpha\gamma} E_{\alpha,1-\alpha\gamma}^{-\gamma} \left(- \bigl( t/\tau_{\star} \bigr)^{\alpha} \right) \right]  \, ,
\end{equation}
coupled with the initial condition $\Psi_{\texttiny{JWS}}(0)=1$.

\subsection{The Kohlrausch-Williams-Watts model} 

The stretched exponential function was first introduced by Kohlrausch in 1854 \cite{Kohlrausch_1854} to describe the discharge relaxation phenomenon in Leiden jar capacitors. Successively, in 1970, this model was rediscovered by Williams and Watt \cite{WilliamsWatt1970} to describe non-symmetric dielectric loss curves showing intermediate shapes between the CC and DC empirical models (see also \cite{WilliamsWattsDevNorth1971}).  It is therefore referred to as the Kohlrausch-Williams-Watts (KWW) model.

\begin{figure}[ht]
\centering
\includegraphics[width=0.46\textwidth]{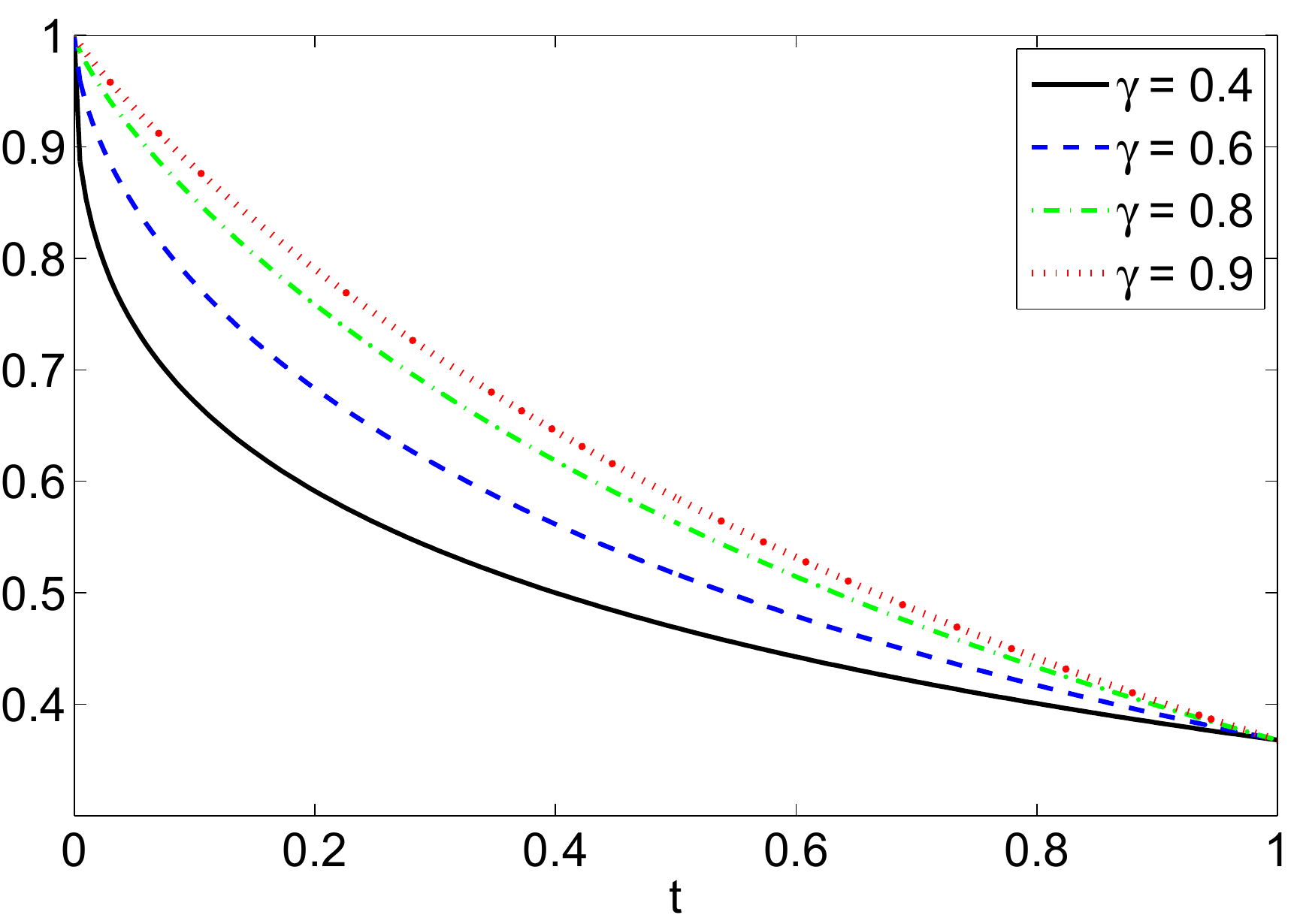}
\includegraphics[width=0.46\textwidth]{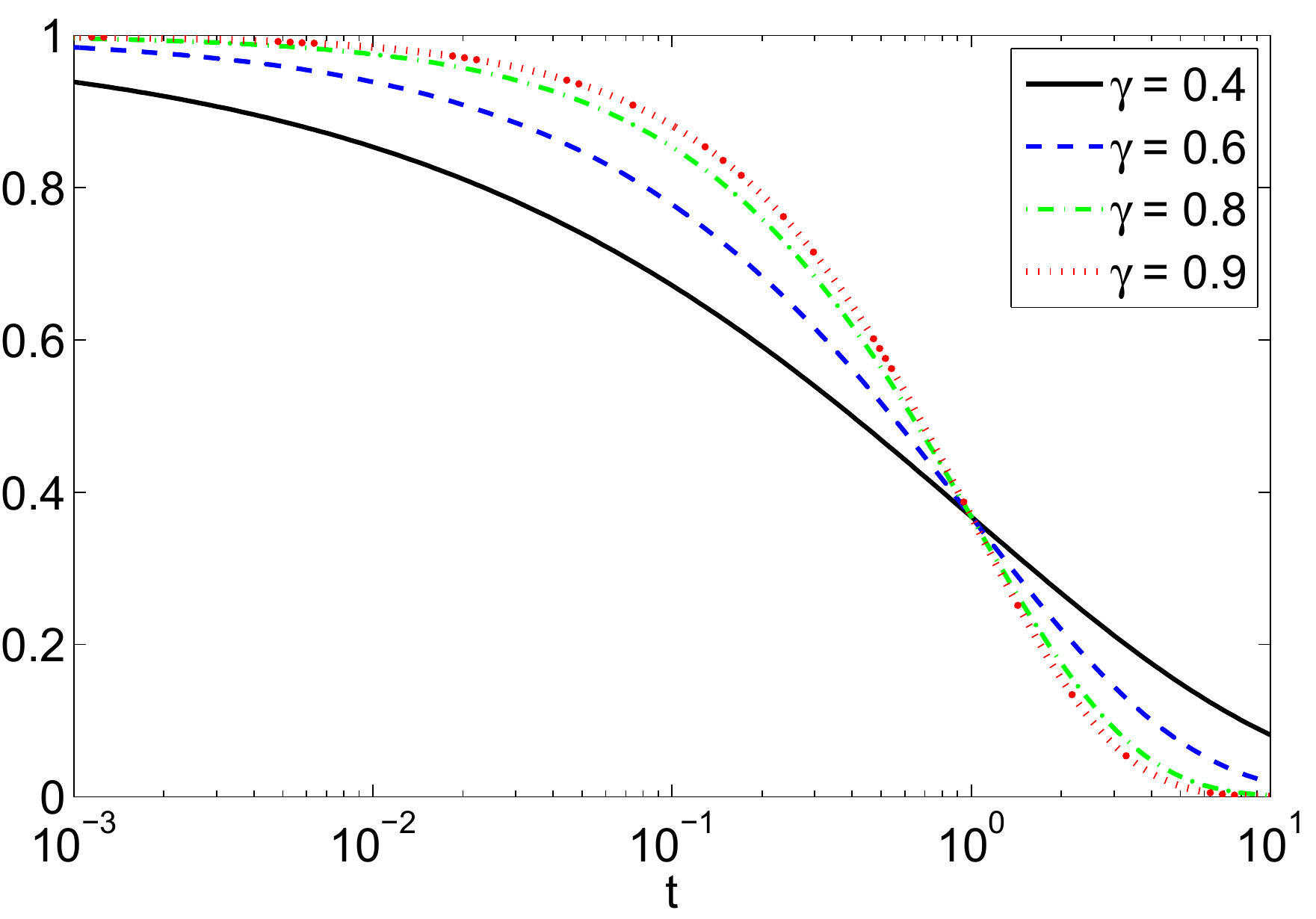}
\caption{Relaxation function $\Psi_{\texttiny{KWW}}(t)$ on varying $\gamma$.}
\label{fig:Rel_KWW}
\end{figure}

Unlike the other models discussed in this paper, the KWW model is introduced starting from its relaxation function
\begin{equation}\label{eq:Relax_KWW}
	\Psi_{\texttiny{KWW}}(t) = \exp\left[-\left({t}/{\tau_{\star}}\right)^{\gamma}\right] , \quad t\ge 0\,, \quad 0<\gamma<1\,,
\end{equation}
which is known as the KWW stretched exponential function. It was used for the first time in luminescence by Werner in 1907, and nowadays has important applications in several fields, as for instance in pharmacy \cite{YoshiokaAsoKojima2001} (we refer to \cite{BerberanSantosBodunovValeur2008} for an historical perspective). The plots of $\Psi_{\texttiny{KWW}}(t)$ for some values of $\gamma$ are presented in Figure \ref{fig:Rel_KWW}.

The response function $\phi_{\texttiny{KWW}}(t)$ can be evaluated as a consequence of the application of (\ref{eq:phi_der_Psi}) to (\ref{eq:Relax_KWW})
\begin{equation}\label{eq:Resp_KWW}
	\phi_{\texttiny{KWW}}(t)
		= \frac{\gamma}{\tau_{\star}}\, \left(\frac{t}{\tau_{\star}}\right)^{\gamma-1}	
			\! \! \! \! \cdot \exp\left[-\left({t}/{\tau_{\star}}\right)^{\gamma}\right]
			, \quad t\ge 0, \quad 0<\gamma<1 \,.
\end{equation}

The spectral density $K^\Psi_{\texttiny{KWW}}(r)$ of the KWW relaxation function can be derived from the Laplace transform of the unilateral and extremal stable density of order $\gamma$ as follows. By recalling the theory of the L{\'e}vy stable densities in probability theory (see  the survey by Mainardi, Luchko and Pagnini \cite{Mainardi-Luchko-Pagnini_2001} where their role of fundamental solutions of the space fractional diffusion equations is discussed in detail), and by following the notation based on the Feller-Takayasu representation, after setting $\tau_{\star}=1$ we get
\begin{equation}
\exp (-t^\gamma) = \int_0^\infty
{\rm e}^{-rt} L_\gamma^{-\gamma}(r)\, dr ,  \quad t\ge 0, \quad
0<\gamma<1\,.
\end{equation}

Here $L$ just denotes  the L{\'e}vy stable density whose expression is provided by a transcendental function of the Wright type, as explained in
the Appendix F of Mainardi's  book \cite{Mainardi2010}, i.e.
\begin{equation}
L_\gamma^{-\gamma}(r)=
\frac{1}{r} \, \sum_{n=1}^\infty \frac{(-1)^n}{n!}\,
\frac{r^{-\gamma n}}{\Gamma(-\gamma n)}, \quad r>0\,.
\end{equation}

In the particular case $\gamma=1/2$, we recover the L{\'e}vy-Smirnov stable density
\begin{equation}
 L_{1/2}^{-1/2}(r)=
   \frac{r^{-3/2}}{2\sqrt{\pi}} \, \exp[-1/(4r)]\,, \; r>0\,.
 \end{equation}

As a consequence of the above result, the spectral density of the relaxation function of the KWW model for $\tau_{\star} \ne 1$ and $0<\gamma<1$ turns out to be
\begin{equation}
	K^\Psi_{\texttiny{KWW}}(r)=
	\tau_{\star}^\gamma \, L_\gamma^{-\gamma}(\tau_{\star}^\gamma r), \quad r>0 \; .
 \end{equation}

\begin{figure}[ht]
\centering
\includegraphics[width=0.46\textwidth]{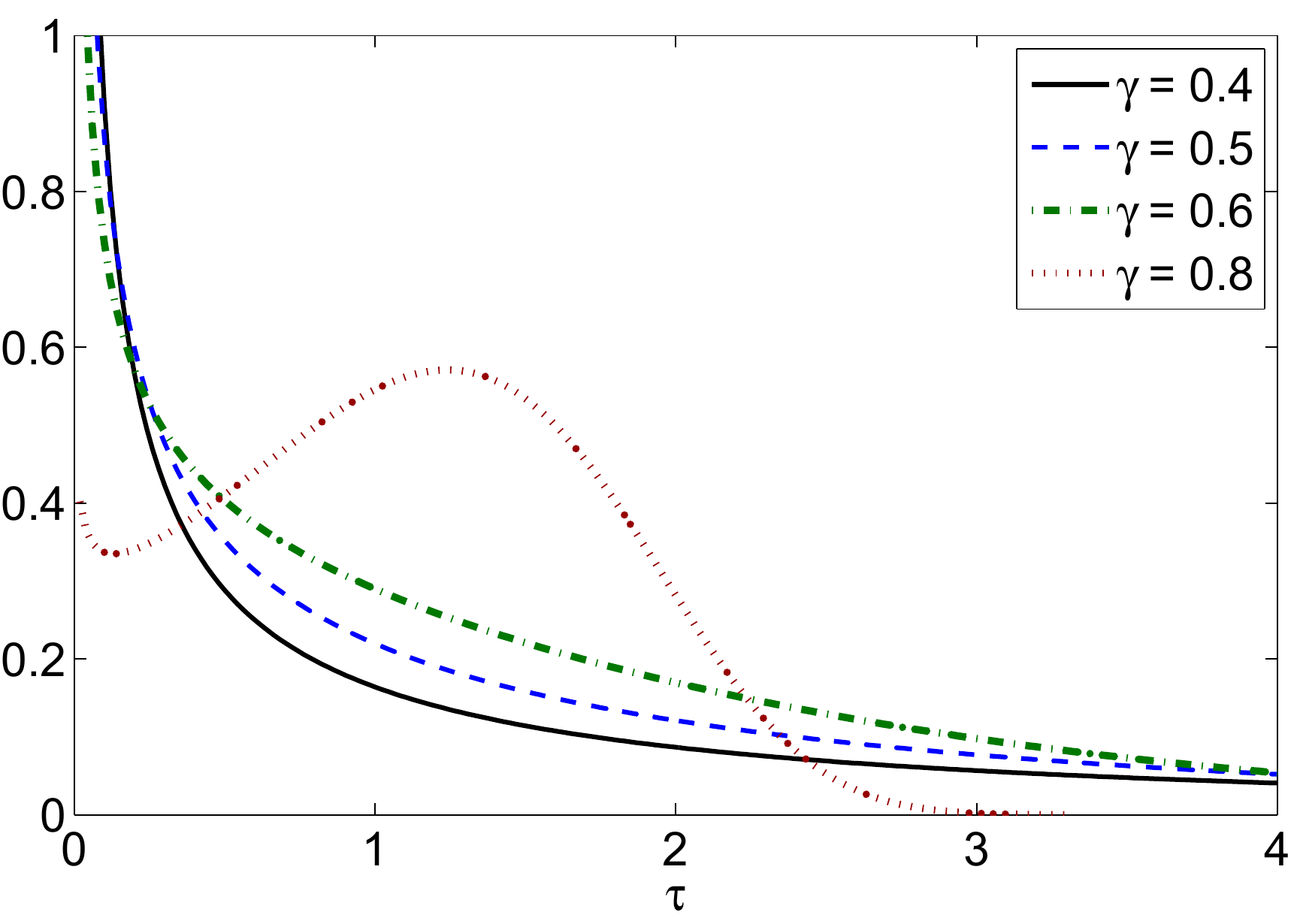}
\includegraphics[width=0.46\textwidth]{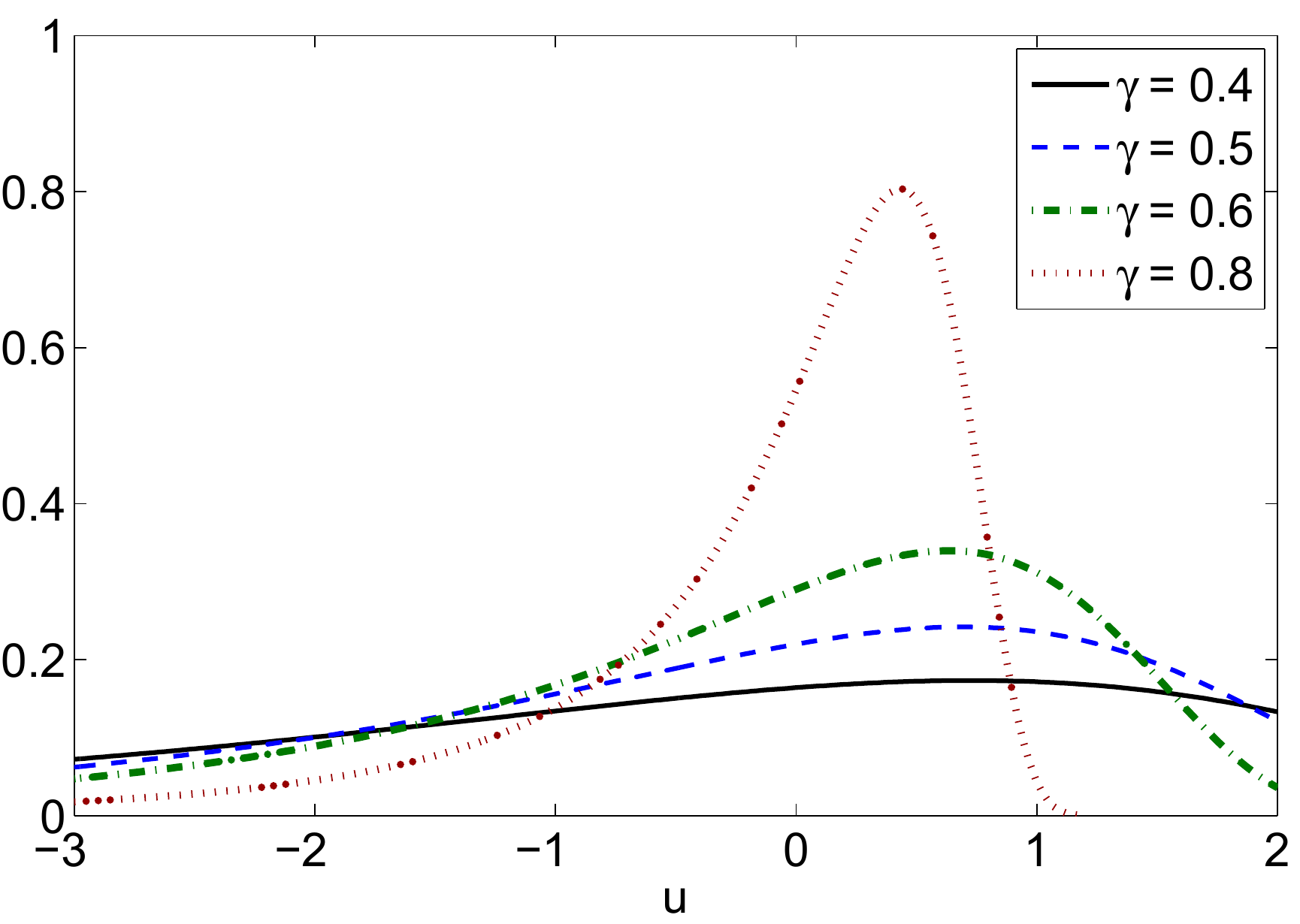}
\caption{Spectral distributions $H^{\Psi}_{\texttiny{KWW}}(\tau)$ (left) and $L^{\Psi}_{\texttiny{KWW}}(u)$ (right).}
\label{fig:Spectral_KWW}
\end{figure}

The corresponding frequency spectral functions (whose plots are presented in Figure \ref{fig:Spectral_KWW}) are hence given by
\begin{equation}
	H^\Psi_{\texttiny{KWW}}(\tau) = \frac{\tau_{\star}^{\gamma}}{\tau^2} L_{\gamma}^{-\gamma}(\tau_{\star}^{\gamma}/\tau)
	, \quad
	L^\Psi_{\texttiny{KWW}}(u) = \eu^{-u} \tau_{\star}^{\gamma} L_{\gamma}^{-\gamma}(\tau_{\star}^{\gamma} \eu^{-u})
\end{equation}

The role of the L{\'e}vy stable density and its connection with the spectral density of the KWW model has been already highlighted, for instance, in \cite{CapelasMainardiVaz2014,JurlewiczWeron1993}. We observe here that for $\gamma \to 1$ the spectral density tends to the generalized Dirac function centred for  $r=1/\tau_{\star}$.

For the KWW model it is straightforward to derive the evolution equation for its relaxation function in the form of a linear differential equation with $0<\gamma <1$
\begin{equation}\label{eq:EvolEq_KWW}
\frac{\du}{\du t} \Psi_{\texttiny{KWW}} =
- \gamma \, (t/\tau_{\star})^{\gamma-1} \, \Psi_{\texttiny{KWW}}(t)\,, \quad t\ge 0 \, ,
\quad \Psi_{\texttiny{KWW}}(0)=1 \, .
\end{equation}

\subsection{The CMV model: a bridge between CC and KWW} 

It is immediate that, with $\gamma=1$, (\ref{eq:EvolEq_KWW}) provides the evolution equation governing the Debye relaxation function. On the other hand, if for $\gamma=1$ one replaces the first order derivative with a Caputo fractional derivative of order $0<\alpha<1$, the evolution equation (\ref{eq:CC_EvolEq_Psi}) for the relaxation function of the CC model is instead obtained.

Moreover, from (\ref{eq:CC_RelaxationFunction}) we remember that $\Psi_{\texttiny{CC}}(t) = E_{\alpha,1}\left(- \bigl( t/\tau_{\star} \bigr)^{\alpha} \right)$ and when $\alpha=1$ the ML function reduces to the exponential;  then the CC model reduces to the Debye model too.

The above considerations induce us to consider a more general   differential equation of fractional order that is expected to govern the relaxation function of a dielectric model including, as particular cases, the CC and KWW models. The aim is to create a link between these two models.

Indeed, this was the purpose of the recent paper by Capelas, Mainardi and Vaz \cite{CapelasMainardiVaz2014} who proposed as relaxation function a transcendental function (of the ML type), known as Kilbas and Saigo function, under the condition of its complete monotonicity. We agree to refer to this model as the CMV model.

This model is based on  the following initial-value problem for the corresponding relaxation function
\begin{equation}\label{eq:EvolEq_CMV}
	\DerCap{0}{\alpha}{t} \Psi_{\texttiny{CMV}}(t) = -\lambda t^\beta  \, \Psi_{\texttiny{CMV}}(t)\,,
	\quad \Psi_{\texttiny{CMV}}(0)=1 \, ,
\end{equation}
where  $\lambda$ is a positive constant and the (dimensionless)
constant $\alpha, \beta$ are subjected to the conditions
\begin{equation}
0 < \alpha \leq 1\,,  \quad
  -\alpha <\beta \le 1 - \alpha\,.
\label{eq:MonotonicityConditions}
\end{equation}

The above conditions  have been  conjectured to  be sufficient  to ensure the existence and complete monotonicity of the  solution $\Psi(t)$ for $t\ge 0$. Then, we recognize that the solution of the initial-value problem (\ref{eq:EvolEq_CMV}) is given by the  Kilbas and Saigo function
\begin{equation}
\begin{array}{ll}
 \Psi_{\texttiny{CMV}}(t) &= E_{\alpha, 1+\frac{\beta}{\alpha},\frac{\beta}{\alpha}}
(-t^{\alpha + \beta}) = \\
&1+\! {\displaystyle \sum_{n=1}^{\infty} (-1)^n \prod_{i=0}^{n-1}
 \!\frac{\Gamma(i(\alpha+\beta) + \beta + 1)}{\Gamma(i(\alpha+\beta) + \alpha +\beta +1)}   t^{(\alpha +\beta) n}}
 \end{array}
 \label{2bis}
\end{equation}
with the conditions (\ref{eq:MonotonicityConditions}).
We find worth to introduce the positive parameter
\begin{equation}
\gamma \equiv \alpha +\beta\,,
\label{2.1bis}
\end{equation}
so the solution reads
\begin{equation}
\begin{array}{ll}
 \Psi_{\texttiny{CMV}}(t) &= E_{\alpha,\frac{\gamma}{\alpha},\frac{\gamma-\alpha}{\alpha}}
(-t^{\gamma}) = \\
& 1 \!+\!{\displaystyle \sum_{n=1}^{\infty} (-1)^n \prod_{i=0}^{n-1}
 \frac{\Gamma(i\gamma + \gamma - \alpha + 1)}{\Gamma(i\gamma + \gamma +1)}
\,  t^{\gamma n}}
\end{array}
 \label{2}
\end{equation}
with  the conditions
\begin{equation}
0<\alpha \le 1\,, \quad
 0 < \gamma  \leq 1.
 \label{2.1.1}
 \end{equation}

We note some particular cases of the CMV model: for $\alpha=\gamma =1, \beta=0$ we recover the Debye model; for $\alpha =1, 0<\gamma<1 , -1<\beta= \gamma -1<0$ we recover the KWW model; for $0<\alpha<1$ and $\beta=\gamma=0$ we recover the CC model.  For more details we refer the reader to  \cite{CapelasMainardiVaz2014}.

\subsection{The excess wing model}\label{SS:ExcessWing} 

In some experiments by dielectric spectroscopy over a large frequency range, it has been observed an excess of contribution at high frequencies in glasses forming glycerol and propylene carbonate in liquid and supercooled-liquid state \cite{LunkenheimeSchneiderBrandLoid2000,SchneiderBrandLunkenheimerLoidl2000_PRL,SchneiderLunkenheimerBrandLoidl1999}.

The presence of an excess wing at high frequencies, in addition to an asymmetric relaxation peak, was initially modelled by using a combination of CC and HN models with different relaxation times. Successively Rudolf Hilfer, from the University of Stuttgart (Germany), proposed a new and simpler model \cite{Hilfer2002_CP,Hilfer2002_JPCM,Hilfer2016} with just one stretching power-law exponent $\alpha$ and two characteristic relaxation times $0 < \tau_1, \tau_2 < \infty$ (see also \cite{Popov-Nigmatullin-Khamzin_2012,KhamzinNigmatullinPopovMurzaliev2013}). This model, which is recognized as the excess wing (EW) model, is described by the dielectric susceptibility
\begin{equation}\label{eq:ExcessWingModel}
	\hat{\chi}_{\texttiny{EW}}(\iu \omega) = \frac{ 1 + (\tau_2 \iu \omega)^{\alpha}}{1 + (\tau_2 \iu \omega)^{\alpha} + \tau_{1} \iu \omega} .
\end{equation}

\begin{figure}[ht]
\centering
\includegraphics[width=.60\textwidth]{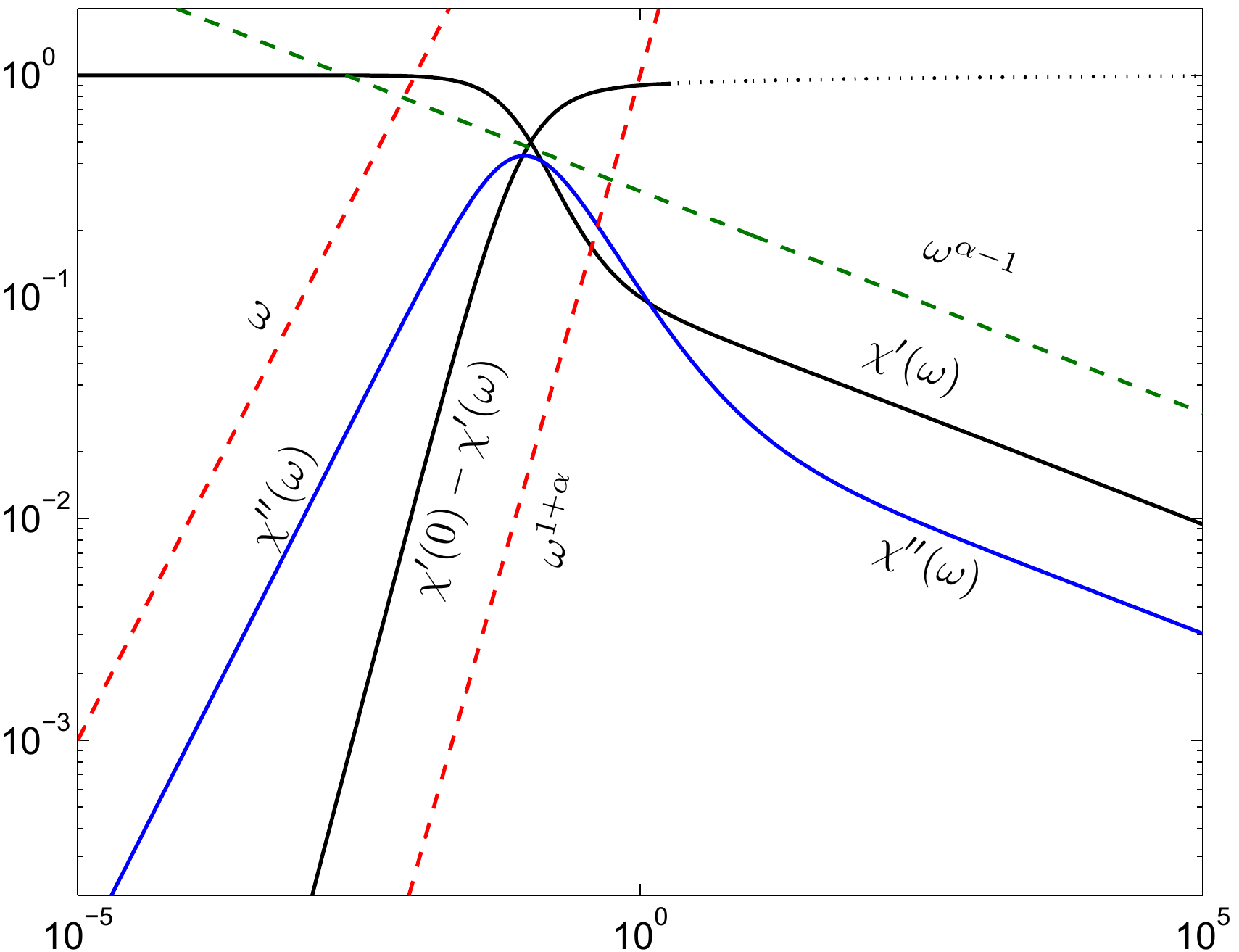}
\caption{EW susceptibility for $\alpha=0.8$, $\tau_1=10$ and $\tau_{2}=1$.}
\label{fig:Chi_URL_EW}
\end{figure}

A similar model was studied in \cite{Mainardi1997,MainardiPironiTampieri1995b,MainardiPironiTampieri1995a} to describe the Basset problem for a sphere accelerating in a viscous fluid. One easily verifies that (\ref{eq:ExcessWingModel}) does not fit the Jonscher's URL since
\[
	\begin{array}{lll}
		\chi'_{\texttiny{EW}}(0) - \chi'_{\texttiny{EW}}(\omega) \sim \tau_{1}\tau_{2}^{\alpha} S_{\alpha} \omega^{\alpha+1}  , \quad &\chi_{\texttiny{EW}}''(\omega) \sim \tau_{1} \omega  , \quad & \omega \tau_1 \ll 1 ,  \\
		\chi'_{\texttiny{JWS}}(\omega) \sim \frac{\tau_{2}^{\alpha}}{\tau_1} S_{\alpha} \omega^{\alpha-1} , \quad &\chi''_{\texttiny{JWS}}(\omega) \sim \frac{\tau_{2}^{\alpha}}{\tau_1} C_{\alpha} \omega^{\alpha-1}  , \quad & \omega \tau_2 \gg 1  ,
	\end{array}
\]
with $C_{\alpha}=\cos(\alpha\pi/2)$ and $S_{\alpha}=\sin(\alpha\pi/2)$, as also illustrated in Figure \ref{fig:Chi_URL_EW} (we assume here $\tau_1 > \tau_2$).

\begin{figure}[ht]
\centering
\includegraphics[width=0.58\textwidth]{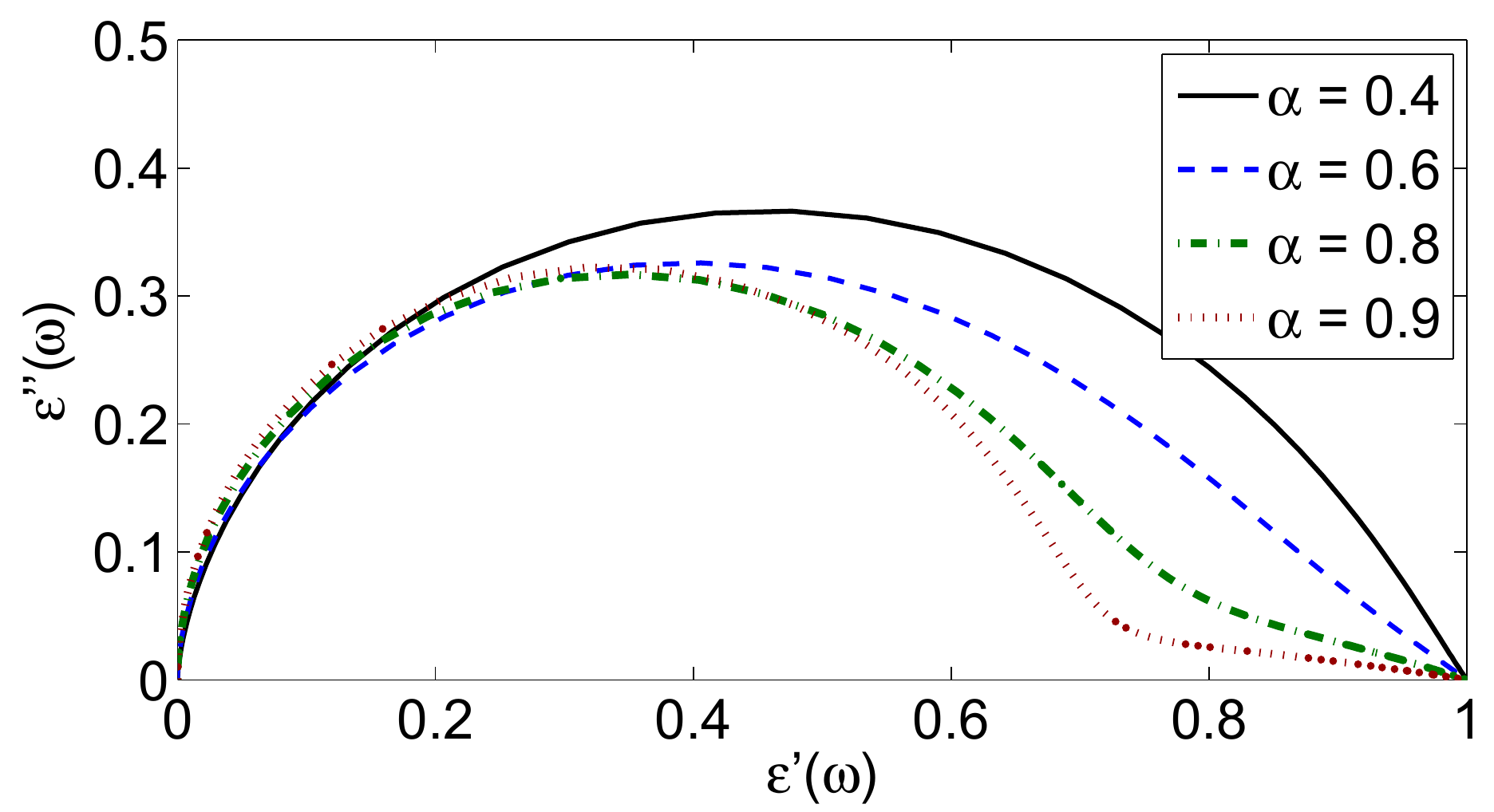}
\caption{Cole-Cole plots for the EW model.}
\label{fig:ColeCole_EW}
\end{figure}

Just for clarity of presentation and to highlight the contribution of both relaxation times, in the Cole-Cole plots of Figure \ref{fig:ColeCole_EW} we have considered two very close relaxation times $\tau_1=2$ and $\tau_2=1$.

The representation of response and relaxation functions by inversion of the Laplace transform poses some additional difficulties. By arranging some terms in different ways, and exploiting expansions of the  form $(1+x)^{-1}=1-x+x^2-x^3+\dots$, it is possible to provide the following (all equivalent) formulations
\begin{eqnarray}
	\frac{ 1 }{1 + (\tau_2 s)^{\alpha} + s \tau_{1}}
	&=& \frac{1}{\tau_1} \sum_{j=0}^{\infty} (-1)^j \left( \frac{\tau_2^{\alpha}}{\tau_1} \right)^j \frac{s^{\alpha j}}{\bigl(s + \tau_{1}^{-1}\bigr)^{j+1}} \\
	&=& \frac{1}{\tau_2^{\alpha}} \sum_{j=0}^{\infty} (-1)^j \left( \frac{\tau_1}{\tau_2^{\alpha}} \right)^j \frac{s^j}{\bigl(s^{\alpha} + \tau_{2}^{-\alpha}\bigr)^{j+1}} \\
	&=& \sum_{j=0}^{\infty} (-1)^j \frac{s^{-\alpha(j+1)}\tau_1^{-(j+1)}}{\bigl(s^{1-\alpha} + \tau_{2}^{\alpha}\tau_{1}^{-1}\bigr)^{j+1}} \
\end{eqnarray}
allowing to represent the inverse Laplace transform in terms of infinite series of ML functions according to
\begin{eqnarray}
	\lefteqn{
	{\mathcal L}^{-1} \left( \frac{ 1 }{1 + (\tau_2 s)^{\alpha} + s \tau_{1}} ; t \right)
	\,=\,  } \hspace{1.2cm} \nonumber & &   \\
	&=& \frac{1}{\tau_1} \sum_{j=0}^{\infty} (-1)^j \left( \frac{t^{1-\alpha}}{\tau_{\star}} \right)^j  E_{1,(1-\alpha) j +1}^{j+1}\left( - t/\tau_1 \right) \label{eq:EW_LT_InverseA} \\
	&=& \frac{1}{\tau_1} \sum_{j=0}^{\infty} (-1)^j \left( \frac{t^{1-\alpha}}{\tau_{\star}} \right)^{-j-1}  E_{\alpha,\alpha(j+1)-j}^{j+1}\left( - \bigl(t / \tau_2\bigr)^{\alpha} \right) \label{eq:EW_LT_InverseB}  \\
	&=& \frac{1}{\tau_1}  \sum_{j=0}^{\infty} (-1)^j \left( \frac{t}{\tau_1} \right)^j
			E_{1-\alpha,j+1}^{j+1}\left( - t^{1-\alpha}/\tau_{\star} \right) , \label{eq:EW_LT_InverseC}  \
\end{eqnarray}
where for notational convenience we put $\tau_{\star} = \tau_1/\tau_2^{\alpha}$ (a further representation in terms of ML function with negative first parameter is also possible but it is of no particular interest).

It is straightforward to provide a representation of the response function $\phi_{\texttiny{EW}}(t)$ and the relaxation function $\Psi_{\texttiny{EW}}(t)$ by inversion of their Laplace transform by using (\ref{eq:Response}) and (\ref{eq:Relaxation}) together with (\ref{eq:EW_LT_InverseA}) and (\ref{eq:EW_LT_InverseC}). However, representations in terms of infinite series of ML functions are clearly of little use especially for practical purposes.

An alternative approach, proposed for the EW model in \cite{CandelaresiHilfer2014} but already explored for similar models in continuum mechanics \cite{MainardiPironiTampieri1995a,MainardiPironiTampieri1995b,Mainardi1997}, consists in first decomposing the dielectric susceptibility in partial fractions and hence proceeding by inversion of the Laplace transform.

By assuming that $\alpha$ is a rational number, i.e. $\alpha = p/q$ with $p,q \in \Nset$, it is possible to express $\widetilde{\chi}_{\texttiny{EW}}(s)$ in terms of the rational function
\begin{equation}
	R(z) = \frac{P(z)}{Q(z)}
	, \quad
	P(z) = 1 + \tau_2^{\alpha} z^{p}
	, \quad
	Q(z) = 1 + \tau_2^{\alpha} z^{p} + \tau_{1} z^{q} ,
\end{equation}
being $\widetilde{\chi}_{\texttiny{EW}}(s) = R\bigl(s^{{1}/{q}})$. Since $\alpha < 1$, it is $0<p<q$ and hence the decomposition in partial fractions of $R(z)$ allows to express $\widetilde{\chi}_{\texttiny{EW}}(s)$ as
\begin{equation}
	\widetilde{\chi}_{\texttiny{EW}}(s) = \sum_{j=1}^{q} \frac{c_{j}}{(s^{\frac{1}{q}}-\lambda_{j})}
\end{equation}
where $\lambda_{j},\lambda_{2},\dots,\lambda_{q}$ are the roots of $Q(z)$ and $c_{1},c_{2},\dots,c_{q}$ the corresponding residues $c_{j} = P(\lambda_{j})/Q'(\lambda_{j})$ (for the sake of simplicity we assume simple roots; the case of multiple roots, which happens very seldom in practice, can be treated with minor changes).

It is now immediate to operate the inversion of the Laplace transform in order to represent the response function in terms of a finite number of ML functions
\begin{equation}
	\phi_{\texttiny{EW}}(t) = {\mathcal L}^{-1} \left( \widetilde{\chi}_{\texttiny{EW}}(s) ; t \right)
	= \sum_{j=1}^{q} c_{j} t^{\frac{1}{q} -1} E_{\frac{1}{q},\frac{1}{q}}\bigl( t^{\frac{1}{q}} \lambda_{j}\bigr) \, .
\end{equation}

In a similar way, since
\[
	\widetilde{\Psi}_{\texttiny{EW}}(s) = \frac{1}{s} - \frac{1}{s} \widetilde{\chi}_{\texttiny{EW}}(s) = \frac{ \tau_1 }{1 + (\tau_2 s)^{\alpha} + s \tau_{1}}
	= \frac{\tau_1}{Q(s^{\frac{1}{q}})},
\]
by putting $d_{j}= \tau_1/Q'(\lambda_{j})$ we obtain the relaxation function
\begin{equation}
	\Psi_{\texttiny{EW}}(t) = {\mathcal L}^{-1} \left( \frac{1}{s} - \frac{1}{s} \hat{\chi}_{\texttiny{EW}}(s) ; t \right)
	= \sum_{j=1}^{q} d_{j} t^{\frac{1}{q} -1} E_{\frac{1}{q},\frac{1}{q}}\bigl( t^{\frac{1}{q}} \lambda_{j}\bigr) \, .
\end{equation}

The roots $\lambda_{j}$ of $Q(z)$ can be real or complex; in the latter case, they however occur in conjugate pairs with corresponding conjugate coefficients $c_{j}$; since $E_{\alpha,\beta}(\bar{z}) = \overline{E_{\alpha,\beta}(z)}$ (as usual, the overbar denotes the complex conjugate), the response and relaxation functions are obviously real-valued functions even if the presence of complex roots
involves a representation as linear combination of complex-valued function.

We present some plots of the relaxation $\Psi_{\texttiny{EW}}(t)$ for different values of $\alpha$ and for the same relaxations times $\tau_1=2$ and $\tau_2=1$ of Figure \ref{fig:ColeCole_EW}.

\begin{figure}[ht]
\centering
\includegraphics[width=0.46\textwidth]{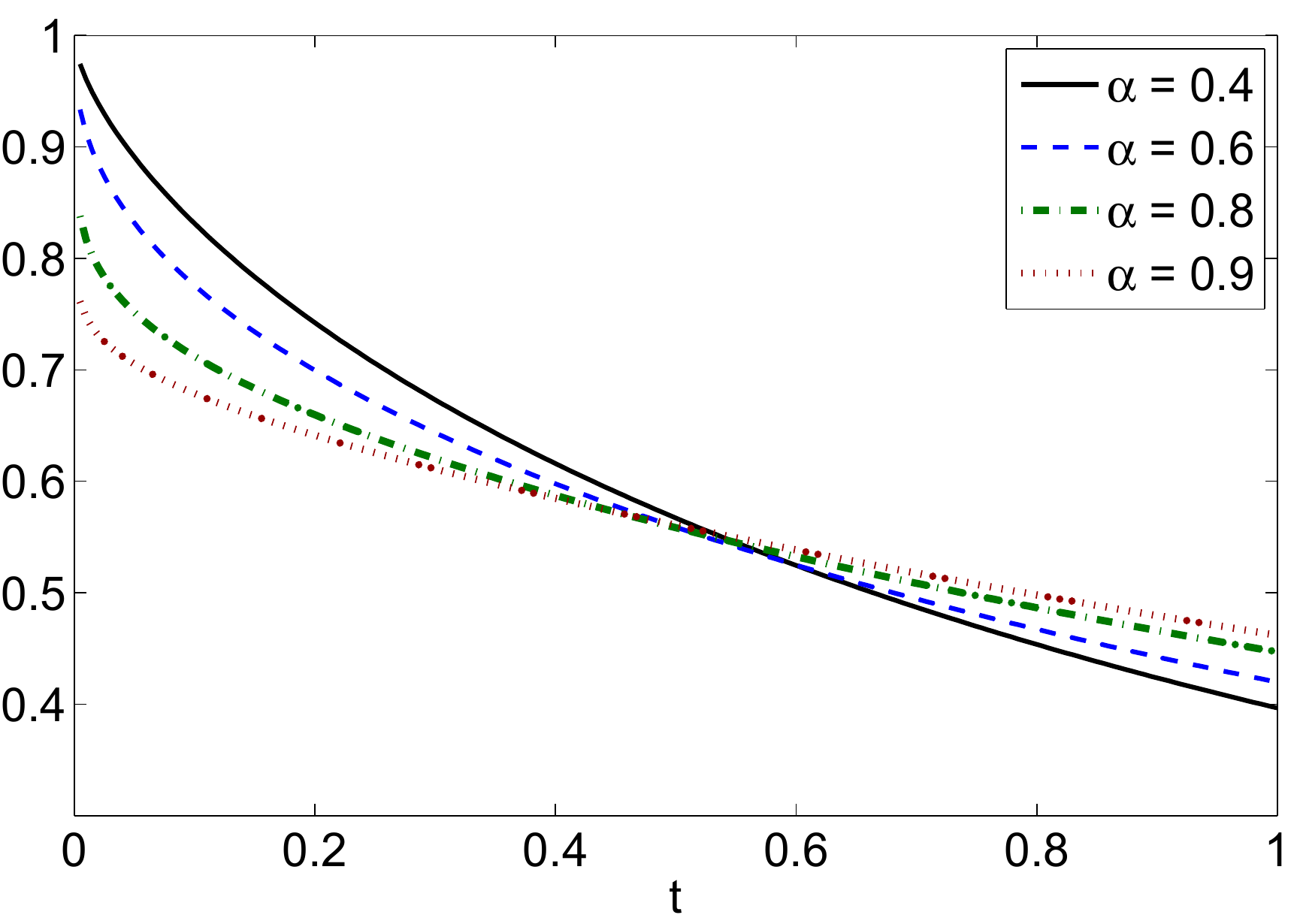}
\includegraphics[width=0.46\textwidth]{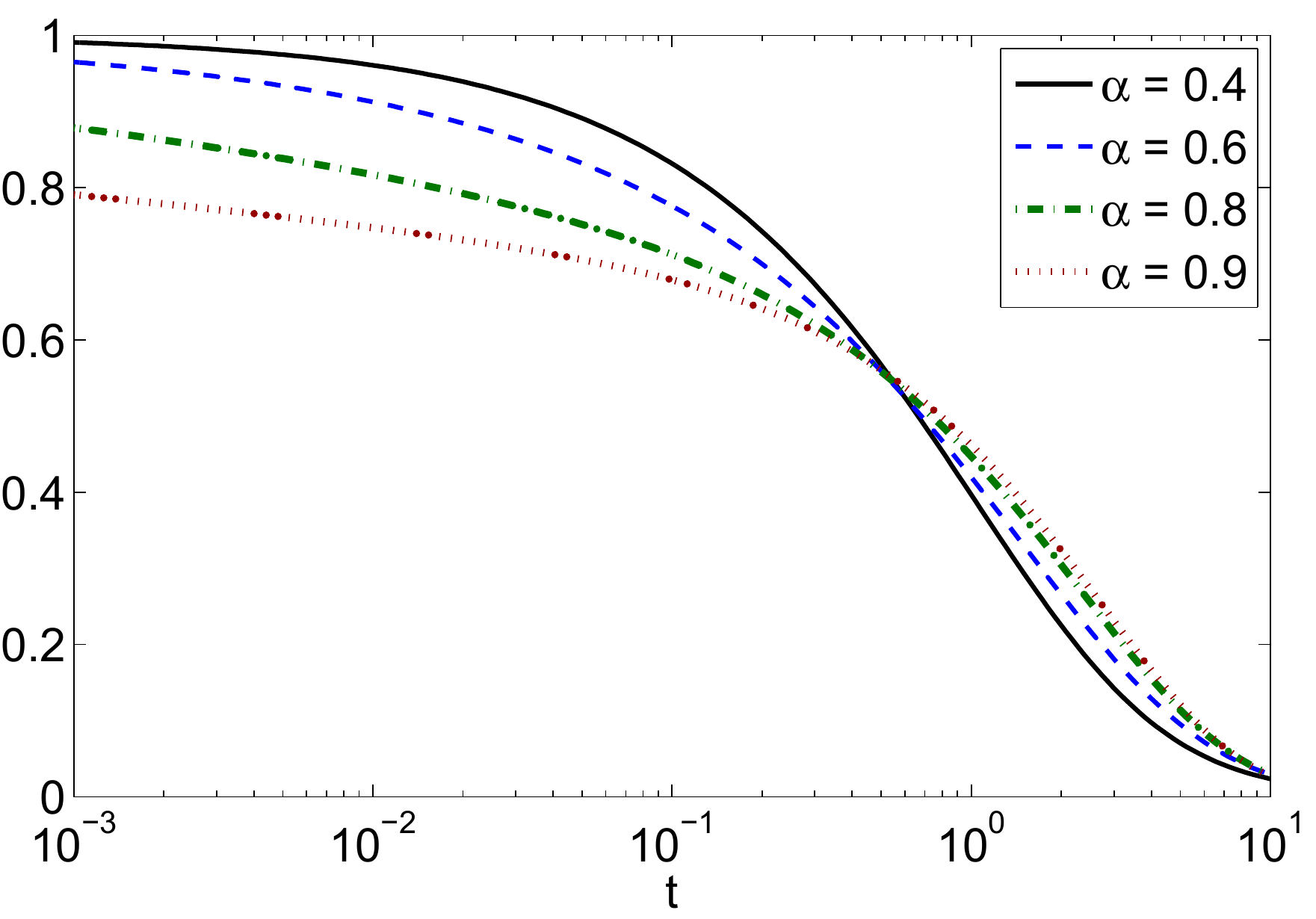}
\caption{Relaxation function $\Psi_{\texttiny{EW}}(t)$ on varying $\alpha$.}
\label{fig:Rel_EW}
\end{figure}

To study the complete monotonicity, we consider the frequency spectral function. To this purpose we observe that, since $\tau_{1},\tau_{2}>0$, $\widetilde{\chi}_{\texttiny{EW}}(s)$ has no poles for $0<\alpha<1$. It is hence possible to use the Titchmarsh formula (\ref{eq:InversionSpectral}) to derive
\begin{equation}
	K^{\phi}_{{\texttiny{EW}}}(r)
	= \frac{1}{\pi} \frac{ \tau_{1} \tau_{2}^{\alpha} r^{\alpha+1} \sin(\alpha \pi) }
				 { (1-\tau_1 r)^2 + 2 (1-\tau_1 r) \tau_{2}^{\alpha} r^{\alpha} \cos \alpha \pi + \tau_{2}^{2 \alpha} r^{2\alpha} }
\end{equation}
and
\begin{equation}
	K^{\Psi}_{{\texttiny{EW}}}(r)
	= \frac{1}{\pi} \frac{ \tau_{1} \tau_{2}^{\alpha} r^{\alpha} \sin( \alpha \pi) }
				 { (1-\tau_1 r)^2 + 2 (1-\tau_1 r) \tau_{2}^{\alpha} r^{\alpha} \cos \alpha \pi + \tau_{2}^{2 \alpha} r^{2\alpha} } \, .
\end{equation}

\begin{figure}[ht]
\centering
\includegraphics[width=0.46\textwidth]{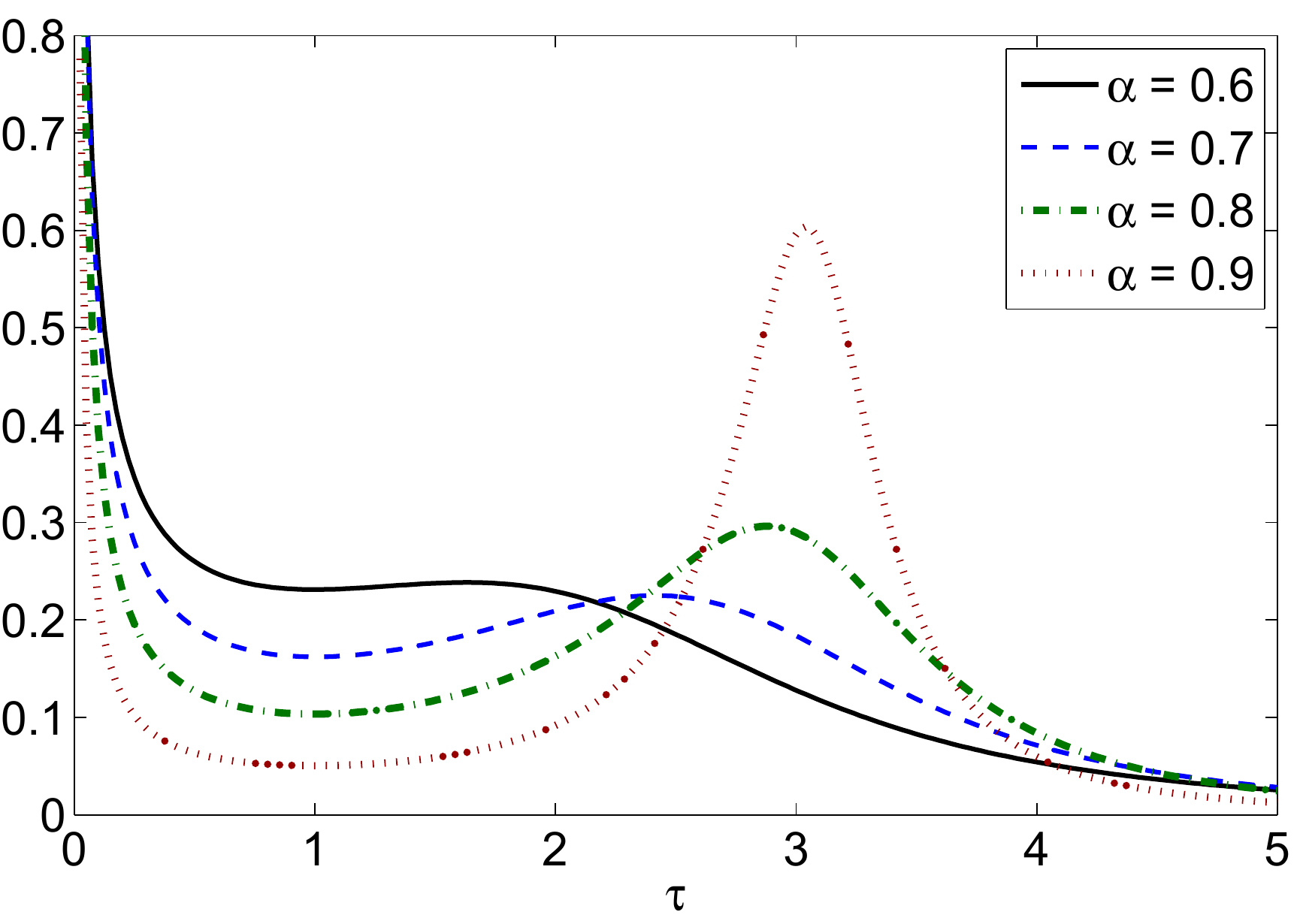}
\includegraphics[width=0.46\textwidth]{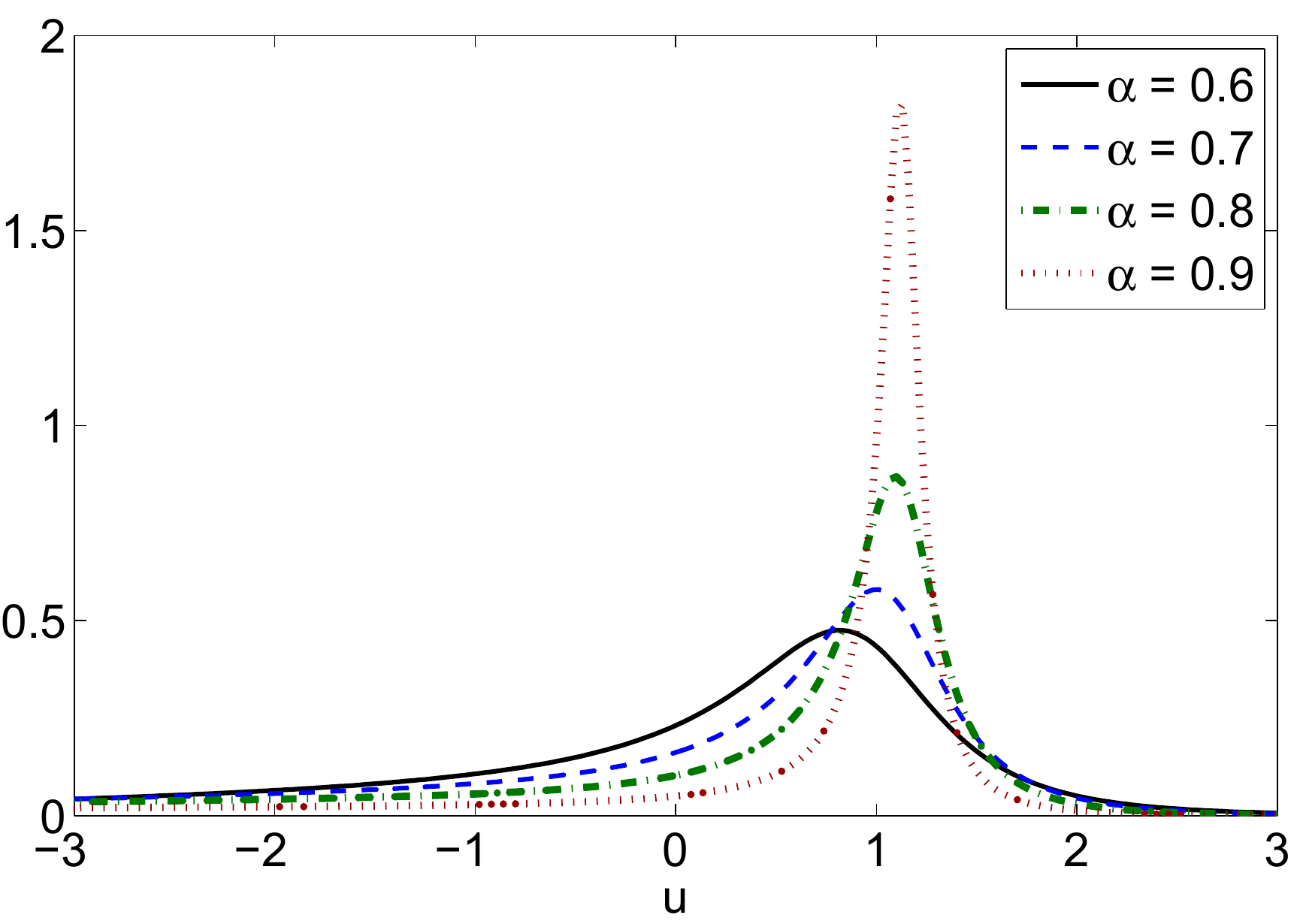}
\caption{Spectral distributions $H^{\Psi}_{\texttiny{EW}}(\tau)$ (left) and $L^{\Psi}_{\texttiny{EW}}(u)$ (right).}
\label{fig:Spectral_EW}
\end{figure}

Since the denominators of $K^{\phi}_{{\texttiny{EW}}}(r)$ and $K^{\Psi}_{{\texttiny{EW}}}(r)$ are always positive, it is immediate to see that $K^{\phi}_{{\texttiny{EW}}}(r)\ge0$ and $K^{\Psi}_{{\texttiny{EW}}}(r)\ge0$ for $0<\alpha\le1$. As usual, the corresponding time spectral functions $H^{\Psi}_{\texttiny{EW}}(\tau)$ and $L^{\Psi}_{\texttiny{EW}}(u)$, obtained from $K^{\Psi}_{{\texttiny{EW}}}(r)$ thanks to (\ref{eq:H_K}) and (\ref{eq:L_K}) are shown in Figure \ref{fig:Spectral_EW}, again for $\tau_1=2$ and $\tau_2=1$.

As already outlined in \cite{Mainardi1997}, thanks to (\ref{eq:K_Psi_K_phi}) the explicit representation of $K^{\phi}_{{\texttiny{EW}}}(r)$ and $K^{\Psi}_{{\texttiny{EW}}}(r)$ allows to define $\phi_{\texttiny{EW}}(t)$ and $\Psi_{\texttiny{EW}}(t)$ in terms of a simple Laplace integral also in the case in which $\alpha$ is not a rational number and without having to deal with an infinite series of of ML functions.

To derive an evolution equation for the EW model, after observing that $\Psi_{\texttiny{EW}}(0)=1$, it is an elementary task to show that the relaxation function $\Psi_{\texttiny{EW}}(t)$ satisfies the multiterm equation with Caputo fractional derivative
\begin{equation}
	D_t \Psi_{\texttiny{EW}}(t) + \frac{\tau_{2}^{\alpha}}{\tau_1} \DerCap{0}{\alpha}{t} \Psi_{\texttiny{EW}}(t) = - \frac{1}{\tau_1} \Psi_{\texttiny{EW}}(t) + \frac{\tau_{2}^{\alpha}}{\tau_1} \frac{t^{-\alpha}}{\Gamma(\alpha)}
	, \, \, \Psi_{\texttiny{EW}}(0) = 1 .
\end{equation}

We finally observe that other models of EW could be obtained from fractional evolution equations of discrete or continuous distributed
order, whose solutions are CM functions, as shown, for example  in \cite{Bazhlekova2015,GorenfloLuchkoStojanovic2013,Kochubei2008,Luchko2011,MainardiMuraGorenfloStojanovic2004}.

\appendix
\renewcommand{\theequation}{\Alph{section}.\arabic{equation}}


\section{Mittag-Leffler functions}\label{S:ML}
\setcounter{equation}{0}
\setcounter{theorem}{0}

The Mittag-Leffler (ML) function is a special function playing a key role in the solution and analysis of fractional differential equations. The first version with just one parameter was introduced in 1902 by the Swedish mathematician Magnus Gustaf Mittag--Leffler \cite{Mittag-Leffler1902} but Wiman \cite{Wiman1905}, few years later, proposed the generalization to two parameters
\begin{equation}\label{eq:ML2}
	 E_{\alpha,\beta} (z) = \sum_{k=0}^{\infty} \frac{z^{k}}{\Gamma(\alpha k + \beta)},
	 \quad z \in \Cset \, .
\end{equation}

In most applications, it is preferable to deal with the Laplace transform of the ML function which has a very simple analytical representation
\begin{equation}
	{\mathcal L} \left( t^{\beta-1} E_{\alpha,\beta}( t^{\alpha} z)   ; s \right) = \frac{s^{\alpha-\beta}}{s^{\alpha} - z}
	, \quad \Re(s)>0 \, \text{ and } \, |s|^{\alpha} > |z|
\end{equation}
highlighting the relationship with the fractional calculus since the presence of fractional powers. For more details, we refer the reader to the recent treatise on functions of the ML type by Gorenflo, Kilbas, Mainardi and Rogosin \cite{GKMR_BOOK2014}.

In 1971, the Indian mathematician Tialk Raj Prabhakar \cite{Prabhakar1971} proposed a further generalization to three parameters of the ML function
\begin{equation}\label{eq:ML3}
	E_{\alpha,\beta}^{\gamma}(z) = \frac{1}{\Gamma(\gamma)} \sum_{k=0}^{\infty} \frac{ \Gamma(\gamma+k) z^{k}}{k! \Gamma(\alpha k + \beta)} .
\end{equation}
and studied integral equations having this function as the kernel. Although Prabhakar considered his work only from a pure and theoretical mathematical point of view, nowadays it is of great importance for the time-domain analysis of the Havriliak-Negami model. Also in this case the Laplace transform has a very simple analytical formulation
\begin{equation}\label{eq:ML3_LT}
	{\mathcal L} \left( t^{\beta-1} E^{\gamma}_{\alpha,\beta}( t^{\alpha} z)   ; s \right) = \frac{s^{\alpha\gamma-\beta}}{(s^{\alpha} - z)^{\gamma}}
	, \quad \Re(s)>0 \, \text{ and } \, |s|^{\alpha} > |z| .
\end{equation}

Important results on the asymptotic behaviour of the standard (one or two parameter) ML function are largely available in the literature (see, for instance, \cite{GKMR_BOOK2014,HauboldMathaiaxena2011,Mainardi2010,Mainardi2014,Paris2002}). Asymptotic expansion of the three parameter ML function are instead less known. An expansion as $t \to +\infty$ has been recently presented  in \cite{MainardiGarrappa2015}
\begin{equation}\label{eq:ML3_exp_large_compl}
	E_{\alpha,\beta}^{\gamma}(-t^{\alpha} ) =
	\left\{ \begin{array}{ll}
		t^{\beta -\alpha \gamma - 1} \displaystyle\sum_{k=0}^{\infty} \binom{-\gamma}{k} \frac{t^{-\alpha k}}{\Gamma(\beta-\alpha\gamma-\alpha k)} \,
		& \text{for } \beta \not=\alpha\gamma \\
		t^{-\alpha \gamma} \displaystyle\sum_{k=1}^{\infty} \binom{-\gamma}{k} \frac{t^{-\alpha k}}{\Gamma(-\alpha k)}
		& \text{for } \beta =\alpha\gamma \
	\end{array} \right.
\end{equation}
and the following asymptotic behaviour can be hence verified
\begin{equation}\label{eq:ML3_exp_large}
	E_{\alpha,\beta}^{\gamma}(-t^{\alpha} ) \sim
	\left\{ \begin{array}{ll}
		\displaystyle\frac{1}{\Gamma(\beta-\alpha \gamma)} t^{-\alpha \gamma} \,
		& \text{for } \beta \not=\alpha\gamma \\
		-\displaystyle\frac{\gamma}{\Gamma(-\alpha)} t^{-\alpha \gamma -\alpha}
		& \text{for } \beta =\alpha\gamma \
	\end{array} \right.
	\quad t \to +\infty
\end{equation}

As a special case (which is of interest in this paper), when $\beta=0$ we have
\begin{equation}\label{eq:ML3_Series0}
	E_{\alpha,0}^{\gamma}(-t^{\alpha} ) = \frac{1}{\Gamma(\gamma)} \sum_{k=1}^{\infty} \frac{\Gamma(\gamma+k) (-1)^k t^{\alpha k}}{k! \Gamma(\alpha k)}
	\sim -\frac{\gamma}{\Gamma(\alpha)} t^{\alpha}
	\quad t \to 0 .
\end{equation}

Results on the complete monotonicity of the Prabhakar function have been recently discussed in \cite{MainardiGarrappa2015,TomovskiPoganySrivastava2014}.

We consider  now another type of three-parameter ML function which differs form the Prabhakar function and has important applications in some fractional differential equations related to phenomena of non standard relaxation studied in this survey paper. These further generalizations  were introduced in 1995 by Kilbas and Saigo for studying the solutions of  non-linear  integral equations of Abel-Volterra type \cite{KilbasSaigo1995a,KilbasSaigo1995b,KilbasSaigo1995c} and are therefore referred to as Kilbas and Saigo functions. The relations between these functions and fractional calculus was presented in \cite{KilbasSaigo1996} and their use for solving, in a closed form, a class of linear differential equations of fractional order was successively discussed in \cite{KilbasSaigo1997,SaigoKilbas2000}. Gorenflo et al. \cite{GorenfloKilbasRogosin1998} presented recurrence relations for these functions and showed the connections  with functions of hypergeometric type for a particular instance of the parameters. The properties of operators in fractional calculus associate with these generalized ML functions were finally investigated in \cite{SaxenaSaigo2005}.

In the complex plane $\mathbb{C}$ we consider the ML type function introduced in \cite{KilbasSaigo1995a} by means of the power series
\begin{equation}
E_{\alpha,m,\ell}(z) =  \sum_{n=0}^{\infty} c_n z^n ,
\,  c_n = \prod_{i=0}^{n-1}
\frac{\Gamma[\alpha(im+\ell)+1]}{\Gamma[\alpha(im+\ell+1)+1]},
\label{1.5}
\end{equation}
with $\alpha,m,\ell \in \mathbb{R}$ such that $\alpha > 0$, $m > 0$ and $\alpha(im+\ell) \neq -1,-2,-3,\ldots$ (an empty product is assumed always equal to one, so that $c_0=1$). Under the above assumptions for the parameters $\alpha$, $m$ and $\ell$, $E_{\alpha,m,\ell}(z)$ can be proved to be an entire function of order $\rho=1/\alpha$ and type $\sigma=m$. As a consequence, for $\epsilon >0$ it is
\begin{equation}  \label {1.6}
 |E_{\alpha,m,\ell}(z)| <
  \exp \left[\left( \frac{1}{m} + \epsilon \right) z^{1/\alpha}\right]\,,
  \quad z \in \mathbb{C}\,.
\end{equation}


\section{Differential operators of non-integer order}\label{S:FractionalIntegralsDerivatives}
\setcounter{equation}{0}
\setcounter{theorem}{0}

In this section, we recall the fractional order operators used throughout the paper. These operators allow to formulate the evolution equations of the various models but also to represent the constitutive law (\ref{eq:ConstitutiveLaw}) in the time domain.

This is obviously not a comprehensive treatment of the subject for which we refer to any of the available textbooks on fractional calculus \cite{Diethelm2010,KilbasSrivastavaTrujillo2006,Mainardi2010,MillerRoss1993,OldhamSpanier1974,Podlubny1999}.

We preliminarily observe that in the following the symbol $*$ will denote the convolution integral between two causal (locally integrable) functions $f(t)$ and $g(t)$, i.e.
$$ f(t) * g(t) \equiv \int_0^t \! f(t-u)\,g(u) \, \du u ,$$
which for classical functions is commutative. Moreover, applying the Laplace transform leads to
$$    f(t) * g(t) = g(t) * f(t) \, \div \, \widetilde f(s) \cdot \widetilde g(s) , $$
where $\div$ denotes the juxtaposition between a time function and its image in the complex  Laplace domain.

\subsection{Riemann-Liouville and Caputo fractional derivatives}

For a casual function $f(t)$ which is assumed absolutely integrable on $\Rset^{+}$, the Riemann-Liouville integral of order $\alpha>0$ is defined as
\begin{equation}\label{eq:RL_Int}
	{}_{0}J_{t}^{\alpha} f(t) \equiv \frac{t^{\alpha-1}}{\Gamma(\alpha)} *  f(t) = \frac{1}{\Gamma(\alpha)}
	\int_{0}^{t} \bigl(t-u\bigr)^{\alpha-1} f(u) \, \du u
	, \quad t \ge 0 ,
\end{equation}

Under the assumption $0<\alpha<1$ (which is reasonable for the models discussed in this paper), the left-inverse of the integral (\ref{eq:RL_Int}) is the Riemann-Liouville fractional derivative
\begin{equation}\label{eq:RL_Der}
	{}_{0} D^{\alpha}_t f(t) = D_{t} \, {}_{0}J_{t}^{1-\alpha}  f(t)
	= \frac{1}{\Gamma(1-\alpha)} \frac{\du}{\du t} \int_{0}^{t} \bigl(t-u\bigr)^{-\alpha} f(u) \, \du u \, .
\end{equation}

An equivalent definition, which is known as the Gr\"{u}nwald-Letnikov derivative, allows to write fractional derivatives by means of fractional differences as
\begin{equation}\label{eq:GL_Der}
	{}_{0}D^{\alpha}_t f(t) = \lim_{h\to0} \frac{1}{h^{\alpha}} \sum_{k=0}^{\infty} \omega_k^{(\alpha)} f(t-kh) ,
\end{equation}
where $h>0$ and $\omega_{k}^{(\alpha)}$'s are the binomial coefficients
\begin{equation}\label{eq:BinomialCoefficients}
	\omega_k^{(\alpha)} = (-1)^{k} \binom{\alpha}{k} = \frac{\alpha(\alpha-1)\cdots(\alpha-k+1)}{k!} \, .
\end{equation}

The interchange of differentiation and integration  in (\ref{eq:RL_Der}) leads to the so-called Caputo  fractional derivative
\begin{equation}\label{eq:C_Der}
	\DerCap{0}{\alpha}{t} f(t) =  {}_{0}J_{t}^{1-\alpha} \, D_{t} f(t)
	= \frac{1}{\Gamma(1-\alpha)}
	\int_{0}^{t} \bigl(t-t'\bigr)^{-\alpha} f'(u) \, \du u
	\, .
\end{equation}

To derive the relationship between RL and Caputo fractional derivatives, it is sufficient to preliminary observe that the Laplace transforms of (\ref{eq:RL_Der}) and  (\ref{eq:C_Der}) are respectively
\begin{equation}\label{eq:LT_RL}
	{\mathcal L} \bigl( {}_{0}D^{\alpha}_t f(t) ; s \bigr) = s^{\alpha} {\mathcal L} \bigl( f(t) ; s \bigr) - \lim_{t \to 0^{+}} {}_{0}J^{1-\alpha}_t f(t)
\end{equation}
and
\begin{equation}\label{eq:LT_C}
	{\mathcal L} \bigl( \DerCap{0}{\alpha}{t} f(t) ; s \bigr) = s^{\alpha} {\mathcal L} \bigl( f(t) ; s \bigr) - s^{\alpha-1} f(0^{+})
	\, ;
\end{equation}
hence, after rewriting
\[
	{\mathcal L} \bigl( \DerCap{0}{\alpha}{t} f(t) ; s \bigr)
	= s^{\alpha} \left( {\mathcal L} \bigl( f(t) ; s \bigr) - \frac{1}{s} f(0^{+}) \right)
	= s^{\alpha} \left( {\mathcal L} \bigl( f(t)-f(0^{+}) ; s \bigr) \right) \, ,
\]
by inverting back to the temporal domain we obtain the well-known relationship
\begin{equation}\label{eq:RL_C_Relat}
	\DerCap{0}{\alpha}{t} f(t) = {}_{0}D^{\alpha}_t \bigl( f(t) - f(0^{+}) \bigr)
\end{equation}
which can be equivalently rewritten as
\begin{equation}\label{eq:RL_C_Relat_bis}
	\DerCap{0}{\alpha}{t} f(t) = {}_{0}D^{\alpha}_t f(t) - \frac{t^{-\alpha}}{\Gamma(1-\alpha)} f(0^{+}) \, .
\end{equation}

These operators, in particular the one of Caputo type, turn out to be useful in order to describe, in the time domain, the constitutive law (\ref{eq:ConstitutiveLaw}) expressing the relationship between the electric and polarization field in some of the discussed models. For instance, for the CC model it is elementary to see that the inversion from the Fourier/Laplace domain, leads to
\begin{equation}\label{eq:CC_ConstLaw_Time}
		\DerCap{0}{\alpha}{t} P_{\texttiny{CC}}(t,x) = - \frac{1}{\tau_{\star}^{\alpha}}P_{\texttiny{CC}}(t,x) + \frac{\Delta \varepsilon}{\tau_{\star}^{\alpha}} E(t,x) , \quad
	P_{\texttiny{CC}}(0,x) = P_{0}(x)  \, ,
\end{equation}
where, for brevity, we denoted $\Delta \varepsilon = \varepsilon_{0}\bigl(\varepsilon_{s}-\varepsilon_{\infty}\bigr)$ and $P(t,x)$ and $E(t,x)$ are the polarization and the electric field respectively, depending also on a space variable $x$. Thanks to the use of the Caputo's derivative, an initial condition of Cauchy type, expressed in terms of a given initial polarization $P_{0}(x)$ at the initial time $t=0$, is coupled to (\ref{eq:CC_ConstLaw_Time}).

Similarly,  for the excess wing model (\ref{eq:ExcessWingModel}) the inversion from the frequency to the time domain does not add particular difficulties because the relationship between the polarization and the electric field can be expressed in terms of the multi-term FDE
\begin{equation*}
	\tau_1 D_t P_{\texttiny{EW}}(t,x) + \tau_{2}^{\alpha} \DerCap{0}{\alpha}{t}  P_{\texttiny{EW}}(t,x) + P_{\texttiny{EW}}(t,x) = \tau_2^{\alpha} \DerCap{0}{\alpha}{t} E(t,x) + E(t,x)
\end{equation*}
in which the standard derivative $D_t$ is combined with the fractional order derivative $\DerCap{0}{\alpha}{t}$. It is not always easy to compute analytic solutions of this equation; however, numerical methods for solving multi-term FDEs are nowadays available (e.g., see \cite{Diethelm2003,DiethelmLuchko2004}).


\subsection{Derivatives of Prabhakar type}\label{SS:OperatorPrabhakar}

Finding suitable differential operators describing, in the time domain, the evolution equations or the constitutive law (\ref{eq:ConstitutiveLaw}) for the DC, HN and JWS models can be less immediate than for CC or EW models.

A heuristic procedure which applies to the HN model (and hence to the special case of the DC model) has been presented by Nigmatullin and Ryabov in \cite{NigmatullinRyabov1997} and successively discussed in \cite{CoffeyKalmykovTitov2006,KalmykovCoffeyCrothersTitov2004}. After introducing the fractional pseudo-differential operator $\bigl({}_{0} D^{\alpha}_{t} + \lambda\bigr)^{\gamma}$ resulting from the inversion of the HN susceptibility (\ref{eq:HN}), by using the binomial expansions, algebra of commutator operators and the Leibiniz formula for the RL derivative, it is possible to reformulate
\begin{equation}\label{eq:HNFracPseudoOp}
	\bigl({}_{0} D^{\alpha}_{t} + \lambda\bigr)^{\gamma}  = \exp \left( - \frac{t\lambda}{\alpha} \, {}_{0}D^{1-\alpha}_t \right) {}_{0}D^{\alpha\gamma}_t \exp \left( \frac{t\lambda}{\alpha} \, {}_{0}D^{1-\alpha}_t  \right) ,
\end{equation}
where the exponentials must be understood as series of factional differential operators. This compound operator is surely helpful for understanding theoretical aspects of the HN model but its practical application for computational purposes appears rather doubtful.

The characterization proposed in (\ref{eq:HNFracPseudoOp}) is however particularly useful, also from the practical point of view, in the special case $\alpha=1$ arising with the DC model because it reduces to
\begin{equation}\label{eq:DC_Operator}
	 \bigl(D_t + \lambda \bigr)^{\gamma} = \eu^{-t\lambda} \, {}_{0}D^{\gamma}_t \, \eu^{t\lambda} \, .
\end{equation}

Hanyga in \cite{Hanyga1999_report,Hanyga1999} proposed the use in (\ref{eq:DC_Operator}) of the Caputo derivative instead of the RL derivative
\begin{equation}\label{eq:DC_Operator_Caputo}
	{}^{{\mathcal C}}{}{\bigl(D_t + \lambda \bigr)^{\gamma}} = \eu^{-t\lambda} \, \DerCap{0}{\gamma}{t} \, \eu^{t\lambda}
\end{equation}
and illustrated the derivations necessary to obtain the Laplace transform of this operator
\begin{equation}\label{eq:DC_Operator_Caputo_LT}
	{\mathcal L} \left( {}^{{\mathcal C}}{}{\bigl(D_t + \lambda \bigr)^{\gamma}} \, ; \, s \right)  = \bigl(s + \lambda\bigr)^{\gamma} \widetilde{f}(s) - \bigl(s + \lambda\bigr)^{\gamma-1}f(0^{+}) \,
\end{equation}
(the reader should note that throughout the paper we use the symbols ``$\mathcal C$'' and ``$\text{C}$'' to distinguish between different ways to regularize fractional operators in the Caputo sense; to this purpose we refer to Remark {\ref{rem:HN_Caputo2}}).

The same approach described in \cite{Hanyga1999_report,Hanyga1999} can be applied, in a straightforward way, to derive also the Laplace transform of (\ref{eq:DC_Operator})
\begin{equation}\label{eq:DC_Operator_LT}
	{\mathcal L} \left( \bigl(D_t + \lambda\bigr)^{\gamma} f(t) \, ; \, s \right)  = \bigl(s + \lambda\bigr)^{\gamma} \widetilde{f}(s) - \lim_{t\to 0^{+}}  {}_{0}J_t^{1-\gamma} \bigl[ \eu^{t\lambda} f(t) \bigr] \, .
\end{equation}

In light of (\ref{eq:RL_C_Relat}) it is possible to verify that the following relationship between (\ref{eq:DC_Operator}) and (\ref{eq:DC_Operator_Caputo}) holds
\begin{equation}
	 {}^{{\mathcal C}}{}{\bigl(D_t + \lambda \bigr)^{\gamma}} f(t) = \bigl(D_t + \lambda \bigr)^{\gamma} \bigl( f(t) - \eu^{-t \lambda} f(0^{+}) \bigr) \, .
\end{equation}

In the time domain, the relationship (\ref{eq:ConstitutiveLaw}) between the electric field and polarization in dielectric of DC type can be therefore expressed as
\begin{equation}
		{}_{0}D^{\gamma}_t \left( e^{t/\tau_{\star}} P_{\texttiny{DC}}(t,x) \right) = \frac{\Delta\varepsilon}{\tau_{\star}^{\gamma}}  e^{t/\tau_{\star}} E(t,x)
\end{equation}
and, as expected, the standard ODE describing the relaxation of Debye type is returned when $\gamma=1$.

In the more general case connected to the HN model, i.e. $\alpha \not=1$, the operator (\ref{eq:HNFracPseudoOp}) seems of little use for computation and presents the same difficulties of the alternative approach proposed in \cite{NovikovWojciechowskiKomkovaThiel2005,WeronJurlewiczMagdziarz_ActaPhysPolB_2005} and consisting in expanding $\left( {}_{0}D^{\alpha}_t  + \tau_{\star}^{-\alpha} \right)^{\gamma}$ by means of an infinite binomial series of fractional RL derivatives
\begin{equation}\label{eq:HN_SeriesExpansion}
	\left( {}_{0}D^{\alpha}_t + \lambda \right)^{\gamma} = \sum_{k=0}^{\infty} \binom{\gamma}{k} \lambda^{k} {}_{0}D^{\alpha(\gamma-k)}_t \, .
\end{equation}

Although the truncation of (\ref{eq:HN_SeriesExpansion}) has been used for numerical computation (see \cite{BiaCaratelliMesciaCicchettiMaionePrudenzano2014}), it presents a major drawback since it is not clear when the above series must be truncated in order to obtain a prescribed accuracy.

An alternative way to introduce operators for the HN model can be devised on the basis of the work presented in \cite{GarraGorenfloPolitoTomovski2014} (successively studied also in \cite{PolitoTomovski2016}) and concerning the so-called Prabhakar integrals and derivatives. These operators are introduced in a similar way as the RL and Caputo operators, after replacing the standard kernel $t^{\alpha-1}/\Gamma(\alpha)$ by the following generalization of the Prabhakar function
\[
	e_{\alpha,\beta}^{\gamma} (t;\lambda) = t^{\beta-1} E_{\alpha,\beta}^{\gamma} (t^{\alpha} \lambda) .
\]

In particular, for a function $f \in L^{1}([0,T])$ the Prabhakar integral of orders $\alpha,\gamma>0$ and parameter $\lambda>0$ can be defined for any $t\in[0,T]$ as
\begin{equation}\label{eq:HN_int}
	\bigl({}_{0}J^{\alpha}_{t} + \lambda\bigr)^{\gamma} f(t) \equiv e_{\alpha,\alpha\gamma}^{\gamma} (t;-\lambda) * f(t)
	= \int_{0}^{t} e_{\alpha,\alpha\gamma}^{\gamma} (t-u;-\lambda) f(u) \, \du u,
\end{equation}
and, since (\ref{eq:ML3_LT}), the corresponding Laplace transform is clearly given by
\[
	{\mathcal L} \left( \bigl({}_{0}J^{\alpha}_{t} + \lambda\bigr)^{\gamma} f(t) \, ; \, s \right)  \equiv \frac{1}{\bigl(s^{\alpha} + \lambda\bigr)^{\gamma}} \widetilde{f}(s) .
\]

Under the assumption $0<\alpha\gamma<1$, the left-inverse of (\ref{eq:HN_int}) is the special derivative
\begin{eqnarray}
	\bigl({}_{0}D^{\alpha}_{t} + \lambda\bigr)^{\gamma} f(t)
	&\equiv& \frac{\du}{\du t} \left(  e_{\alpha,1-\alpha\gamma}^{-\gamma} (t;-\lambda) * f(t)  \right) \nonumber \\
	&=& \frac{\du}{\du t} \int_{0}^{t} e_{\alpha,1-\alpha\gamma}^{-\gamma} (t-u;-\lambda) f(u) \, \du u , \label{eq:HN_der} \
\end{eqnarray}
and it is a simple exercise to verify that the corresponding Laplace transform is
\begin{equation}\label{eq:HN_der_LT}
	{\mathcal L} \left( \bigl({}_{0} D^{\alpha}_{t} + \lambda\bigr)^{\gamma} f(t) \, ; \, s \right)
	= \bigl(s^{\alpha} + \lambda\bigr)^{\gamma} \widetilde{f}(s) - \lim_{t\to0^+} {\mathbf E}_{\alpha,1-\alpha\gamma,-\lambda,0^{+}}^{-\gamma} f(t) ,
\end{equation}
where the operator ${\mathbf E}_{\rho,\mu,\omega,0^{+}}^{\gamma}$, introduced in \cite{GarraGorenfloPolitoTomovski2014}, is the convolution integral whose kernel is $t^{\mu-1}E_{\rho,\mu}^{\gamma}(\omega t^{\rho})$, i.e.
\begin{equation}
	{\mathbf E}_{\rho,\mu,\omega,0^{+}}^{\gamma} f(t) = \int_0^t (t-u)^{\mu-1}E_{\rho,\mu}^{\gamma}(\omega (t-u)^{\rho}) f(u) \, \du u \, .
\end{equation}

We must note that the definition of the derivative $\bigl({}_{0}D^{\alpha}_{t} + \lambda\bigr)^{\gamma}$ on the basis of the integral (\ref{eq:HN_int}) can appear a bit difficult to handle since the presence in the kernel of the Prabhakar function, whose evaluation is usually quite difficult (some methods have been recently proposed in \cite{Garrappa2015,StanislavskyWeron2012}). This kind of definition is however interesting because it allows to introduce a regularization of the same type of the regularization (\ref{eq:RL_C_Relat}) introduced for the Caputo derivative (see also \cite{Kochubei2011}). As proposed in \cite{GarraGorenfloPolitoTomovski2014}, it is indeed possible to exchange integrals and derivatives in (\ref{eq:HN_der}), thus to introduce, for an absolutely continuous function $f$, the operator
\begin{eqnarray}\label{eq:HN_Caputo}
	{}^{\text{\tiny{C}}}{}{\bigl({}_{0}D^{\alpha}_{t} + \lambda \bigr)^{\gamma}}  f(t)
	&\equiv& e_{\alpha,1-\alpha\gamma}^{-\gamma} (t;-\lambda) * \frac{\du}{\du t} f(t)   \nonumber \\
	&=& \int_{0}^{t} e_{\alpha,1-\alpha\gamma}^{-\gamma} (t-u;-\lambda) f'(u) \, \du u ,  \label{eq:HN_der_Cap} \
\end{eqnarray}
where the letter ``C'' indicates that (\ref{eq:HN_Caputo}) can be considered as the counterpart of the Caputo approach for the derivative (\ref{eq:HN_der}). In this case, we observe that in the Laplace transform domain it is
\begin{equation}\label{eq:HN_der_Cap_LT}
	{\mathcal L} \left( {}^{\text{\tiny{C}}}{}{\bigl({}_{0}D^{\alpha}_{t} + \lambda \bigr)^{\gamma}} f(t) \, ; \, s \right)
	= \bigl(s^{\alpha} + \lambda\bigr)^{\gamma} \widetilde{f}(s) - s^{-1}\bigl(s^{\alpha} + \lambda\bigr)^{\gamma}  f(0^+)   				
\end{equation}
and hence by moving back to the temporal domain it is
\begin{equation}
	{}^{\text{\tiny{C}}}{}{\bigl({}_{0}D^{\alpha}_{t} + \lambda \bigr)^{\gamma}} f(t)
	= \bigl({}_{0} D^{\alpha}_{t} + \lambda\bigr)^{\gamma}  f(t) - e_{\alpha,1-\alpha\gamma}^{-\gamma}(t;-\lambda)  f(0^+)
\end{equation}
or, equivalently,
\begin{equation}\label{eq:HN_RL_Caputo_Relat}
	{}^{\text{\tiny{C}}}{}{\bigl({}_{0}D^{\alpha}_{t} + \lambda \bigr)^{\gamma}} f(t)
	= \bigl({}_{0}D^{\alpha}_{t} + \lambda\bigr)^{\gamma} \left( f(t) - f(0^+) \right) ,
\end{equation}
thus establishing relationships between $\bigl({}_{0} D^{\alpha}_{t} + \lambda\bigr)^{\gamma}$ and its regularized version in the Caputo sense ${}^{\text{\tiny{C}}}{}{\bigl({}_{0}D^{\alpha}_{t} + \lambda \bigr)^{\gamma}}$, which are the analogous of the relationships (\ref{eq:RL_C_Relat}) and (\ref{eq:RL_C_Relat_bis}) holding between RL and Caputo derivatives.

\begin{remark}\label{rem:HN_Caputo}
The operator ${}^{\text{\tiny{C}}}{}{\bigl({}_{0}D^{\alpha}_{t} + \lambda \bigr)^{\gamma}}$ cannot be intended as the $\gamma$ power of the Caputo derivative shifted by $\lambda$, namely ${}^{\text{\tiny{C}}}{}{\bigl({}_{0}D^{\alpha}_{t} + \lambda \bigr)^{\gamma}} \not= \bigl(\DerCap{0}{\alpha}{t} + \lambda \bigr)^{\gamma}$. This difference is better perceived in the limit case $\gamma=1$ if we observe, by applying first (\ref{eq:HN_der_Cap_LT}) and hence (\ref{eq:LT_C}), that
\begin{equation}
	{}^{\text{\tiny{C}}}{}{\bigl({}_{0}D^{\alpha}_{t} + \lambda \bigr)} f(t) = \DerCap{0}{\alpha}{t} f(t) + \lambda \bigl( f(t) - f(0^{+})\bigr)
\end{equation}
and hence ${}^{\text{\tiny{C}}}{}{\bigl({}_{0}D^{\alpha}_{t} + \lambda \bigr)} f(t) \not= \bigl(\DerCap{0}{\alpha}{t} + \lambda \bigr) f(t)$. Actually, in light of (\ref{eq:HN_RL_Caputo_Relat}) it is possible to conclude that the regularizing effect of ${}^{\text{\tiny{C}}}{}{\bigl({}_{0}D^{\alpha}_{t} + \lambda \bigr)}$ does not act just on the fractional derivative but also on the identity operator which returns $f(t)-f(0^+)$ instead of $f(t)$. To treat the evolution equation (\ref{eq:CC_EvolEq_Psi}) for the relaxation function of the CC model as the particular case for $\gamma=1$ of the evolution equation (\ref{eq:HN_Evolution_Psi}) of the HN model, it would be necessary to introduce a Caputo regularization affecting only the fractional derivative and not the identity operator as well.
\end{remark}

\begin{remark}\label{rem:HN_Caputo2}
The approach followed by Hanyga in \cite{Hanyga1999_report,Hanyga1999} to regularize (in the Caputo's sense) the operator $\bigl(D_t + \lambda \bigr)^{\gamma}$, and consisting in replacing ${}_{0}D^{\alpha}_t$ by $\DerCap{0}{\alpha}{t}$ in (\ref{eq:DC_Operator}), could be followed, at least hypothetically, also for $\bigl({}_{0} D^{\alpha}_{t} + \lambda\bigr)^{\gamma}$ in (\ref{eq:HNFracPseudoOp}). Both the derivation process and the series expansion of the exponentials would however require very strong assumptions for the functions to which apply the operator, thus severely restricting its feasibility. It is clear that the Caputo's regularization of operators for HN models is still an open problem which deserves further investigation. In this paper, we limit ourselves to denote with different symbols the operators obtained by Hanyga (for which the symbol ``$\mathcal C$'' is indeed used) and the one (\ref{eq:HN_der_Cap}) obtained by interchanging integration and derivation (for which the symbol ``$\text{C}$'' is instead used).
\end{remark}

We also mention that in \cite{Garrappa2016_CNSNS} it has been derived a representation of $\bigl( {}_{0}D^{\alpha}_t + \lambda \bigr)^{\gamma}$ in terms of fractional differences of Gr\"{u}nwald-Letnikov type according to
\begin{equation}\label{eq:HN_GL}
	\bigl( {}_{0}D^{\alpha}_t + \lambda \bigr)^{\gamma} f(t)
	=	\lim_{h\to 0} \frac{(1 + h^{\alpha} \lambda)^{\gamma}}{h^{\alpha \gamma}}	
	\sum_{k=0}^{\infty} \Omega_{k}^{(\alpha,\gamma)} f(t-kh) ,
\end{equation}
where the coefficients $\Omega_{k}^{(\gamma)}$ are given by
\begin{equation}
	\Omega_{0}^{(\alpha,\gamma)} = 1
	, \quad
	\Omega_{k}^{(\alpha,\gamma)}
	= \displaystyle \frac{1}{1 + h^{\alpha} \lambda}
		\sum_{j=1}^{k} \omega_{j}^{(\alpha)} \left( \frac{(1+\gamma)j}{k}-1\right)
				 \Omega_{k-j}^{(\alpha,\gamma)} ,
\end{equation}
with $\omega_{j}^{(\alpha)}$ the binomial coefficients (\ref{eq:BinomialCoefficients}). As observed in \cite{Garrappa2016_CNSNS}, the coefficients $\Omega_{k}^{(\alpha,\gamma)}$ are a generalization of the binomial coefficients $\omega_{k}^{(\alpha)}$ and, indeed, it is $\Omega_{k}^{(\alpha,1)}=\omega_{k}^{(\alpha)}$. Thus, for $\gamma=1$ and $\lambda=0$ the operator (\ref{eq:HN_GL}) gives back the differences (\ref{eq:GL_Der}) and hence it can be considered as a generalization of the Gr\"{u}nwald-Letnikov derivative.

By using the regularized derivative ${}^{\text{\tiny{C}}}{}{\bigl({}_{0}D^{\alpha}_{t} + \lambda \bigr)^{\gamma}}$ it is now possible to express the constitutive law (\ref{eq:ConstitutiveLaw}) for HN models as
\begin{equation}
	{}^{\text{\tiny{C}}}{}{\bigl({}_{0}D^{\alpha}_{t} + \tau_{\star}^{-\alpha} \bigr)^{\gamma}} P_{\texttiny{HN}}(t,x) = \frac{\Delta \varepsilon}{\tau_{\star}^{\alpha \gamma}} E(t,x)
\end{equation}
which can be completed by an initial condition $P_{\texttiny{HN}}(t,x) = P_{0}(x)$. Note that the use of (\ref{eq:HN_GL}), together with (\ref{eq:HN_RL_Caputo_Relat}), provides a tool for the discretization of this equation. A similar approach can be followed for the JWS model and indeed it is easy to evaluate
\begin{equation*}
	{}^{\text{\tiny{C}}}{}{\bigl({}_{0}D^{\alpha}_{t} + \tau_{\star}^{-\alpha} \bigr)^{\gamma}} P_{\texttiny{JWS}}(t,x)
	= \Delta\varepsilon {}^{\text{\tiny{C}}}{}{\bigl({}_{0}D^{\alpha}_{t} + \tau_{\star}^{-\alpha} \bigr)^{\gamma}} E(t,x)
	- \Delta\varepsilon \DerCap{0}{\alpha}{t} E(t,x) \, .
\end{equation*}


\section{Biographical notes}

We conclude this survey by presenting brief biographical notes on some of the authors who distinguished in dielectric studies and introduced the models today named after them. Their names are surely familiar among physicists, chemists, engineers and applied mathematics but in some cases very few is known about them.

\medskip 
{\bf Donald West Davidson} was born in 1925 and died on 2 August 1986 in Ottawa (Canada). He got BSc and MSc degrees at the University of New Brunswick (Canada). During the PhD at the Brown University of Providence in Rhode Island (USA) he conducted studies \cite{DavidsonCole1951_JCP} on dielectric relaxation under the supervision of R.H. Cole. He joined in 1953 the Division of Applied Chemistry at the National Research Council in Ottawa (Canada) where he continued his dielectric studies on molecular motion in liquids \cite{Ripmeester1990}. 

\medskip 
{\bf Petrus (Peter) Josephus Wilhelmus Debye} was born on March 24, 1884 at Maastricht (the Netherlands) and died on November 2, 1966 at Ithaca (USA). He got a degree in electrical engineering in 1905 at the Technische Hochschule in Aachen (Germany) and completed his doctoral program in Munich (Germany) in July 1908. He was appointed as Professor of Theoretical Physics at the University of Zurich in 1911 and at the University of Utrecht in 1912. Successively he worked at the the Physics Institute of G\"{o}ttingen, at the Physics Laboratory of the Eidgen\"{o}ssische Technische Hochschul in Zurich, at  the University of Leipzig, at the Max Planck Institute in Berlin-Dahlem and at the University of Berlin. In 1936 he was awarded of the Nobel Prize in Chemistry ``for his contributions to our knowledge of molecular structure through his investigations on dipole moments and on the diffraction of X-rays and electrons in gases''. He moved to USA in 1940 to became Professor of Chemistry and, later, also chairman of the Department of Chemistry, at the Cornell University at Ithaca (New York, USA) by becoming emeritus in 1950 \cite{Bio_Debye}.

\medskip 
{\bf Kenneth Stewart Cole} was born on July 10, 1900 at Ithaca in New York (USA) and died on April 18, 1984. He studied Physics at Oberlin College in Ohio (USA) and obtained the PhD at the Cornell University (New York, USA) under the supervision of F.K. Richtmyer in 1926. He obtained in 1926 a postdoctoral fellowship by the National Research Council to study the membrane capacity of sea-urchin eggs at Harvard. He joined in 1929 the Department of Physiology of the Columbia University of New York  and in 1946  was appointed as Professor of Biophysics and Physiology and head of the Institute of Radiobiology and Biophysics at the University of Chicago. Successively directed laboratories of the Naval Medical Research and of the National Institutes of Health. Among other academic honours, Dr. Cole received in 1967 the U.S. Medal of Science and was honoured by foreign membership in the Royal Society of London (UK) in 1972 \cite{Bio_ColeKS,Valerio2014}. 

\medskip 
{\bf Robert Hugh Cole} was born on October 26, 1914 in Oberlin, Ohio (USA) and died in Providence, Rhode Island (USA), on November 17, 1990. As his brother Kenneth S. Cole (with which he conducted an intensive collaboration over the years), he graduated (in 1935) at the Oberlin College in Ohio (USA) and, after the PhD earned in 1940 at the Harvard University, he became an Instructor in Physics at the same University. In 1946 R.H. Cole became Associate Professor of Physics at the University of Missouri but one year later he accepted an Associate Professorship at the Brown University where in 1949 assumed the Chairmanship of the Chemistry Department. He received several prestigious awards (among them a Guggenheim Fellowship in 1956 and the Irving Langmuir Prize in 1975) and was appointed as John Howard Appleton Lecturer by the Brown Chemistry Department in 1975 \cite{BioColeRH1985,Valerio2014}. 

\medskip 
{\bf Stephen J. Havriliak} was born on June 30, 1931 and passed away on February 11, 2010. He studied at the American International College of Springfield in Massachusetts (USA) and got his PhD in Chemistry from the Brown University of Providence in Rhode Island (USA). He worked at the Rohm and Haas Research Laboratories of Philadelphia in Pennsylvania (USA). 

\medskip 
{\bf Andrzej Karol Jonscher} was born in Warsaw (Poland) on 1921 and died in London (UK) in 2005. He graduated in Electrical Engineering at Queen Mary College (University of London) in 1949 and there obtained his PhD in 1952. In 1951 he joined the GEC Research Laboratories in Wembley, later named Hirst Research Center, where he worked on physical principles of semiconductor devices. In 1960 appeared his monograph ``Principles of Semiconductors Device Operation''. He joined Chelsea College, University of London, in 1962 as reader and in 1965 Professor of Solid State Electronics in 1965, where interest in amorphous semiconductors gradually led him to studies of the dielectric properties of solids, with special emphasis on the ``universality'' of relaxation processes. In 1983 appeared his monograph “Dielectric Relaxation in Solids” and in 1996 the companion monograph ``Universal Relaxation Law''. Under Jonscher's guidance, the Chelsea Dielectric Group was started in the 1970's at Chelsea College: it grew up and became one of the leading research groups specialising in low frequency response down to milli-Hertz and in the corresponding time-domain behaviour. In 1987 Jonscher became Emeritus Professor at Royal Holloway, University of London, where he has been continuing his work up to his death.

\medskip 
{\bf Friedrich Wilhelm Georg Kohlrausch} was born in Rinteln (Germany) on October 14, 1840 and died in Marburg (Germany) on January 17, 1910. He studied at the university of G\"ottingen where he become professor of physics from 1866 to 1870. Successively he worked at the universities of Darmstadt, W\"urzburg, Strassburg and Berlin \cite{Bio_Kohlrausch}.

\medskip 
{\bf Shinchi Negami} get his B.S. at the Yokohama National University (Japan) in 1957 and moved to USA in 1960 where he received the MS degree from the Lehigh University of Bethlehem in Pennsylvania (USA) after discussing a thesis on ``Dynamic Mechanical Properties of Synovial Fluid''. He worked at the Kent State University in Ohio (USA) and as a research chemist for Rohm and Haas Research Laboratories of Philadelphia in Pennsylvania (USA). 

\vspace*{-5pt}
\section*{Acknowledgements}

The work of RG has been partially supported by the INdAM-GNCS and partially by the COST Action CA15225.
The work of FM has been carried out in the framework of the activities  of the National Group of Mathematical Physics (INdAM-GNFM) and of the Interdepartmental Center ``L. Galvani'' for integrated studies of Bioinformatics, Biophysics and Biocomplexity of the University of Bologna.
The work of GM is supported by the COST Action CA15225.

The authors are deeply indebted to Prof. Andrzej Hanyga, Prof. John Ripmeester and Prof. Karina Weron for providing useful material for the realization of this work and to Prof. Virginia Kiryakova for her useful suggestions and the editorial assistance.


\def\cprime{$'$} \def\cprime{$'$}

\bigskip \smallskip

 \it

 \noindent

$^1$ Roberto Garrappa \\
Dept. of Mathematics, University of Bari \\
Via E.Orabona 4, I -- 70126 Bari, ITALY \\
  e-mail: roberto.garrappa@uniba.it \\ [4pt]

$^2$ Francesco Mainardi \\
Dept. of Physics, University of Bologna\\
Via Irnerio 46, I -- 40126 Bologna, ITALY \\
  e-mail: francesco.mainardi@bo.infn.it \\[4pt]

$^3$ Guido Maione \\
Dept. of Electrical and Information Engineering, Politecnico di Bari\\
Via E.Orabona 4, I -- 70125 Bari, ITALY  \\
  e-mail: guido.maione@poliba.it \\[4pt]	

\hfill 

\end{document}